\documentclass[11pt,a4paper]{article}
\usepackage{jcappub}

\usepackage[utf8]{inputenc}
\usepackage[T1]{fontenc}
\usepackage{aas_macros}
\usepackage{amsmath}
\usepackage{amsfonts}
\usepackage{mathrsfs}
\usepackage{amssymb}
\usepackage{graphicx}
\usepackage{geometry}
\usepackage{hyperref}
\usepackage{xcolor}
\usepackage{float}
\usepackage[title]{appendix}

\newgeometry{vmargin={25mm}, hmargin={20mm}}

\newcommand{\s}[1]{\textrm{#1}}
\newcommand{\st}[1]{\textrm{\tiny#1}}
\newcommand{\mc}[1]{\mathcal{#1}}
\newcommand{\sv}{\langle \sigma_\s{ann} v \rangle}

\newcommand{\citeeq}[1]{Eq.~\eqref{#1}}
\newcommand{\citefig}[1]{Fig.~\ref{#1}}
\newcommand{\citetab}[1]{Tab.~\ref{#1}}
\newcommand{\citesec}[1]{Sec.~\ref{#1}}
\newcommand{\citeapp}[1]{Appendix~\ref{#1}}

\begin{document}
	
	\title{Testing the predictions of axisymmetric distribution functions of galactic dark matter with hydrodynamical simulations}
	
	\author[a]{Mihael Petač,}
	\author[a]{Julien Lavalle,} 
	\author[b,c]{Arturo Núñez-Castiñeyra,}
	\author[b]{Emmanuel Nezri}
	
	\affiliation[a]{Laboratoire Univers et Particules de Montpellier (LUPM), Universit\'e de Montpellier (UMR-5299) \& CNRS, Place Eug\`ene Bataillon, F-34095 Montpellier Cedex 05, France}
	\affiliation[b]{Aix Marseille Univ, CNRS, CNES, Laboratoire d'Astrophysique de Marseille (LAM), 38 rue F. Joliot-Curie, 13388 Marseille Cedex 13, France}
	\affiliation[c]{Université de Paris and Université Paris Saclay, CEA, CNRS, AIM, F-91190 Gif-sur-Yvette, France}
	
	\emailAdd{petac@lupm.in2p3.fr, lavalle@in2p3.fr, arturo.nunez@lam.fr, emmanuel.nezri@lam.fr}
	
	\date{\today}
	
	\abstract{
		Signal predictions for galactic dark matter (DM) searches often rely on assumptions on the DM phase-space distribution function (DF) in halos. This applies to both particle (e.g. $p$-wave suppressed or Sommerfeld-enhanced annihilation, scattering off atoms, etc.) and macroscopic DM candidates (e.g. microlensing of primordial black holes). As experiments and observations improve in precision, better assessing theoretical uncertainties becomes pressing in the prospect of deriving reliable constraints on DM candidates or trustworthy hints for detection. Most reliable predictions of DFs in halos are based on solving the steady-state collisionless Boltzmann equation (e.g. Eddington-like inversions, action-angle methods, etc.) consistently with observational constraints. One can do so starting from maximal symmetries and a minimal set of degrees of freedom, and then increasing complexity. Key issues are then whether adding complexity, which is computationally costy, improves predictions, and if so where to stop. Clues can be obtained by making predictions for zoomed-in hydrodynamical cosmological simulations in which one can access the true (coarse-grained) phase-space information. Here, we test an axisymmetric extension of the Eddington inversion to predict the full DM DF from its density profile and the total gravitational potential of the system. This permits to go beyond spherical symmetry, and is a priori well suited for spiral galaxies. We show that axisymmetry does not necessarily improve over spherical symmetry because the (observationally unconstrained) angular momentum of the DM halo is not generically aligned with the baryonic one. Theoretical errors are similar to those of the Eddington inversion though, at the 10-$20\%$ level for velocity-dependent predictions related to particle DM searches in spiral galaxies. We extensively describe the approach and comment on the results.
	}

	
	\maketitle
	
	\tableofcontents
	
	\section{Introduction}
	\label{sec:intro}
	
	Many searches for dark matter (DM) candidates target processes or phenomena that depend on the DM velocity distribution. This is, for instance, the case for thermally produced cold DM (CDM) candidates \cite{ZeldovichEtAl1975,LeeEtAl1977a,GunnEtAl1978,DolgovEtAl1981,BinetruyEtAl1984a,SrednickiEtAl1988,SteigmanEtAl2012}, like weakly-interacting massive particle (WIMP) DM \cite{PrimackEtAl1988,jungman_supersymmetric_1996,ArcadiEtAl2017,LeaneEtAl2018}, either in searches based on direct detection techniques \cite{goodman_detectability_1985,DrukierEtAl1986,FreeseEtAl2013} or related to the capture of WIMPs in astrophysical bodies \cite{PressEtAl1985,KraussEtAl1985,SalatiEtAl1989,LopesEtAl2021}, or searches based on indirect detection techniques \cite{Bergstroem2000,BringmannEtAl2012c,LavalleEtAl2012}, which may also be sensitive to velocity-dependent processes like $p$-wave-suppressed \cite{McDonaldEtAl2001,EssigEtAl2013a,BoddyEtAl2018,BoudaudEtAl2019a,LiuEtAl2020} or Sommerfeld-enhanced annihilation \cite{HisanoEtAl2004,HisanoEtAl2005,BoddyEtAl2018}. In the former situation, making predictions requires information on the local velocity distribution at the relevant position, while in the latter, one needs to access the velocity distribution function all over the targeted DM halo. There are actually many other DM candidates' signatures for which the velocity distribution matters, for instance, to make detailed predictions of the microlensing event rate induced by primordial-black-hole (PBH) DM \cite{Paczynski1986,Griest1991,green_astrophysical_2017}. All this generically demands for both theoretical and observational ways to infer the velocity distribution of DM in haloes, for which uncertainties should ideally be controlled.
	
	Since there is no firmly established reliable tracer of the DM velocity distribution function in virialised structures (though see some attempts in \cite{LisantiEtAl2012,Herzog-ArbeitmanEtAl2018}), one usually has to predict or constrain it by relying on dynamical consistency, and therefore on solutions to the steady-state collisionless Boltzmann equation applied to gravitational systems. This actually characterizes a minor part of a vast research program dedicated to the understanding of galactic dynamics \cite{binney_galactic_2008}. In this context, although many past studies have often assumed Gaussian (Maxwellian) velocity distributions based on the isothermal approximation \cite{PressEtAl1985,Green2003} (which provide rough order-of-magnitude estimates), it is well established that {\em predictions} in this framework do not reliably compare with the phase-space content of DM structures \cite{binney_galactic_2008,VogelsbergerEtAl2008,lacroix_predicting_2020}. Maxwellian template functions can still be used to {\em fit} velocity distributions in cosmological simulations of galaxies, with adjustable free parameters not grounded from first principles, but this procedure cannot be considered as a physical {\em prediction} per se \cite{LingEtAl2010,BozorgniaEtAl2016}. However, the increasing precision of experimental or observational data relevant to DM searches within the Milky Way or in its neighborhood \cite{Strigari2013} presses for theoretical improvement. In particular, the advent of the Gaia mission \cite{GaiaCollab2016} and its impressive catalog of the Milky Way and its neighbors' stars \cite{GaiaCollab2016a,GaiaCollab2018,GaiaCollab2020} provides us with an unprecedented sample of stars as viewed in phase-space coordinates, which allows us to more strongly constrain the Milky Way DM content and its overall gravitational potential (see e.g.~\cite{CatenaEtAl2010a,McMillan2011,CatenaEtAl2012,Read2014,PifflEtAl2014,FornasaEtAl2014,McMillan2017,CautunEtAl2020,petac_equilibrium_2020} for a series of works showing the growing impact of kinematic data). Yet, many theoretical inference methods are based upon assumptions that are themselves not systematically tested, and that could induce uncontrolled theoretical errors. The subject of this paper is precisely to estimate the theoretical errors associated with some inference proposals.

	A tractable self-consistent theoretical approach is based on the Eddington inversion method \cite{eddington_distribution_1916,Widrow2000,binney_galactic_2008,UllioEtAl2001a,VergadosEtAl2003,FerrerEtAl2013,lacroix_anatomy_2018}, which allows one to translate a constrained DM density profile and a constrained overall gravitational potential (comprising all components of the system, including baryons), into a prediction for the DM phase-space distribution function (PSDF) at each point of the system. This method relies on the assumption of maximal symmetry, i.e. spherical symmetry and an isotropic velocity field. Although it does not apply to arbitrary combinations of DM and baryonic configurations (see \cite{lacroix_anatomy_2018} for a detailed review), it was recently tested against cosmological simulations and shown to be surprisingly reliable given the strong assumptions made \cite{lacroix_predicting_2020}. Theoretical errors were estimated at the $\lesssim 10$-20\% level along most of the radial evolution of the positive or negative $n^{\rm th}$ moments of the velocity and relative velocity distributions, $\langle v^{\pm n}\rangle$ and $\langle v_{\rm rel}^{\pm n}\rangle$. These moments are those relevant in the predictions of velocity-dependent DM signals. This makes this approach rather powerful because poorly demanding in terms of computational time, while reasonably precise.
	
	The Eddington inversion can actually be extended to anisotropic velocity fields \cite{osipkov_spherical_1979,merritt_spherical_1985,Cuddeford1991,BozorgniaEtAl2013,FornasaEtAl2014,lacroix_anatomy_2018}, still under the assumption of spherical symmetry. Other theoretically much finer approaches rely on action-angle methods, which somewhat represents the state-of-the-art developments in the interpretation of kinematic data \cite{BinneyEtAl2015,SandersEtAl2016,PifflEtAl2015}. Still, action-angle methods do not, for the moment, show a decisive improvement over Eddington-like approaches. Moreover, distribution functions for the actions are difficult to derive from first principle and are either rather empirical or based on some calibration procedures \cite{CallinghamEtAl2020,HattoriEtAl2020}. In this paper, we propose to test an extension of the Eddington inversion that allows not only to account for an anisotropic velocity field, but also to deal with axisymmetric systems. Adding more degrees of freedom and going beyond spherical symmetry may somewhat capture more information when applied to axisymmetric systems like spiral galaxies, or systems potentially exhibiting significant oblateness or prolateness. This improved method, based on an inversion similar to the Eddington inversion, was originally proposed in \cite{hunter_two-integral_1993}. It was then further developed and applied to DM searches in \cite{petac_two-integral_2019}. More recently, it was used in a thorough analysis of the Milky Way's kinematic data to get constraints on the DM halo and to make predictions for direct DM searches \cite{petac_equilibrium_2020}. In this paper, we want to test the reliability of this improved inversion itself with a study similar to the one performed in \cite{lacroix_predicting_2020}, by comparing predictions for the velocity distribution derived from knowledge of the matter content of cosmological simulations with the velocity distributions directly measured in the same simulations. We will use zoom-in cosmological simulations of Milky Way analogs presented in Refs.~\cite{mollitor_baryonic_2015} and \cite{nunez-castineyra_cosmological_2021}.
	
	The paper is organised as follows. In \citesec{sec:methods} we first introduce the inversion methods designed for computing the equilibrium PSDFs of collisionless self-gravitating systems. In \citesec{sec:simulations} we present the deployed set of hydrodynamical simulations and provide the details regarding their analysis. \citesec{sec:results} contains our main results concerning the accuracy of the inversion methods in predicting the velocity distribution of DM. In \citesec{sec:searches} we asses the corresponding uncertainties in the predictions for the astrophysical factors that enter direct and indirect DM searches. We conclude in \citesec{sec:conclusions}.
	
	\section{Phase-space structure of DM haloes}
	\label{sec:methods}
	
	The structure of DM haloes can be conveniently described in terms of the PSDF, $f$, which is proportional to the number of single species particles, $N$, in a given volume of the six-dimensional position-velocity space:
	\begin{align}
		\label{eqn:psdf}
		m \cdot dN = f(\textbf{r}, \textbf{v}) \; \s{d}^3r \; \s{d}^3v \,
	\end{align}
	where $\textbf{r}$ and $\textbf{v}$ denote the corresponding position and velocity vectors, while $m$ is the mass of individual particle. In the past, crude estimates of $f$ were often justified by poor observational constraints and the lack of highly accurate experimental searches for signals associated with DM particles. Therefore, the phase-space structure of DM haloes was often approximated by factorizing $f(\textbf{r}, \textbf{v})$ into the spatial DM density distribution, $\rho(r)$, and Maxwell-Boltzmann (MB) velocity distribution, $P_\st{MB}(v)$:
	\begin{align}
	\label{eqn:MB}
	f(\textbf{r}, \textbf{v}) = \rho(r) \cdot P_\st{MB}(v) \, .
	\end{align}
	While the above approximation usually leads to the correct order-of-magnitude estimates for the astrophysical factors that enter the interpretation of DM searches, such a simplifying assumption suffers from a number of drawbacks. To begin with, it is not clear how to appropriately choose the velocity dispersion associated with the MB distribution and several different choices have been suggested throughout the literature -- for their comparison see, e.g.,~\cite{lacroix_anatomy_2018,lacroix_predicting_2020}. Beside the ambiguity in the associated velocity dispersion, the MB velocity distribution predicts non-zero probability for particles with velocities well beyond the typical escape velocity, $v_\s{esc}$, of galaxies. To address this issue, a sharp truncation of $P_\st{MB}(v)$ beyond $v_\s{esc}$ has often been used, however, this still leads to spurious results for probes that are particularly sensitive to the high-velocity tail of the distribution. Finally, even if correct velocity dispersion and escape velocity are chosen, the ansatz from \citeeq{eqn:MB} represents a stationary solution of the Boltzmann equation only if the associated DM density profile is the one of the \textit{singular isothermal sphere}, i.e. $\rho(r) \propto r^{-2}$, and there is no additional contributions to the total gravitational potential of the system. However, the presence of baryons together with various astronomical observations, as well as cosmological simulations, which provide strong evidence that the DM density slope is shallower in the centres of galaxies and steeper in the outskirts, generically imply a deviation from the simple Maxwellian velocity distribution.
	
	The above-described shortcomings of models based on the MB velocity distribution, as well as the availability of increasingly precise observations that can be used to constrain the dynamical structure of galaxies, motivate us to go beyond the simplest approximations. In the following we will describe two alternative models that are by construction equilibrium solutions to the Boltzmann equation, but differ in the assumed level of symmetry. Firstly, we will take a closer look at the spherically symmetric Eddington inversion and briefly mention its anisotropic generalizations. This will be followed by a short review of the axisymmetric inversion method, that has been applied in the context of DM only recently, but offers a more accurate approach for modelling the phase-space distribution of DM in disc galaxies, which are characterised by their axisymmetric morphology.

	\subsection{Eddington inversion}
	\label{sec:methods_edd}
	
	It has long been recognised that it is possible to obtain explicit solutions of the collisionless Boltzmann equation for self-gravitating systems with maximal symmetry. More precisely, Eddington showed that for a collection of collisionless particles with a given radial density profile, $\rho(r)$, embedded in a spherical gravitational potential, $\Psi(r)$,~\footnote{Throughout this work $\Psi(\textbf{r})$ denotes the \textit{relative gravitational potential} that is defined as $\Psi(\textbf{r}) \equiv -\Phi(\textbf{r}) + \Phi_0$, where $\Phi(\textbf{r})$ is the standard gravitational potential and $\Phi_0$ a constant such that $\Psi(\textbf{r})$ vanishes at the boundary of the system. In the following we focus on isolated systems for which $\Phi_0 = 0$. On the other hand, for truncated objects \citeeq{eqn:psdf_edd}, as well as \citeeq{eqn:psdf_axi_even} and \citeeq{eqn:psdf_axi_odd}, contain additional boundary term -- for more detailed discussion see, e.g.,~\cite{lacroix_anatomy_2018,petac_velocity-dependent_2018}.} an unique spherically symmetric and isotropic PSDF can be computed as follows~\cite{eddington_distribution_1916}:
	\begin{align}\label{eqn:psdf_edd}
		f(\mc{E}) = \frac{1}{\sqrt{8} \pi^2} \cdot \int_0^\mc{E} \frac{\s{d} \Psi}{\sqrt{\mc{E} - \Psi}} \cdot \frac{\s{d}^2 \rho}{\s{d} \Psi^2} \; .
	\end{align}
	In the above expression, $\mc{E} \equiv \Psi(r) - v^2/2$ is the \textit{relative energy} that fully parametrizes the class of PSDFs describing such systems. While the Eddington inversion allows one to obtain a DM phase-space distribution consistent with the presence of several galactic components that enter the total gravitational potential, it is limited to spherically symmetric configurations. Furthermore, in its original formulation it is restricted to ergodic, and hence isotropic, distribution functions. There exist several generalizations to anisotropic systems, in which case the PSDF additionally depends on the magnitude of angular momentum, $|\textbf{L}|$, i.e. $f \equiv f(\mc{E}, |\textbf{L}|)$. However, the existing extensions rely on relatively strong assumptions regarding the velocity anisotropy profile -- the latter is either assumed to be a constant~\cite{binney_galactic_2008}, have the Osipkov-Merritt form~\cite{osipkov_spherical_1979,merritt_spherical_1985} or a combination thereof~\cite{Cuddeford1991,wojtak_distribution_2008,BozorgniaEtAl2013}). Furthermore, the anisotropy profile has to be specified a priori, but is in practice impossible to constrain from the existing observations. An additional problem of such anisotropic generalizations of the Eddington approach is the fact that they often lead to unphysical solutions, as was recently investigated in~\cite{lacroix_anatomy_2018}. Therefore for the purpose of this work, we focus only on the original Eddington's formulation of the inversion method.
	
	\subsection{Axisymmetric inversion method}
	\label{sec:methods_aim}
	
	By assuming that the system under consideration is axisymmetric, the corresponding PSDF can be written as a function of two integrals of motion, namely $f = f(\mc{E}, L_z)$, where $\mc{E}$ is the aforementioned relative energy and $L_z$ is the specific angular momentum around the axis of symmetry, i.e. $L_z \equiv R v_\phi$, where $\{R, \phi, z\}$ denote to the standard set of cylindrical coordinates. In this case, a generalization of the Eddington inversion formula can be obtained, allowing one to compute $f(\mc{E}, L_z)$ for an arbitrary axisymmetric density-potential pair. In particular, we will adopt the numerical approach developed by Hunter \& Qian~\cite{hunter_two-integral_1993,qian_axisymmetric_1995}, that relies on theoretical foundations previously laid out by Lynden-Bell~\cite{lynden-bell_stellar_1962}. Until recently, the method was applied only to stellar systems, while it was for the first time systematically studied in the context of DM in~\cite{petac_two-integral_2019}. We refer the reader to the above references for the proof of the method and detailed discussions regarding its numerical implementation.~\footnote{In this work we evaluate the axisymmetric PSDF using our numerical implementation, which is publicly available at \url{https://github.com/mpetac/AIM}} In the remainder of this section, we will provide a short review of the method.
	
	Under the assumptions stated above, the PSDF can be decomposed in two parts, $f_+$ that is even in $L_z$ and $f_-$ that is odd:
	\begin{align}
	f(\mc{E}, L_z) = f_+(\mc{E}, |L_z|) + f_-(\mc{E}, L_z) \; ,
	\end{align}
	The even part contains all the information regarding the density distribution, while the odd part describes the rotational properties of the considered system. Hunter \& Qian~\cite{hunter_two-integral_1993,qian_axisymmetric_1995} showed that the $L_z$-even part of PSDF can be computed by providing an analytic continuation of the density-potential pair in the complex plane and evaluating the following contour integral:
	\begin{align} \label{eqn:psdf_axi_even}
	f_+(\mc{E}, |L_z|) = \frac{1}{4 \pi^2 i \sqrt{2}} \oint_{C(\mc{E})} \frac{\s{d} \xi}{\sqrt{\xi - \mc{E}}} \left. \frac{\s{d}^2 \rho(R^2, \Psi)}{\s{d} \Psi^2} \right|_{\substack{\Psi = \xi \;\;\;\;\;\;\;\;\;\; \\ R^2=\frac{L_z^2}{2(\xi - \mc{E})}}} \; .
	\end{align}    
	In the above expression $C(\mc{E})$ refers to a path which tightly wraps around the real axis between the value of the potential at spatial infinity and a value corresponding to a circular orbit with relative energy $\mc{E}$. Additionally, $\rho$ is considered as a function of the radial coordinate and the total gravitational potential, which is in principle always possible for monotonic $\Psi(R^2 ,z^2)$. However, in the great majority of practical situations one cannot express the density profile as an explicit function of the total gravitational potential and one is forced to perform the derivative implicitly, using the $z$-coordinate:
	\begin{align} \label{eqn:derivative_expansion}
	\frac{\s{d}^2 \rho(R^2, \Psi)}{\s{d} \Psi^2} & =  \frac{\s{d}^2 \rho(R^2, z^2)}{\s{d} (z^2)^2} \left(\frac{\s{d} \Psi(R^2, z^2)}{\s{d} z^2} \right)^{-2} \nonumber \\ & \;\;\;\;\;\;\;\;\;\;
	- \frac{\s{d} \rho(R^2, z^2)}{\s{d} z^2} \frac{\s{d}^2 \Psi(R^2, z^2)}{\s{d} (z^2)^2}\left(\frac{\s{d} \Psi(R^2, z^2)}{\s{d} z^2} \right)^{-3} \; ,
	\end{align}
	and evaluate it at $z^2$ such that $\Psi(R^2, z^2) = \xi$. Values of $z^2$ fulfilling the latter equality typically need to be found via numerical minimization routines. The $L_z$-odd part of PSDF can be computed analogously, using the following expression:
	\begin{align} \label{eqn:psdf_axi_odd}
	f_-(\mc{E}, L_z) = \frac{\textrm{sign}(L_z)}{8 \pi^2 i} \oint_{C(\mc{E})} \frac{\s{d} \xi}{{\xi - \mc{E}}} \left. \frac{\s{d}^2 \left( \rho \bar{v}_\phi \right)}{\s{d} \Psi^2} \right|_{\substack{\Psi = \xi \;\;\;\;\;\;\;\;\;\; \\ R^2=\frac{L_z^2}{2(\xi - \mc{E})}}} \; .
	\end{align}
	It is important to note that, in order to evaluate $f_-$ one needs to specify also the rotation profile, $\bar{v}_\phi(R^2,z^2)$, which is in the case of DM haloes essentially unconstrained by the observations~\footnote{There have been several studies that find correlations between the spin parameter and the environment, as well as merger history, of individual DM haloes~\cite{bullock_universal_2001,sharma_angular_2005,teklu_connecting_2015,zavala_link_2016,zjupa_angular_2017}. However, such information can be used only to provide indirect constraint on $\bar{v}_\phi$, since the spin parameter is related merely to the total angular momentum of the halo.}. However, as will be discussed in \citesec{sec:results_velocity}, numerical simulations show that $\bar{v}_\phi$ is typically much smaller than the velocity dispersion in the azimuthal direction and hence $f_-$ provides only a subdominant contribution to $f$. For this reason, as well as possible issues with obtaining physical solutions, we will in the following consider only the $\bar{v}_\phi(R^2, z^2) = 0$ case.

	\section{Simulations}
	\label{sec:simulations}
	
	Within the past decades, we have witnessed rapid improvements in the capabilities of simulating the formation and evolution of galaxies. While the initial studies were exclusively focused on the gravitational dynamic of collisionless DM particles, subsequent works began including baryonic physics with ever-increasing accuracy. This gave rise to the so-called \textit{hydrodynamical simulations}, which treat gas as a fluid, while the corresponding sub-grid physics is handled by a number of prescriptions that are carefully calibrated against various observational constraints.
	
	Today, sophisticated algorithms coupled with modern computational capabilities allow us to simulate galaxies with unprecedented realism and address questions which were until recently inaccessible to existing methods. These improvements are of great importance also for studies of the DM phase-space distribution within spiral galaxies, which we address in this work. In order to adequately resolve the full six-dimensional distribution of DM in the position-velocity space, one needs to track millions of individual particles that self-consistently evolve in the presence of baryons. Despite the aforementioned improvements, a satisfying resolution can only be reached by performing dedicated simulations focused on a single object, commonly referred to as zoom-in simulations. The aim of this work is to carry out a detailed comparison of the DM phase-space distribution as predicted by the inversion methods, described in the previous section, with the results of the high-resolution zoom-in simulations. In particular, we first extract the quantities that can be, at least in principle, inferred from astronomical observations and, subsequently, obtain the corresponding phase-space distribution of DM using the Eddington and axisymmetric inversion methods. Finally, the predicted DM phase-space distributions can be compared with the ones directly extracted from the simulated objects. This allows us to assess the systematic uncertainties induced by the assumptions upon which the inversion methods are built, namely the hypothesis of dynamical equilibrium and the symmetries of the system.
	
	In the following, we first give a brief description of the deployed set of hydrodynamical simulations. Subsequently, we provide a detailed discussion regarding our approach of extracting the baryonic gravitational potentials and the DM density profiles of the simulated objects, as they represent the crucial ingredients for reconstructing the corresponding PSDFs via the inversion methods.
	
	\subsection{Simulation setup}
	\label{sec:simulations_setup}
	
	We perform the outlined comparison of the inversion methods with the simulations on three distinct hydrodynamical runs (the same as used in \cite{lacroix_predicting_2020}), which tracked the formation and evolution of individual spiral galaxies. All three runs were performed using the Eulerian hydrodynamical adaptive mesh code RAMSES~\cite{teyssier_cosmological_2002}, while their initial conditions were randomly generated using the MUSIC package~\cite{hahn_multi-scale_2011} for a similar set of fiducial cosmological parameters. Furthermore, all three simulations used the same prescriptions for the sub-grid baryonic physics, though with one of the run with improved spatial resolution, hence a slightly different tuning of the subgrid control parameters~\cite{lacroix_predicting_2020}.
	
	Our main benchmark simulation is dubbed ``Mochima'' and has been originally introduced in~\cite{nunez-castineyra_cosmological_2021}, where it served as the control run for comparing different implementations of baryonic physics. It possesses slightly higher resolution than the other two runs, which we refer to as ``Halo B'' and ``Halo C'', that have been presented and thoroughly studied in~\cite{mollitor_baryonic_2015}. As previously mentioned, all three simulations rely on identical baryonic prescriptions. For gas they use the conventional cooling, ultraviolet background and self-shielding recipes, while the star formation is modelled using the Schmidt law with adapted density threshold and efficiency. The supernova (SN) feedback is implemented according to the ``delayed cooling'' prescription, which relies on non-thermal injection of energy $\sim 10$ Myr after the birth of star particles, generated with a Chabrier initial mass function, with the energy of $10^{51}$ erg per SN event. Further details regarding the implementation of baryonic physics in Mochima and Halo B \& C can be found in~\cite{nunez-castineyra_cosmological_2021} and~\cite{mollitor_baryonic_2015}, respectively.
	
	Within the Mochima simulation, the galaxy lies at the centre of 36 Mpc cosmological box with an approximate resolution (estimated by the minimal cell size) of $l_\s{sm} \sim 35$ pc. The masses of individual DM and star particles are $M_\st{DM} = 1.95 \cdot 10^5 M_\odot$ and $M_\star = 1.57 \cdot 10^4 M_\odot$, respectively. On the other hand, Halo B and Halo C both implement a 20 Mpc cosmological box with approximate resolution of $l_\s{sm} \approx 150$ pc, while the masses of individual DM and star particles are $M_\st{DM} = 2.31 \cdot 10^5 M_\odot$ and $M_\star = 2.87 \cdot 10^4 M_\odot$, respectively. Apart from the resolution, the galactic DM halo found in the Mochima simulation notably differs from the ones in Halo B and Halo C by having a steep central cusp, while the other two runs exhibit cores. This is presumably due to a weaker SN feedback~\cite{lacroix_predicting_2020}, which is a consequence of tuning the free parameters of the delayed cooling prescription to the values appropriate for the higher resolution of the Mochima simulation~\cite{onorbe_how_2014,nunez-castineyra_cosmological_2021}. On the other hand, all three runs exhibit a contraction of the DM halo around the disc scale length in comparison with the corresponding DM-only simulations. Apart from the above properties, all three hydrodynamical runs resulted in DM haloes that are to a good approximation spherically symmetric and close to dynamical equilibrium -- for quantitative analysis see~\cite{nunez-castineyra_cosmological_2021}. Finally, it is worth noting that in the case of Halo B there is a significant, $\mc{O}(1 \, \s{kpc})$, displacement between the centres of baryons and the gravitational potential~\cite{mollitor_baryonic_2015,nunez-castineyra_cosmological_2021}.

	\subsection{Parametric fits of the galactic components}
	\label{sec:simulations_fitting}
	
	The predictive power of inversion methods crucially depends on accurate measurements of the relevant input quantities. As explained in \citesec{sec:methods}, these consist of the total gravitational potential of the system, $\Psi$, and density distribution of DM, $\rho$, but possibly also other physical quantities, such as DM's velocity anisotropy in the case of anisotropic generalizations of Eddington approach or DM's rotational profile in the case of axisymmetric inversion method. In practice, accurate determinations of $\rho$ and $\Psi$ are limited by observational uncertainties, however, since the main purpose of this work is to assess the systematic uncertainties that arise due to the modelling assumptions of the inversion methods, we neglect the observational errors and fit the gravitational potential and DM density profile directly to the particle grids produced by the simulations.
	From the observational perspective, the situation is even worse for the orbital anisotropy and rotational profile of the DM halo -- these can not be observed through the existing probes and hence represent irreducible sources of systematic uncertainties. While we have chosen not to study anisotropic generalizations of the Eddington approach due to the several associated issues mentioned in \citesec{sec:methods_edd}, we have explicitly checked that rotational profiles of the DM haloes provide a negligible correction with respect to the non-rotating case. Therefore, we restrain from performing an extensive study of the systematics which arise from unobservable anisotropy and rotational profiles.
	
	In the following we specify the assumed parametric forms for the physical quantities that enter Eddington and axisymmetric inversion methods, as well as describe how they were extracted from the simulations. For the purpose of this work, we decompose the total gravitational potential into baryonic and DM component. As the latter is in one-to-one correspondence with the DM density distribution, we first turn our attention to the gravitational potential sourced by the baryons only, and subsequently address the DM density distribution.
	
	\subsubsection{Baryonic gravitational potential}
	
	The distribution of baryons within galaxies can be modelled in a number of different ways, which do not necessarily lead to equivalent results. For example, stellar and gas components are often treated separately since they can be constrained by different sets of observations. Moreover, the stars can be further subdivided in distinct classes, depending on their spatial distribution, kinematical properties and/or spectra. For each of these constituents special observationally-motivated density distributions have been suggested throughout the literature, e.g. truncated triaxial distributions for the bulge and various double exponential distributions for different disc components (for comprehensive overview see, e.g.,~\cite{binney_galactic_2008}), whose gravitational potentials typically have to be computed through numerical quadrature. However, for the purpose of obtaining the phase-space distribution of DM through the inversion methods, one is interested only in the overall baryonic gravitational potential. Therefore, we in this work resort to a simpler approach, where the total baryonic potential is approximated by a combination of two analytic functions -- these greatly reduce the computational cost of performing the inversions and have a well-defined continuation in the complex plane, which is required by the axisymmetric method. In particular, to accommodate the central bulge we adopt a spherical Hernquist potential~\cite{hernquist_analytical_1990}:
	\begin{align}
		\Psi_\st{H}(R, z) = \frac{G M_\st{H}}{\sqrt{R^2 + z^2} + a_\st{H}} \; ,
	\end{align}
	while the disc is approximated by a single Miyamoto-Nagai~\cite{miyamoto_three-dimensional_1975} (MN) component:
	\begin{align}
		\label{eqn:MN}
		\Psi_\st{MN}(R,z) = \frac{G M_\st{MN}}{\sqrt{R^2 + \left(a_\st{MN} + \sqrt{z^2 + b_\st{MN}^2}\right)^2}} \; .
	\end{align}
	Even though the above model is much simpler than separately inferring the density distributions of multiple disc components, we in the following demonstrate that it performs very well in fitting the over all baryonic gravitational potential. For the purpose of Eddington inversion, however, the above axisymmetric ansatz needs to be converted to a spherically symmetric approximation, $\Psi_\s{sph}(r)$. We achieve this by demanding that the corresponding rotation curve, which is typically the most constraining observation regarding the galactic potential, remains unchanged:
	\begin{align}
		\label{eqn:potential_sph}
		\frac{\partial \Psi_\s{sph}(r)}{\partial r} \equiv \left. \left( \frac{\partial \Psi_\st{H}(R,z)}{\partial R} + \frac{\partial \Psi_\st{MN}(R,z)}{\partial R} \right) \right|_{\substack{R = r \\ z=0}} \,  .
	\end{align}
	
	The first step in our analysis is to match the aforementioned sum of Hernquist and MN potentials to the baryonic gravitational potential inferred from the simulations. The latter is computed as a sum of Keplerian potentials of the individual baryonic particles:~\footnote{We explicitly checked that the method is numerically stable, namely that the result does not change upon removing the particle with the smallest $|\boldsymbol{r} - \boldsymbol{r}_k|$.}
	\begin{align}
		\Psi_\s{sim}(\boldsymbol{r}) = G \sum_{k \, \in \, \s{stars}, \s{gas}} \frac{m_k}{|\boldsymbol{r} - \boldsymbol{r}_k|} \; ,
	\end{align}
	where $G$ is the gravitational constant, while $m_k$ and $\boldsymbol{r}_k$ are the mass and position vector of the $k^{\rm th}$ gas cell or stellar particle.
	However, the above expression does not automatically lead to an axisymmetric potential, which is required by the adopted parametrization. Therefore, $\Psi_\s{sim}(\boldsymbol{r})$ has to be averaged over the azimuthal angle, $\phi$, to obtain the average axisymmetric potential $\bar{\Psi}_\s{sim}(R, z)$, while for the purpose of fitting we also compute the corresponding standard deviation $\sigma_{\bar{\Psi}}(R, z)$. In particular, they are obtained by evaluating $\Psi_\s{sim}(\boldsymbol{r})$ in $N_\phi = 100$ points, corresponding to a randomly chosen values of $\phi$ at a given pair $(R,z)$ of coordinates:
	\begin{align}
		\bar{\Psi}_\s{sim}(R,z) & = \frac{1}{N_\phi} \sum_{i=1}^{N_\phi} \Psi_\s{sim}(\boldsymbol{r}(R,z,\phi_i)) \, , \nonumber \\
		\sigma^2_{\bar{\Psi}}(R,z) & = \frac{1}{N_\phi - 1} \sum_{i=1}^{N_\phi} \left( \Psi_\s{sim}(\boldsymbol{r}(R,z,\phi_i)) - \bar{\Psi}_\s{sim}(R,z) \right)^2 \, .
	\end{align}
	
	The fitting of the parametric functions is performed over a grid of 400 points, denoted as $\mc{P}_\Psi$, where $R$ and $z$ are logarithmically spaced in the interval $R,z \in [0.1 \, \s{kpc}, 20 \, \s{kpc}]$, while we explicitly checked that varying these choices does not substantially change our results. We proceed by minimizing the following $\chi^2$ test statistics with respect to the free parameters $\boldsymbol{\theta} \equiv \{M_\st{H}, a_\st{H}, M_\st{MN}, a_\st{MN}, b_\st{MN} \}$:
	\begin{align}
		\chi^2(\boldsymbol{\theta}) = & \sum_{i \, \in \, \mc{P}_\Psi} \left( \frac{\Psi_\st{H}(R_i, z_i \, | \, M_\st{H}, a_\st{H}) + \Psi_\st{MN}(R_i, z_i \, | \, M_\st{MN}, a_\st{MN}, b_\st{MN}) - \bar{\Psi}_\s{sim}(R_i, z_i)}{\sigma_{\bar{\Psi}(R_i, z_i)}} \right)^2 \, .
	\end{align}
	The resulting best-fit parameters, $\boldsymbol{\theta}_\s{min}$, for the baryonic gravitational potentials of the three simulated objects studied in this work are reported in \citetab{tab:potential}. We assess the goodness of the fits by computing the corresponding reduced chi-squared, $\chi^2_\s{red} \equiv \chi^2(\boldsymbol{\theta}_\s{min}) / (N - P)$, where $(N - P)$ is the difference between the number of points used in the fit and the number free parameters. For all the studied objects it amounts to $\chi^2_\s{red} \ll 1$, which indicates excellent agreement between our parametrization and the values extracted from the simulations. This can also be appreciated from the relative difference between our best-fit models and the azimuthally averaged baryonic gravitational potential, which is shown left-hand panel in \citefig{fig:mochima_potential} for Mochima, while the corresponding plots for Halo B and Halo C can be found in \citefig{fig:halobc_potential} of \citeapp{app:halo_bc}. From the figures it can be observed that our ansatz for the baryonic potential leads to relative differences of less than 10\% over the entire considered $R$-$z$ range in all three simulated objects. In the case of the spherical gravitational potential, the residuals are notably larger at $z \gtrsim 2$ kpc and the relative difference can exceed 20\%, as can be seen from the corresponding plot in the right-hand side panel of \citefig{fig:mochima_potential}. While this could be improved by performing an independent fit for the spherical model, the approach of approximating it through \citeeq{eqn:potential_sph} ensures that the galactic gravitational potential is consistent with the associated rotation curve, which typically provides the strongest observational constraint on a galactic mass model. Furthermore, as we show in \citeapp{app:comparison_zhao} by comparing our results with those of~\cite{lacroix_predicting_2020}, where the baryonic gravitational potential was derived directly from the enclosed baryonic mass, these different choices can induce at most a few percent difference in the inferred moments of the DM velocity distribution.
	
	\begin{table}
		\centering
		\begin{tabular}{l||c|c|c|c|c}
			Simulation \qquad & \quad $M_\st{H}$ [$M_\odot$] \quad & \quad $a_\st{H}$ [kpc] \quad & \quad $M_\st{MN}$ [$M_\odot$] \quad & \quad $a_\st{MN}$ [kpc] \quad & \quad $b_\st{MN}$ [kpc] \\ \hline
			Mochima & $1.9 \cdot 10^{10}$ & $0.61$ & $5.7 \cdot 10^{10}$ & $15$ & $0$ \\
			Halo B & $2.1 \cdot 10^{9}$ & $0.41$ & $1.3 \cdot 10^{11}$ & $4.4$ & $1.3$ \\
			Halo C & $4.6 \cdot 10^{10}$ & $3.4$ & $8.3 \cdot 10^{10}$ & $3.1$ & $0.55$
		\end{tabular}
		\caption{The best-fit parameters for the baryonic gravitational potential obtained for the three simulated objects studied in this work.}
		\label{tab:potential}
	\end{table}
	
	\begin{figure}
		\centering
		\includegraphics[width=0.49\linewidth]{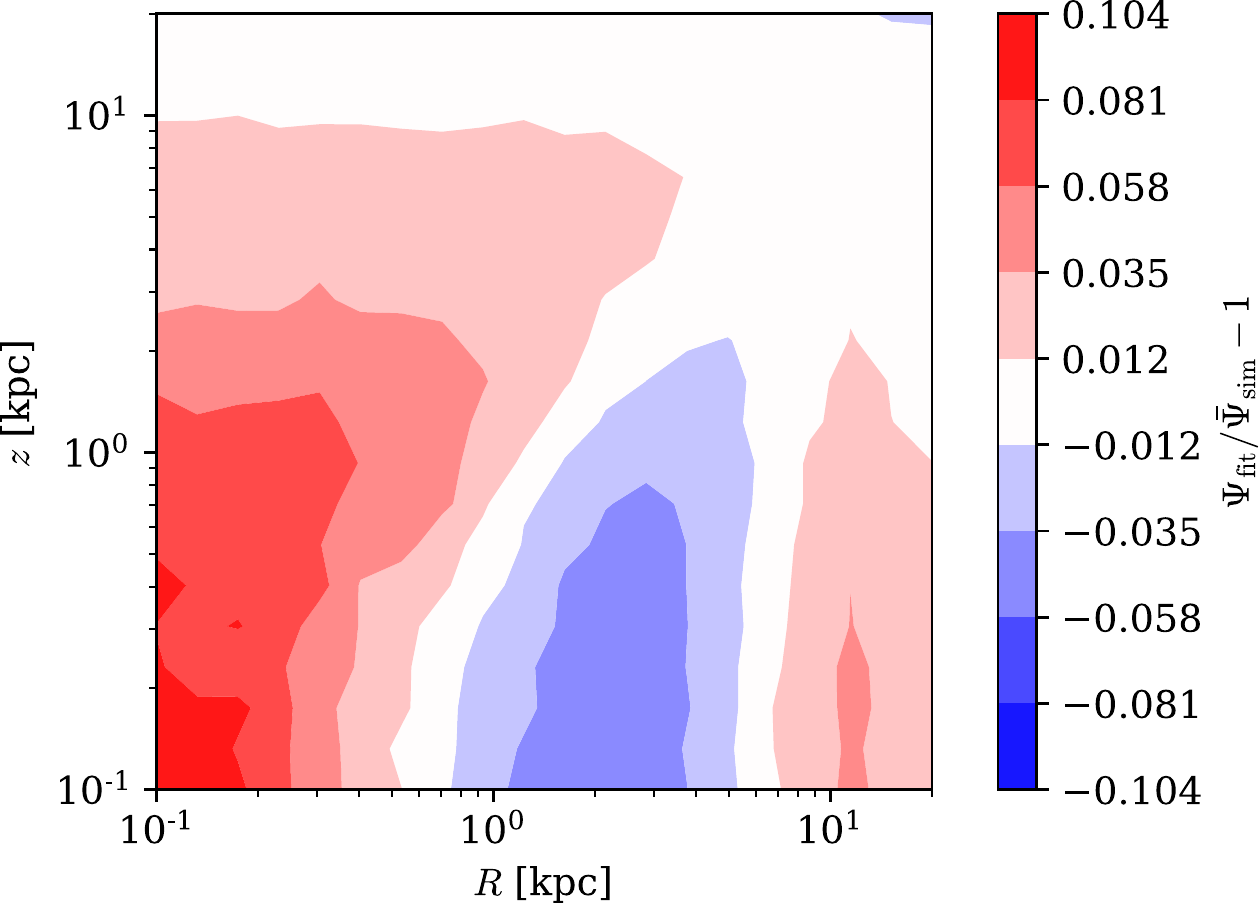}
		\includegraphics[width=0.49\linewidth]{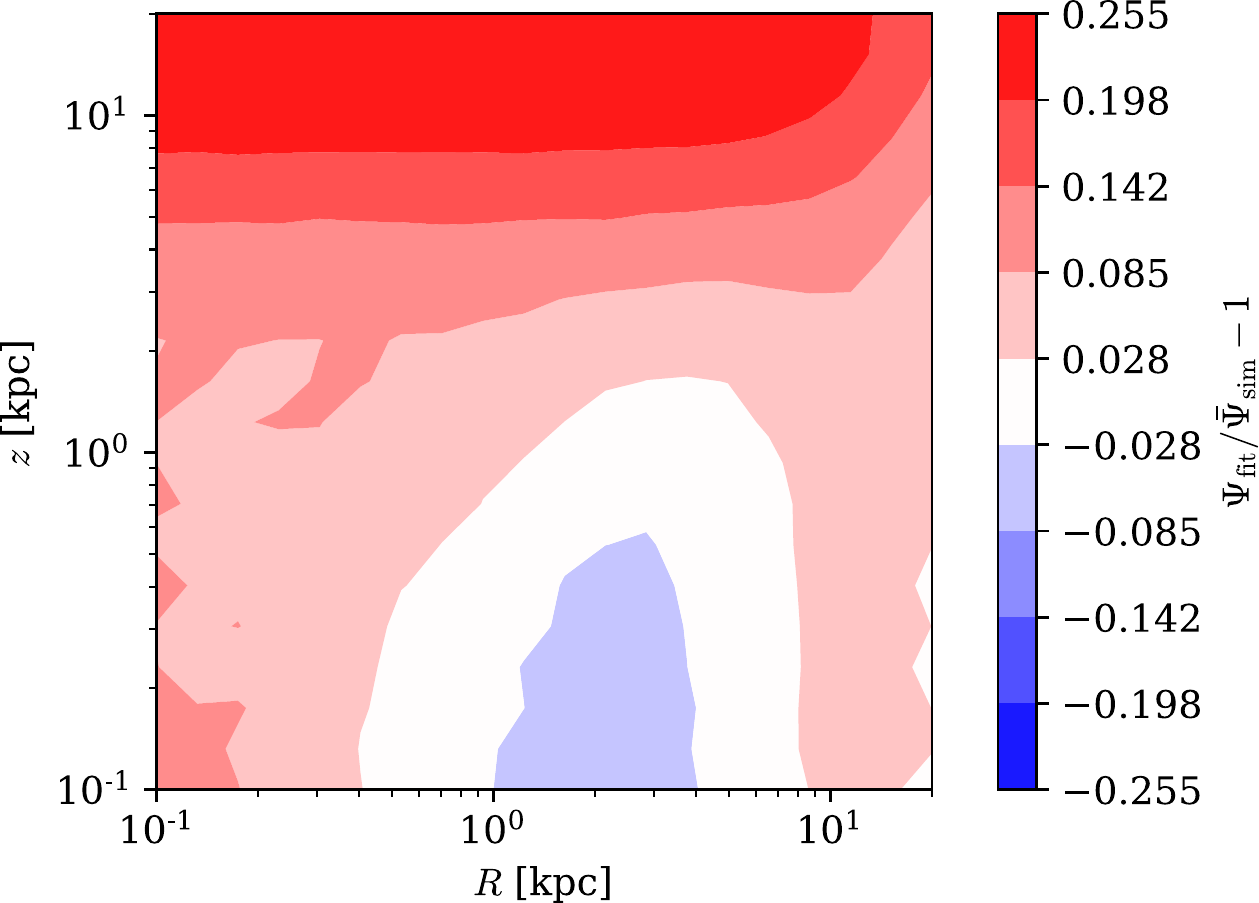}
		\caption{Relative difference between our best-fit and the azimuthally-averaged baryonic gravitational potential for the Mochima simulation. In the left-hand side panel we show the results for the axisymmetric parametrization, while the corresponding spherical solution (obtained according to \citeeq{eqn:potential_sph}) is shown in the right-hand side panel.}
		\label{fig:mochima_potential}
	\end{figure}

	\subsubsection{DM density profile}
	\label{sec:simulations_fitting_density}
	
	The other key quantity for obtaining the phase-space distribution of DM through the inversion methods is the DM density profile, which at the same time also determines the DM's contribution to the total gravitational potential. Throughout the literature, there exist several different parametric density profiles that have been motivated either by numerical simulations or observations. The amount of resolved substructure in high-resolution simulations makes it inconceivable to fully capture the DM distribution of individual objects, however, it is still possible to obtain reasonably good approximations using simple parametric forms. In this work, we limit ourselves to the commonly used cuspy Navarro-Frank-White (NFW)~\cite{navarro_universal_1997} and cored Burkert (BUR)~\cite{burkert_structure_1995} density profiles:
	\begin{align}
		\rho_\st{NFW}(r) & = \frac{\rho_s}{r/r_s \left(1 + r/r_s\right)^2} \, , \\
		\rho_\st{BUR}(r) & = \frac{\rho_s}{\left(1 + r/r_s \right) \left(1 + r^2/r_s^2\right)} \, ,
	\end{align} 
	which depend only on two free parameters, namely the characteristic density, $\rho_s$, and radius, $r_s$. The choice of these two profiles is additionally motivated by the fact that they allow for closed analytical expressions for the corresponding gravitational potentials, which greatly reduces the computational cost of performing the phase-space inversions. As we will show at the end of this section, the NFW profile turns out to be a reasonable approximation for the DM halo found within the Mochima simulation, while Halo B and Halo C exhibit cores that are significantly better fit by the Burkert profile. We have explicitly checked that allowing for spheroidal shape of the halo (i.e. substituting $r \rightarrow \sqrt{R^2 + z^2 / q^2}$ in the above formulas, with $q$ controlling the flattening of the density profile) does not lead to significant improvements in the fits -- for Mochima and Halo C we find $q \sim 1$, while Halo B prefers somewhat flattened halo, however, this is primarily due to a smearing effect related to a significant displacement between the centres of DM and baryons~\cite{mollitor_baryonic_2015}. To further improve the fits, one could, for example, consider the Zhao parameterization~\cite{zhao_analytical_1996}, where the DM density slopes at different parts of the profile are allowed to vary freely. These generalized profiles were actually used in tests of the Eddington inversion on the same simulations in~\cite{lacroix_predicting_2020}. While such a parametrization leads to noticeable improvements in the goodness of the fit, especially in the case of Mochima due to particularly steep central cusp, obtaining the corresponding PSDF is much more computationally demanding since the associated gravitational potential needs to be evaluated numerically. Furthermore, we found that Zhao's family of density profiles can often lead to non-physical solutions within the scope of axisymmetric inversion method. On the other hand, allowing for more flexible parametrization does not significantly affect the resulting moments of the DM velocity distribution, which are of primary interest for DM searches -- see \citeapp{app:comparison_zhao} for the comparison of our results with those of~\cite{lacroix_predicting_2020}.
	
	To obtain the best-fit parameters of the NFW and Burkert density profiles we rely on the corresponding expressions for the enclosed DM mass:
	\begin{align}
		M_\st{NFW}(r) & = 4 \pi \rho_s r_s^3 \left[ \log \left(1 + \frac{r}{r_s}\right) - \frac{r}{r + r_s} \right] \, , \\
		M_\st{BUR}(r) & = \pi \rho_s r_s^3 \left[ \log \left( \left(1 + \frac{r}{r_s} \right)^2 \cdot \left(1 + \frac{r^2}{r_s^2}\right) \right) - 2 \arctan \left( \frac{r}{r_s} \right) \right] \, .
	\end{align}
	The reason for fitting the enclosed DM mass, instead of the DM density profiles directly, is the fact that extracting $M(r)$ from the simulations is much more numerically stable. In particular, inferring the DM density distribution is highly susceptible to computational artefacts related to the finite resolution, while the DM mass enclosed within radius $r$ can be reliably computed by simply summing over the mass of individual DM particles that are located within a sphere of the same radius, $\mc{S}(r)$:
	\begin{align}
		M_\s{sim}(r) = \sum_{i \in \mc{S}(r)} M_i \, .
	\end{align}
	After computing $M_\s{sim}(r)$ in $N_r = 100$ evenly distributed radial points in the range $r \in [1 \, \s{kpc}, 100 \, \s{kpc}]$, the following loss function is minimised with respect to the free parameters, $\boldsymbol{\theta} = \{\rho_s, r_s\}$:
	\begin{align}
		\mc{L} (\boldsymbol{\theta}) = \sum_{i=1}^{N_r} \left(\frac{M_\st{NFW/BUR}(r_i | \, \boldsymbol{\theta})}{M_\s{sim}(r_i)} - 1 \right)^2 \; .
	\end{align}
	As already mentioned above, the DM halo found in Mochima simulation exhibits a central cusp and is, therefore, better fit by an NFW density profile. The latter performs well at $r \gtrsim 5 \, \s{kpc}$, where the relative difference in the enclosed DM mass does not exceed 10\%. On the other hand, in the inner few kpc the fit leads to an increasing deficit of DM, which implies that the simulated halo has a significantly steeper central cusp. This can be clearly seen in \citefig{fig:mochima_mass}, where we show the true and best-fit NFW mass profiles, as well as their relative difference. For comparison we also include the DM mass profile corresponding to the Zhao parametrization obtained in~\cite{lacroix_predicting_2020}, which performs notably better with residuals at the percent level at all radii. In case of Halo B and Halo C, for which the analogous plots can be found in \citefig{fig:halobc_mass} of \citeapp{app:halo_bc}, the cored Burket profile provides a decent match, with relative errors of less than 10\% over most of the considered radial range. We report the best-fit parameters for all three simulations in \citetab{tab:fit_density}.
	
	\begin{figure}
		\centering
		\includegraphics[width=0.45\linewidth]{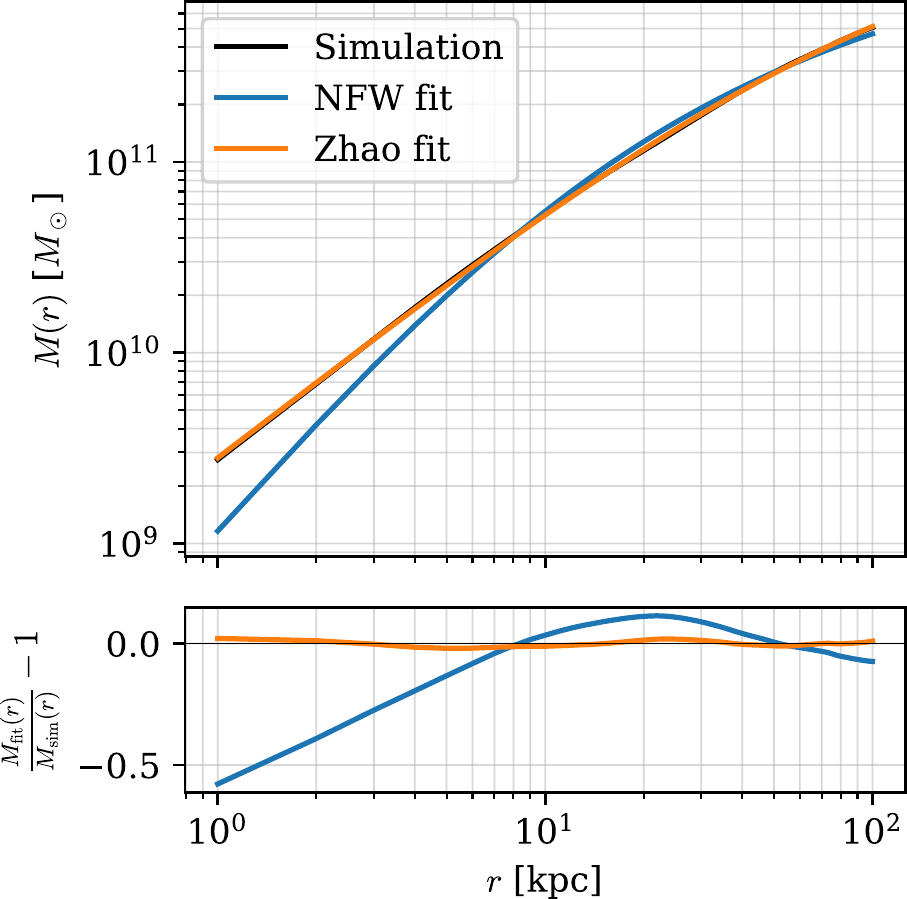}
		\caption{Upper panel shows the mass profile extracted from Mochima simulation and the corresponding NFW and gNFW fits, while the lower panel shows their relative difference, as a function of radius.}
		\label{fig:mochima_mass}
	\end{figure}

	\begin{table}
		\centering
		\begin{tabular}{c||c|c|c|c}
			Simulation \qquad & \quad Profile \quad & \quad $\log_{10} \left( \rho_s \, [M_\odot / \s{pc}^3] \right)$ \quad & \quad $r_s$ [kpc] \quad \\ \hline
			Mochima & NFW & 7.26 & 11.5 \\
			Halo B & Burkert & 7.82 & 5.67 \\
			Halo C & Burkert & 7.81 & 5.94
		\end{tabular}
		\caption{The appropriate (NFW or Burkert) density profiles and their corresponding best-fit values of the parameters for the three objects studied in this work.}
		\label{tab:fit_density}
	\end{table}

	\section{Comparison of the predictions of inversion methods with the simulations}
	\label{sec:results}
	
	In the following we explore the level of agreement between the predictions of the inversion methods and the actual phase-space distribution of DM within the simulations. Since the inversion methods relie on the DM density profile as an input, all the relevant predictions are fully encoded within the position-dependent velocity distribution of DM. Throughout this section, we will take a closer look at the latter, as well as various related quantities, such as the DM velocity moments and orbital anisotropy. In doing so, our main goal is to highlight the key differences that arise between the Eddington and axisymmetric inversion and assess the associated systematic uncertainties by comparing the resulting predictions to the values extracted directly from the simulations. In the discussion of our results, we will primarily focus on the Mochima simulation, which offers the highest resolution among the available runs, while we will also briefly comment on our findings for Halo B and Halo C, for which the corresponding plots can be found in \citeapp{app:halo_bc}.
	
	Before entering a detailed comparison of the predictions of the inversion methods with the simulations, we briefly summarize the results obtained in our fits of baryonic gravitational potentials and DM density profiles. The DM halo found in the Mochima simulation has a cuspy density profile with a best-fit NFW scale radius of $r_s = 11.5 \; \s{kpc}$. On the other hand, Halo B and Halo C favour a Burkert DM density profile with a core radius of $\sim 6 \; \s{kpc}$. Further differences can be observed in the obtained fits of the baryonic gravitational potentials. Mochima and Halo C feature a relatively massive bulge, which is, however, in the case of the latter not as dense due to the unusually large scale length, $a_\st{H} = 3.4 \; \s{kpc}$. We find the least pronounced disc component in the case of Mochima, as it has the lowest mass, $M_\st{MN} = 5.7 \cdot 10^{10} \; M_\odot$, and particularly large scale length, $a_\st{MN} = 15 \; \s{kpc}$. Halo C possesses slightly more massive disc, but with a considerably smaller disc scale length, which is shorter than the one of the accompanying bulge component, i.e. $a_\st{MN} = 3.1 \; \s{kpc} \lesssim a_\st{H}$. On the other hand, for Halo B we find the least massive bulge and the most massive disc component, with $M_\st{MN} = 1.3 \cdot 10^{11} \; M_\odot$ and $a_\st{MN} = 4.4 \; \s{kpc}$. Even though the disc in Halo B features somewhat larger scale height, $b_\st{MN} = 1.3 \; \s{kpc}$, we non-the-less expect that the axisymmetric inversion will bring the most significant improvements over the Eddington method for this simulation due to the most pronounced disc component.
	
	\subsection{Velocity distributions}
	\label{sec:results_velocity}
	
	The most direct comparison between the predictions of the inversion methods and the simulations can be performed by examining the probability distribution for DM velocity at different positions within the DM halo. Due to the symmetry assumptions, the Eddington approach leads to identical velocity distribution over the entire spherical shell associated with some galactocentric radius $r$, while the axisymmetric inversion method distinguishes among different positions along the meridional ($\hat{R}$-$\hat{z}$) plane. Furthermore, the original Eddington's formulation, where the phase-space distribution can be expressed as a function of only the relative energy, necessarily results in an isotropic velocity distribution, whereas the two-integral method leads to distinct predictions for the velocity distributions along the azimuthal direction and the meridional plane -- i.e., $f(\mc{E}, L_z)$ explicitly depends on $v_\phi$ through $L_z$, while $\mc{E}$ depends only on the velocity magnitude. On the other hand, the simulations are not restricted by any symmetry assumptions and, in principle, further distinct components of the velocity distribution as well as their dependence on the three-dimensional position vector could be studied. However, in order to perform a direct comparison with the predictions of the inversion methods, as well as maintain sufficient number of particles to adequately sample the velocity distributions, we in this work restrict our attention to the probability density functions for the velocity magnitude, $P(v)$, (also referred to as the speed distribution), meridional velocity, $P_\st{M}(v_\st{M})$ and azimuthal velocity, $P_\phi(v_\phi)$. All of these were extracted from the simulations by binning the velocities of particles that fall within a torus, centred at a given pair of $\{R, z\}$ coordinates, with radius $\delta(R, z) = \delta_\s{min} + \varepsilon \cdot \sqrt{R^2 + z^2}$. For each simulation $\delta_\s{min}$ and $\varepsilon$ were chosen such that each torus contained at least 1000 particles, while we explicitly checked that varying these hyper-parameters does not significantly affect our results.
	
	In the following we first perform a qualitative comparison of the predictions under different symmetry assumptions by inspecting the aforementioned velocity distributions at several galactocentric distances. This is followed by a quantitative comparison based on relative entropy measure and the moments of velocity distribution, as well as the resulting orbital anisotropy, over the entire DM halo.
	
	\subsubsection{Direct comparison}
	\label{sec:results_velocity_direct}
	
\begin{figure}[h]
		\centering
		\includegraphics[width=\textwidth]{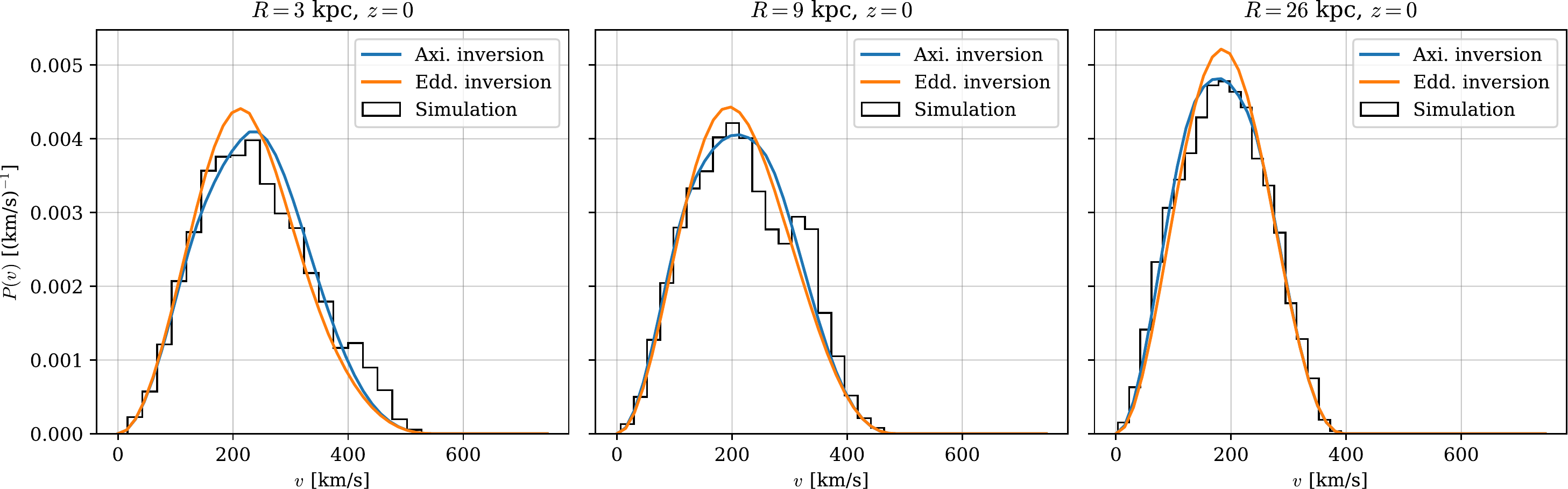}
		\caption{Comparison of the speed distribution extracted from the Mochima simulation with the corresponding predictions of the axisymmetric and Eddington inversion methods at three different radial distances along the galactic plane.}
		\label{fig:mochima_vel}
	\end{figure}

\begin{figure}[h]
		\centering
		\includegraphics[width=\textwidth]{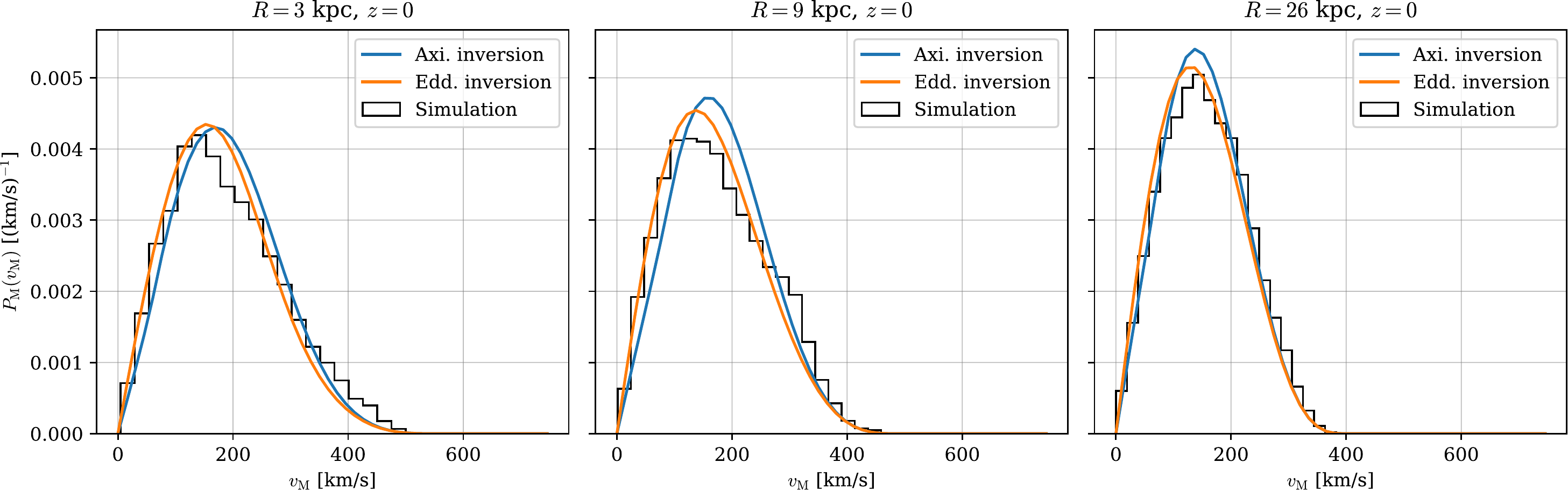}
		\caption{Comparison of the meridional velocity distribution extracted from the Mochima simulation with the corresponding predictions of the axisymmetric and Eddington inversion methods at three different radial distances along the galactic plane.}
		\label{fig:mochima_vel_merid}
	\end{figure}

\begin{figure}[h]
		\centering
		\includegraphics[width=\textwidth]{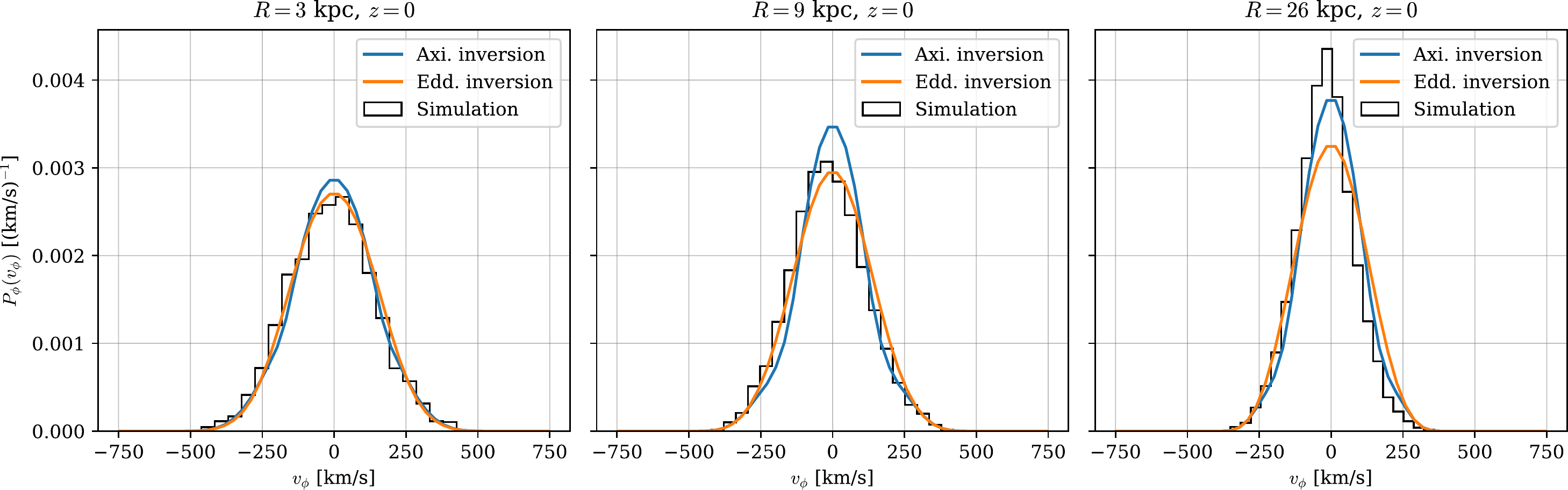}
		\caption{Comparison of the azimuthal velocity distribution extracted from the Mochima simulation with the corresponding predictions of the axisymmetric and Eddington inversion methods at three different radial distances along the galactic plane.}
		\label{fig:mochima_vel_azim}
	\end{figure}

	In the following we present the comparison of velocity distributions predicted by the axisymmetric and Eddington inversion methods with the ones extracted from Mochima simulation at three different radial distances along the galactic plane. As can be observed from the corresponding plots -- shown in Figures~\ref{fig:mochima_vel},~\ref{fig:mochima_vel_merid} and~\ref{fig:mochima_vel_azim}, for speed, meridional and azimuthal velocity distribution, respectively -- the inversion methods provide fairly good approximations to the true velocity distributions. However, some further comments are in place to emphasize the key differences between the Eddington and axisymmetric inversion methods. To begin with, it should be noted that the axisymmetric approach in the presence of flattened gravitational potential (which in our case arises due to the baryonic disc, but could also be sourced by a flattened DM halo) results in kinematically warmer particles along the meridional plane, while the velocity dispersion along the azimuthal direction is decreased. This leads to shifts in the peaks of the speed distributions, shown in \citefig{fig:mochima_vel}, to slightly larger velocities, which indeed provides a better match with the speed distributions observed in the simulation. The indicated trend is most evident in the plot corresponding to $R = 3 \, \s{kpc}$, where the contribution of baryonic disc to the total gravitational potential is the most significant, while it slowly diminishes with increasing $R$. On the other hand, it is quite surprising to see that the predictions for the meridional velocity distribution, shown in \citefig{fig:mochima_vel_merid}, do not show the same improvement. While the most probable meridional velocity in the simulation seems to be in better agreement with the predictions of Eddington inversion, one can observe a significant excess of power in the true $P_\st{M}(v_\st{M})$ at larger velocities. The axisymmetric inversion method generally shows the correct trend, i.e. slightly increased probability density for high $v_\st{M}$, the match in the actual shape of the high-velocity tails of the distributions is rather poor. Further unexpected discrepancies are observed in the azimuthal velocity distribution, $P_\phi(v_\phi)$, shown in \citefig{fig:mochima_vel_azim}. As can be seen from the plots, the performance of the inversion methods strongly depends on the radial distance. At $R = 3 \, \s{kpc}$ both of the methods predict $P_\phi(v_\phi)$ reasonably well, at $R = 9 \, \s{kpc}$ the Eddington prediction performs better due to the aforementioned larger spread in the azimuthal velocity distribution, while at $R = 26 \, \s{kpc}$ the axisymmetric inversion method provides the best match due its significantly more peaked distribution. As a final remark, the plots of $P_\phi(v_\phi)$ also clearly show that accounting for the halo rotation provides only a small correction to the predicted velocity distribution. While non-vanishing $\bar{v}_\phi$ is present within the simulation, leading to slightly asymmetric $P_\phi(v_\phi)$ with respect to $\bar{v}_\phi = 0$, it is evident that correctly predicting the width of the distribution remains the primary challenge of the inversion methods.
	
	In the case of Halo B and Halo C, for which the analogous plots can be found in Figures~\ref{fig:halob_vel} and~\ref{fig:haloc_vel} of \citeapp{app:halo_bc}, we find larger discrepancies between the predictions of inversion methods and the true velocity distributions, as well as bigger differences between the spherical and axisymmetric models themselves. This can be most clearly seen at $R = 3 \, \s{kpc}$, where the speed and meridional velocity distributions of Halo B are much better approximated by the axisymmetric inversion, while in the case of Halo C the Eddington method performs better. At larger radii, where the differences between the two methods become smaller, we observe similar trends as in the case of Mochima simulation, namely the axisymmetric inversion generally leads to slightly more accurate predictions for the speed distribution, while the Eddington approach performs better in predicting the meridional velocity distribution. On the other hand, the azimuthal velocity distribution tends to be better approximated by the Eddington prediction at small $R$ and axisymmetric inversion method at large $R$ in both simulations .
	
	Since the Eddington inversion is only a subclass of solutions allowed by the more general axisymmetric method, one would naively expect that the latter should always provide a more accurate description of the studied system. However, as it is evident from the above results, this is not always the case. The apparent inconsistency can be resolved by noting that the simulated objects are only approximately axisymmetric and that the net angular momentum of the DM halo is not necessarily normal to the galactic plane. Such departures from the assumptions of the axisymmetric model can lead to less accurate predictions for the components of the velocity distribution, or in certain cases even the speed distribution, than the simpler spherically symmetric approach. Upon more careful inspection of Halo C, we indeed find a $53^\circ$ misalignment between the angular momentum of the DM halo and the normal vector of the disc plane, which explains the poor accuracy of the axisymmetric method when contrasted with the predictions of the Eddinton inversion. On the other hand, in the case of Mochima and Halo B the misalignment is less significant, $\sim 20^\circ$, and, hence, the axisymmetric method manages to provide more accurate results for the speed distribution, but not for the meridional and azimuthal components. However, as we demonstrate in \citeapp{app:misalignment}, when the angular momentum of the DM halo is chosen as the principal axis of the system, the axisymmetric inversion leads to equally or more accurate predictions than the Eddington inversion even for the individual components of the velocity distribution. Nevertheless, we for the remainder of this work adapt the coordinate system in which the $z$-axis is perpendicular to the baryonic disc, since in practice the orientation of halo's angular momentum is not accessible through the existing observations.

	\subsubsection{Similarity in terms of the relative entropy}
	\label{sec:results_velocity_DKL}
	
	The above discussion highlights the key qualitative differences between the velocity distributions extracted from the simulations and the corresponding predictions of the two inversion methods. By simple visual inspection it is possible to appreciate the improvements in the predictions for the speed distribution introduced by the axisymmetric approach, whereas the situation is much less clear for $P_\st{M}(v_\st{M})$ and $P_\phi(v_\phi)$. Moreover, one would ideally like to assess the match between the predictions and the simulations over the entire radial range, instead of inspecting it only at several hand-picked values of $R$. For these reasons, we resort to the relative entropy, also known as Kullback-Leibler divergence $D_\st{KL}$~\cite{kullback_information_1951}, which is commonly used as a metric for quantifying the similarity between two probability distributions. Given an observed discrete (i.e. binned) probability distribution $P$ and the corresponding theoretical prediction $Q$, the value of $D_\st{KL}$ can be computed as follows:
	\begin{align}
		\label{eqn:relative_entropy}
		D_\st{KL} (P|Q) \equiv \sum_{x \in \mc{X}} P(x) \log\left(\frac{P(x)}{Q(x)}\right) \,
	\end{align}
	where $\mc{X}$ spans the common probability space -- according to the standard convention, we set the terms for which $P(x) = 0$ to zero, while we limit $\mc{X}$ to a range of values where $Q(x) \neq 0$. As can be appreciated from \citeeq{eqn:relative_entropy}, the relative entropy vanishes if the two probability distributions are exactly equal, i.e. $D_\st{KL}(P|Q=P) = 0$, while it takes increasing positive values with increasing difference between the compared probability distribution.~\footnote{Strictly speaking, the latter is true only if the domains of $P$ and $Q$ overlap. However, in the context of comparing the predicted velocity distributions with the simulations, this is always the case. A mismatch in the domains can only arise if the velocity distributions have different escape velocities. However, since the latter are determined by the fits of DM density profile and the baryonic gravitational potential or possible truncation of the object (and not the inversion methods per se), they are not particularly important for assessing the accuracy of the predictions of inversion methods, especially since the dominant contribution towards $D_\st{KL}$ arises from the bulk of the distribution.}
	
	In \citefig{fig:mochima_entropy} we show the obtained values of $D_\st{KL}$ for the speed distribution, as well as its meridional and azimuthal components, as a function of radial distance $R$ along the galactic plane of the Mochima simulation. The plots confirm our qualitative observations based on the visual inspection of the velocity distributions. In particular, the speed distribution predicted by the axisymmetric inversion method is notably closer to the one extracted from the simulation over most of the considered radial range. On the other hand, the meridional velocity distribution seems to be more accurately approximated by the Eddington inversion, however, at several values of $R$ the predictions of the two methods come very close and occasionally the axisymmetric method leads to an even better match. This agrees with our qualitative observations from the previous section, where we noted that the Eddington approach tends to predict more accurately the most probable velocity, while the axisymmetric method better accounts for the abundance of high-velocity particles. Regarding the azimuthal velocity distribution, the relative entropy shows that the two methods perform roughly the same up to $R \sim 5 \, \s{kpc}$, where the Eddington prediction becomes more accurate. This remains true up to $R \sim 15 \, \s{kpc}$, beyond which the axisymmetric inversion leads to a better agreement with the simulation. Additionally, from \citefig{fig:mochima_entropy} we can see a clear trend that the predictions of inversion methods are the most accurate around $R \sim 25 \, \s{kpc}$, with slowly degrading performance towards the center and large fluctuations at greater $R$. The latter is caused by the presence of numerous DM substructures in the outskirts of the halo, while the increasing differences in the inner parts might be implying that a more elaborate modelling is required to fully capture the intricate interplay between DM and baryons. Finally, we note that the predictions of the axisymmetric method can be substantially improved by choosing the principal axis of the system to be aligned with the angular momentum of the DM halo -- for comparison see \citefig{fig:mochima_entropy_tilted} of \citeapp{app:misalignment}.
	
	\begin{figure}[h]
		\centering
		\includegraphics[width=\textwidth]{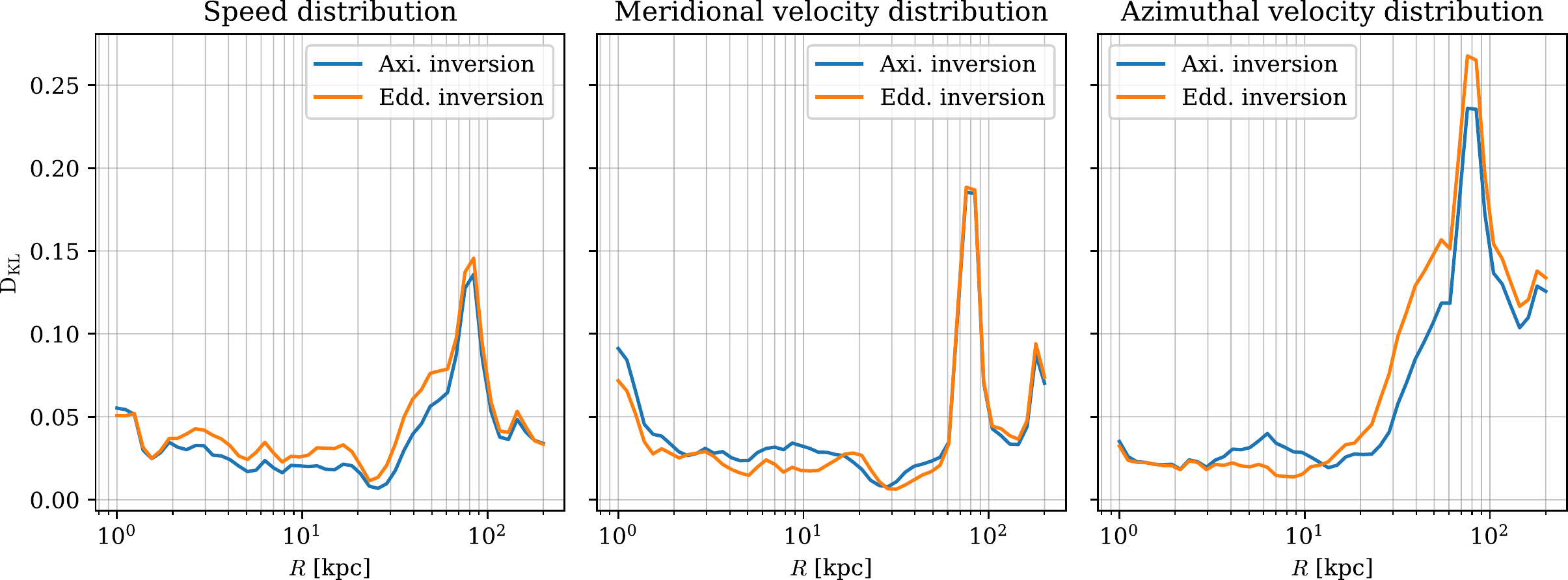}
		\caption{Relative entropy between the true velocity distributions and the predictions of axiymmetric and Eddington inversion methods as a function of the radial distance along the galactic plane.}
		\label{fig:mochima_entropy}
	\end{figure}

	The analogous results for the relative entropy profiles of Halo B and Halo C, shown in Figures~\ref{fig:halob_entropy} and~\ref{fig:haloc_entropy} of \citeapp{app:halo_bc}, again conform with our qualitative findings from the previous section. Most notably, the axisymmetric method performs significantly better than the Eddington inversion in the central part of Halo B, which was expected due to the particularly massive baryonic disc. The opposite is true in the inner few kpc of Halo C, where due to the aforementioned misalignment between the galacitc disc and the angular momentum of the DM halo the Eddington inversion provides more accurate results. At $R \gtrsim 5 \, \s{kpc}$ the predictions for the speed distributions in Halo B and Halo C improve for both inversion methods, although they are still somewhat less accurate than what we found in the case of Mochima simulation. At intermediate distances the meridional and azimuthal velocity distributions tend to be more accurately predicted by the Eddington inversion, which is again related to the departures from the assumed axial symmetry. At large distances the predictions of the two models become very similar, however, their agreement with the velocity distributions observed in the simulations degrades, which is linked to the presence of significant DM substructures in the outskirts of the halo.

	\subsection{Velocity moments}
	\label{sec:results_moments}
	
	The above comparison of the predicted vs. ``measured'' velocity distributions allows us to test the predictive power of the inversion methods at the most fundamental level. However, for practical purposes one is often interested only in the moments of the speed distribution that typically enter the interpretation of results of DM searches. Furthermore, the velocity moments can also provide us with additional insights regarding the accuracy of the inversion methods' predictions, since positive and negative moments are particularly sensitive to the high and low velocity tails of the distribution, respectively, which were not of great importance for the discussion in the previous section.
	
	Motivated by the above considerations, we show in the left-hand side panel of \citefig{fig:mochima_mom} the first two negative and positive moments of the speed distributions along the galactic plane for the Mochima simulation. In the right-hand side panel of the same figure we show the corresponding relative differences between the two inversion methods and the simulation for easier assessment of the (dis)agreement. The first thing to notice is the fact that both inversion methods lead to velocity moments that provide rather good match to the ones extracted from the simulation. The relative error in the first and second moments is smaller than 7\% and 15\%, respectively, at all $R$, with the exception of $\langle v^{-2} \rangle$ which can differ up to 40\%. In general, for negative moments the inversion methods have a bias towards larger values at $R \lesssim 20 \, \s{kpc}$ and smaller values at $R \gtrsim 30 \, \s{kpc}$. On the contrary, positive moments are somewhat under predicted at $R \lesssim 30 \s{kpc}$, while they on average agree with the simulation beyond that radius. However, it should be noted that at large radii the moments extracted from the simulation strongly fluctuate, which is most likely related to the presence of DM substructures. The other important observation is that the axisymmetric method tends to provide more accurate predictions only in the range of $2 \, \s{kpc} \lesssim R \lesssim 10 \, \s{kpc}$. At first glance this is perhaps surprising, given the fact that according to the relative entropy metric the axisymmetric method performs better over the entire radial range. However, it should be noted that velocity moments are particularly sensitive to the tails of the velocity distribution, while for $D_\st{KL}$ the position of the peak is significantly more important. Similarly, due to the sensitivity to different tails of the speed distribution, the positive and negative moments exhibit different behaviours. In particular, the improvement provided by the axisymmetric method is much more significant for positive moments, as could be expected from our discussion of the shape of the speed distribution in the previous section.
	
	Analogous results regarding the moments of speed distribution for Halo B and Halo C are shown in Figures~\ref{fig:halob_mom} and~\ref{fig:haloc_mom} of \citeapp{app:halo_bc}. From there it can be immediately observed that the predictions for these two simulations are less accurate and larger errors arise in the inner few kpc within the scope of Eddington method for Halo B and axisymmetric method for Halo C. These trends could have again been anticipated from our observations at the level of velocity distributions. At $R \gtrsim 5 \, \s{kpc}$ the relative errors become much smaller and both of the inversion methods predict the velocity moments with accuracy better than 20\%, up to the values of $R$ where fluctuations due to DM substructure become significant. The general trends are similar to those of the Mochima simulation, namely the negative (positive) moments are over (under) predicted in the inner parts of the haloes and vice-versa in the outskirts. On the other hand, the improvement of the axisymmetric method over the Eddington inversion is less clear, as it depends on the particular halo and velocity moment that is being considered.
	
	\begin{figure}[H]
		\centering
		\includegraphics[width=\textwidth]{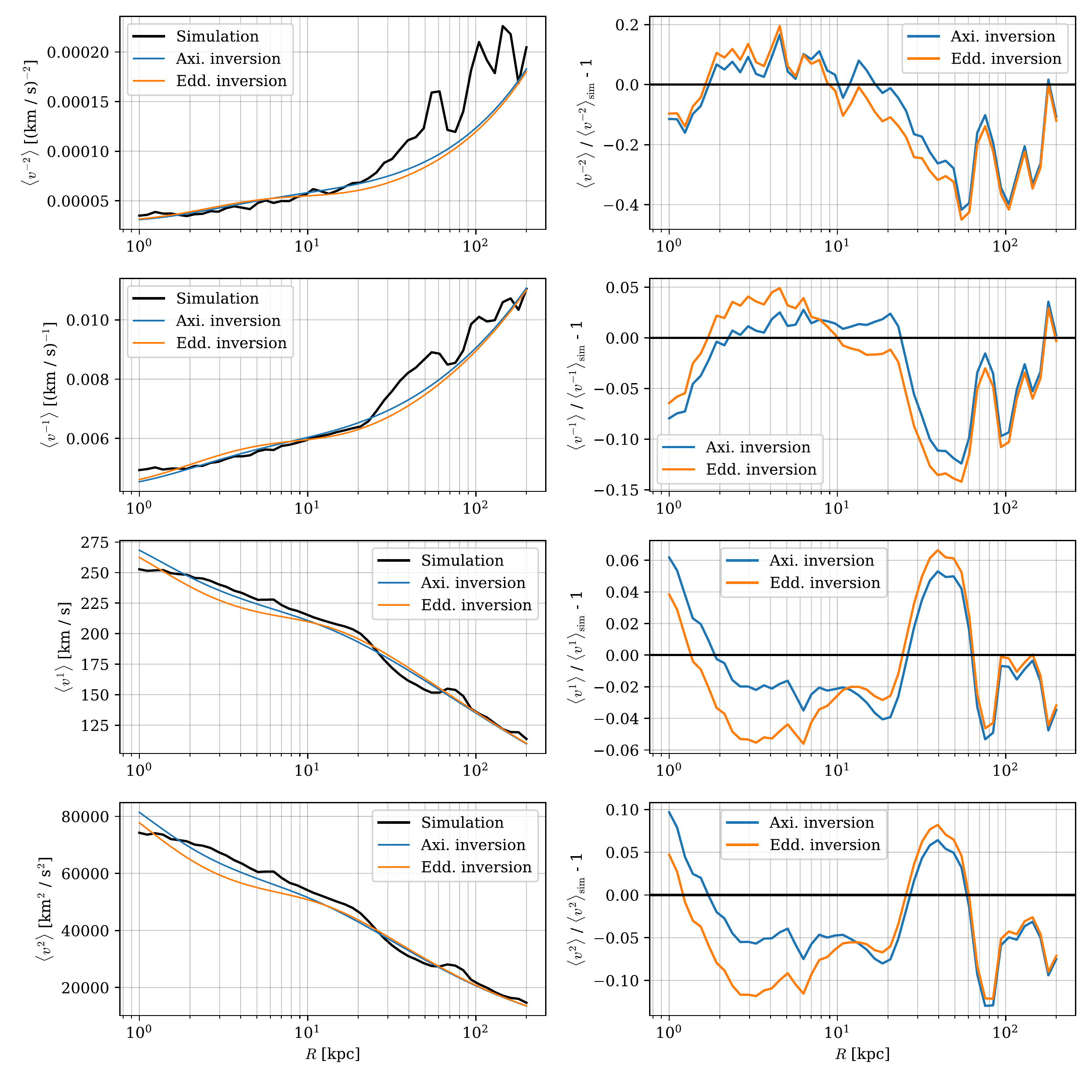}
		\caption{Values of the first two positive and negative moments of the speed distribution as a function of the radial distance along the galactic plane for the Mochima simulation. The left-hand side panels shows the values extracted directly from the simulation as well as the corresponding predictions of the inversion methods, while the right-hand side panels show the relative differences between the predictions and the true values.}
		\label{fig:mochima_mom}
	\end{figure}

	\subsection{Velocity anisotropy}
	\label{sec:results_anisotropy}
	
	Finally, we turn our attention to the orbital anisotropy of the DM haloes. While this quantity is not directly observable and can lead to notable effects only in the scope of direct detection experiments, it is still interesting from the theoretical point of view. As briefly mentioned in \citesec{sec:methods_edd}, there exist anisotropic generalizations of the Eddington inversion method, in which orbital anisotropy has to be specified beforehand. However, as thoroughly investigated in~\cite{lacroix_anatomy_2018}, such models often lead to non-physical solutions and, hence, we chose not to include them in our analysis. On the other hand, in the context of axisymmetric inversion method, the orbital anisotropy is a prediction of the model and is primarily determined by the total gravitational potential and DM density distribution -- while in principle the orbital anisotropy is also sensitive to the unobservable rotational velocity of the DM halo, it turns out that for realistic rotation velocities its contribution is negligible. To benchmark the agreement between the predictions of the inversion methods and the simulations, we address in the following the velocity anisotropy along the galactic plane, which can be defined as:~\footnote{Note the difference in the definition with the respect the standard definition of velocity anisotropy, $\beta(r) \equiv 1 - \sigma_\st{T}(r) / (2 \sigma_r(r))$ where $\sigma_\st{T}(r)$ and $\sigma_r(r)$ are the tangential and radial velocity dispersions, which was conceived in the context of the spherically symmetric models. The two definitions coincide only within the galactic plane.}
	\begin{align}
		\beta_{\hat{R}}(R) \equiv \frac{1}{2} - \left. \frac{\sigma^2_\phi(R, z)}{2\sigma^2_\st{M}(R, z)} \right|_{z=0} \,
	\end{align}
	where $\sigma_\phi(R,z)$ and $\sigma_\st{M}(R,z)$ are the velocity dispersions along the azimuthal direction and meridional plane, respectively. This choice is motivated by the fact that the orbital anisotropy of axisymmetric model is maximal along the galactic plane, while it vanishes along the $z$-axis. As a side note, by recalling that the rotational properties of the halo are fully encoded in the $L_z$-odd part of the PSDF, defined in \citeeq{eqn:psdf_axi_odd}, it is easy to verify that the contribution of halo rotation to $\beta_{\hat{R}}$ enters only through $\sigma_{\phi}^2 = \langle v_\phi^2 \rangle - \langle v_\phi \rangle^2$ and, hence, is suppressed by $\bar{v}_\phi^2 / \langle v_\phi^2 \rangle$.
	
	In \citefig{fig:mochima_aniso} we show the orbital anisotropy along the disc plane for the two inversion methods, as well as $\beta_{\hat{R}}(R)$ extracted from Mochima simulation. While in the Eddington case $\beta_{\hat{R}}(R) = 0$ by construction, the axisymmetric method agrees reasonably well with the simulation in the inner $\sim 10 \, \s{kpc}$, where it predicts increasing positive values of $\beta_{\hat{R}}$. However, in the outskirts the predicted anisotropy diminishes much faster than what is observed in the simulation. Since the baryonic disc becomes sub-dominant beyond its scale length, the axisymmetric model slowly approaches the spherically symmetric, and hence isotropic, configuration. On the other hand, within the simulation $\beta_{\hat{R}}$ keeps on growing up to $\sim 100 \, \s{kpc}$, and only after that begins to decline. This implies that our models are not able to fully capture the dynamical properties of the DM halo. In the case of the axisymmetric method it would be possible to accommodate this trend by enforcing an oblate DM halo (i.e. by setting $q > 1$ in the spheroidal generalization of the NFW density profile, mentioned in \citesec{sec:simulations_fitting_density}). However, during the fitting procedure we have found no evidence for oblateness of the halo and hence we interpret this as a hint that a more general approach, such as action-angle modelling, could be required to faithfully reproduce the observed velocity anisotropy.
	
	\begin{figure}[H]
		\centering
		\includegraphics[width=0.6\textwidth]{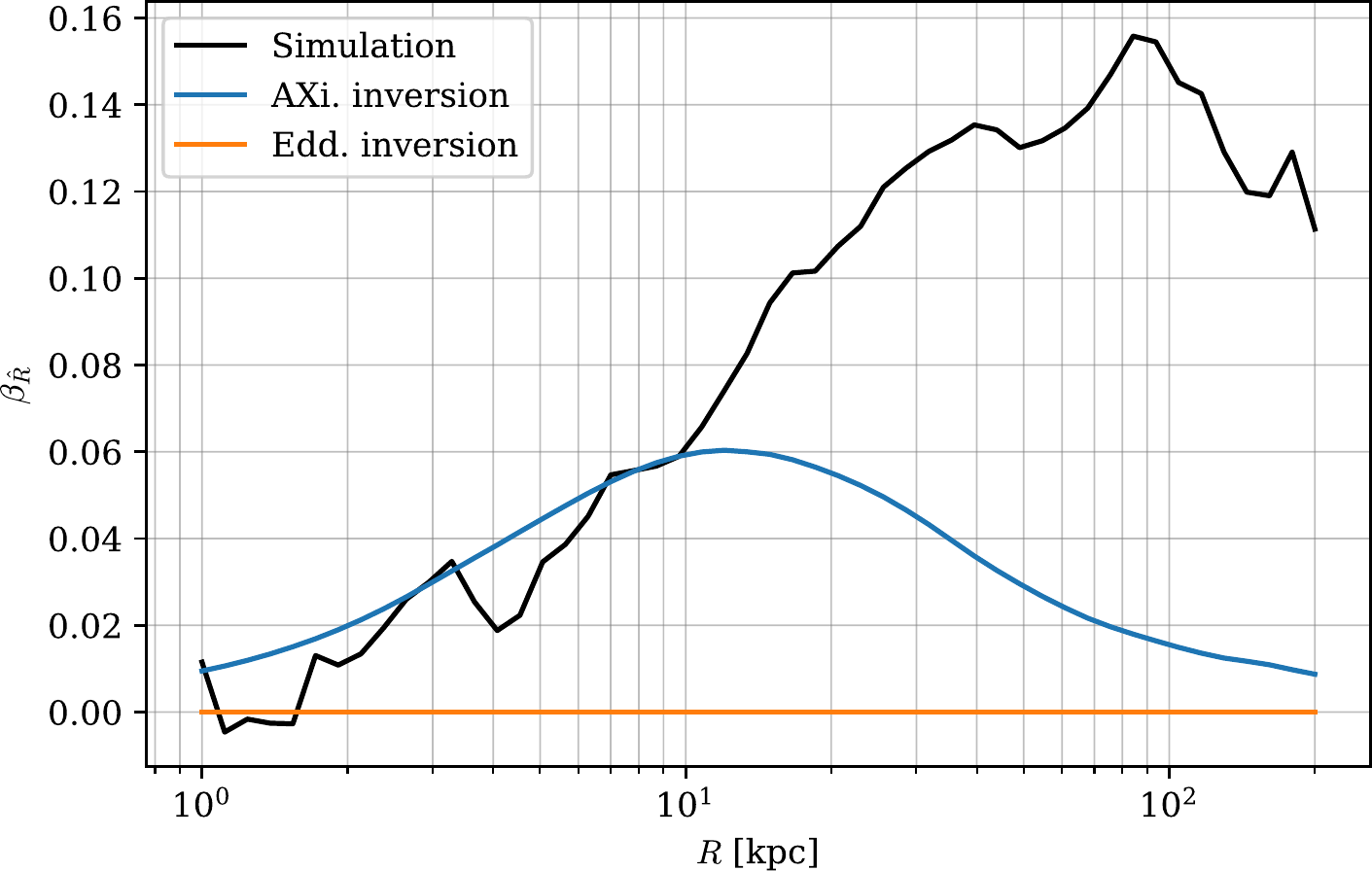}
		\caption{Velocity anisotropy along the galactic plane as a function of the radial distance obtained from the Mochima simulation and the corresponding predictions of the inversion methods.}
		\label{fig:mochima_aniso}
	\end{figure}

	In comparison with the Mochima simulation, the matching between the measured and predicted $\beta_{\hat{R}}(R)$ is notably worse for Halo B and Halo C, as can be seen from the corresponding plots shown in \citefig{fig:halobc_aniso} of \citeapp{app:halo_bc}. While the predictions of the axisymmetric inversion method are similar to the case of Mochima, i.e. $\beta_{\hat{R}}(R)$ rises in the inner part, peaks around the disc's scale length and then slowly falls towards zero, the orbital anisotropies obtained from the simulations show different trends. In the case of Halo B, $\beta_{\hat{R}}(R)$ is much lower than predicted by the axisymmetric model for $R \lesssim 20 \, \s{kpc}$, while it sharply rises beyond that radius, peaks near $100 \, \s{kpc}$ and than rapidly declines. For Halo C the orbital anisotropy grows more gradually, which is in somewhat better agreement with the prediction of the axisymmetric method, but it does not seem to decrease significantly even beyond its peak at $R \sim 100 \, \s{kpc}$.
	
	\section{Implications for DM searches}
	\label{sec:searches}
	
	One of the key motivations for developing accurate phase-space models for the distribution of DM within galaxies comes from numerous experimental searches for hypothetical particle candidates. In the past, efforts have been primarily focused on WIMPs~\cite{PrimackEtAl1988,SrednickiEtAl1988,jungman_supersymmetric_1996,SteigmanEtAl2012,ArcadiEtAl2017,LeaneEtAl2018}, which can be efficiently searched for through probes that rely on vast quantities of DM within the Milky Way or other nearby galaxies. Such endeavours can be broadly categorised as direct and indirect searches, where the former aim to detect nuclear recoils due to the scattering of the Galactic DM with target atoms in highly sensitive laboratory experiments, while the latter focus on detecting emissions associated with DM annihilations (or decays) in nearby regions with high DM densities, such as the centres of the Milky Way or its satellite galaxies. Due to the absence of tangible signals, significant efforts have been put into diversifying the search strategies to probe a broader range of theoretically motivated DM candidates -- to mention some, these include axion-like particles, non-thermally produced massive particles (e.g., super-WIMPs or FIMPs), various extensions of the Higgs sector or even primordial black holes (for a comprehensive review see, e.g.,~\cite{feng_dark_2010,lin_tasi_2019}). Despite the staggering diversity of theoretical models, many of them can be strongly constrained by the existing direct and indirect detection experiments or other astrophysical probes, which likewise require accurate knowledge of the DM distribution within galaxies.
	
	In the following we will explore the systematic uncertainties which arise in direct and indirect DM searches as a consequence of the modelling assumptions regarding the DM phase-space distribution. In particular, we focus on the so-called astrophysical factors that enter the interpretation of direct detection experiments, as well as the velocity boost factors that are needed for accurately predicting the velocity-dependent annihilation rates in indirect searches.
	
	\subsection{Direct detection}
	\label{sec:searches_direct}
	
	Direct detection (DD) experiments provide a unique probe for investigating possible interactions between dark and ordinary matter at low energies. By using large exposures of selected target materials they are capable of setting strong limits on the scattering rate of atomic nuclei of the Galactic DM particles. Due to their high sensitivities, it is crucial to have accurate predictions for the expected DM signals, but also robust methods of rejecting various backgrounds, which both require accurate models for the distribution of DM within the Milky Way. This will become even more important for the next-generation detectors, as they are excepted to become sensitive to new types of backgrounds, such as solar and atmospheric neutrinos, as well as observational signatures, e.g. directional distribution of the events. In order to correctly interpret their results, accurate models for the DM signatures -- e.g., the energy spectrum, yearly variations or spatial distribution of the expected events, which all crucially depend on the local velocity distribution of DM -- will become paramount.
	
	For a DM candidate with a given differential DM-nucleus cross-section, $\s{d}\sigma / \s{d}E_r$, the expected differential recoil rate per target nucleus can be computed as~\cite{goodman_detectability_1985,DrukierEtAl1986,FreeseEtAl2013}:
	\begin{align}
	\frac{\s{d}R}{\s{d}E_r} = \frac{\rho_\odot}{m_A m_\chi} \cdot \int_{|\textbf{v}| > v_\textrm{min}} \s{d}^3v \; P_\st{LAB}(v) \cdot v \cdot \frac{\s{d} \sigma}{\s{d} E_r} \label{eqn:rate_dif} \\
	\textrm{with} \;\;\; v_{\textrm{min}} = \sqrt{\frac{m_A E_r}{2 \mu^2_{A \chi}}} \; \; \; , \; \; \; \mu_{A \chi} = \frac{m_A m_\chi}{m_A + m_\chi} \; , \nonumber
	\end{align}
	where $E_r$ is the nuclear recoil energy, $m_{A/\chi}$ the target nucleus/DM mass, $\rho_\odot$ the local DM density and $P_\st{LAB}(v)$ the local velocity distribution of DM particles in the detector's (LAB) frame. To perform the mapping from Galactic rest frame to the LAB frame one needs to correctly account for the local circular velocity, Sun's peculiar motion an Earth's circular velocity around the Sun. However, since in the simulations there is no clear analogue of the solar system, we simplify the mapping by accounting only for the dominant contribution which comes from the local circular velocity,  $\textbf{v}_\s{circ}(R_\odot)$, assuming that the detector is located at solar galactocentric distance $R_\odot = 8.1 \; \s{kpc}$.~\footnote{The chosen value of $R_\odot$ is to a large degree arbitrary since the stellar discs in simulations significantly differ in many respects from the one of the Milky Way. Therefore, no particular importance should be given to the absolute values of the astrophysical factors obtained in this work, but rather to the relative differences between the two theoretical predictions and the values extracted directly from the simulations.} In this case, the LAB velocity distribution can be related to the DM speed distribution, $P(v)$, as follows:
	\begin{align}
		P_\st{LAB}(v) = \int d^3 v' \; P(v') \cdot \delta\left(v - |\textbf{v}' - \textbf{v}_\s{circ}|\right) \, .
	\end{align}
	On the other hand, in simulations we first select all DM particles that lie within a torus centred at $(R,z) = (R_\odot, 0)$ with radius $\delta = 0.5 \; \s{kpc}$, which provides us with a sample of more than 5000 particles. Subsequently, we subtract $\textbf{v}_\s{circ}$ from the individual particle velocities and obtain $P_\st{LAB}(v)$ by binning them according to their velocity magnitudes.
	
	For spin-independent (SI) interactions obeying isospin symmetry, which correspond to coherent DM scattering with all the nucleons in the target atoms, the differential cross-section can be expressed as:
	\begin{align}
	\frac{\s{d} \sigma}{\s{d} E_r} = \frac{m_A \sigma^{\textrm{SI}}_{n}}{2 \mu^2_{A \chi} v^2} A^2 F^2(E_r) \; ,
	\end{align}
	where $\sigma^{\textrm{SI}}_{n}$ is the SI DM-nucleon cross-section at zero momentum transfer, $A$ the mass number of the target nucleus and $F(E_r)$ a nuclear form factor. As can been seen from the above expression, the SI differential cross-section introduces an additional factor of $v^{-2}$ within the integral of \citeeq{eqn:rate_dif}, which also appears in the case of spin-dependent (SD) interactions, however, this is not always true for more general scattering operators. Nonetheless, on can factorize \citeeq{eqn:rate_dif} into a term determined by the specific particle physics model under consideration and an astrophysical factor, which is the convolution of the process's velocity dependence with the DM's velocity distribution. For SI and SD case the relevant integral takes the following form:
	\begin{align}
	\label{eqn:g}
	g(v_\textrm{min}) \equiv  \int_{|\textbf{v}| > v_\s{min}} \s{d}^3v \; \frac{P_\st{LAB}(\textbf{v})}{v} \; .
	\end{align}
	However, it is often desired to go beyond the simplest scattering operators, since there are many other ways in which DM can couple to the nucleons. In order to address the wide range of possibilities, a fully general set of non-relativistic effective scattering operators has been assembled -- for their systematic treatment see~\cite{fitzpatrick_effective_2013,anand_model-independent_2013,dent_general_2015,bishara_chiral_2017}. For many phenomenologically interesting models the leading order contribution to the differential cross-section can be velocity independent, hence, it is useful to additionally define:
	\begin{align}
	\label{eqn:h}
	h(v_\textrm{min}) \equiv  \int_{|\textbf{v}| > v_\s{min}} \s{d}^3v \;\; P_\st{LAB}(\textbf{v}) \cdot v \; .
	\end{align}
	It turns out that the above functions, $g(v_\s{min})$ and $h(v_\s{min})$, cover the velocity dependencies of all possible non-relativistic effective scattering operators expanded up to the quadratic order in momentum transfer and relative velocity. Therefore, their accurate determination is of great importance for correctly interpreting the direct detection constraints on DM-nucleus interactions.
	
	In \citefig{fig:mochima_dd} we present the comparison of the inversion methods' predictions and the corresponding true values of $g(v_\s{min})$ and $h(v_\s{min})$ obtained directly from the Mochima simulation. For $g(v_\s{min})$, displayed in the left-hand side plot, the predictions of axisymmetric method notably differ from the Eddington prediction at low $v_\s{min}$, which is a consequence of its significantly lower azimuthal velocity dispersion. Since the LAB is approximately moving with the local circular velocity in the azimuthal direction, one expects significantly less scatterings with $v_\s{min} \lesssim 100 \, \s{km/s}$ within the axisymmetric model.	On the other hand, above $v_\s{min} \sim 100 \, \s{km/s}$ the trend is reversed (i.e. $g(v_\s{min}$) associated with axisymmetric model becomes larger), while for $v_\s{min} \gtrsim 300 \, \s{km/s}$ the two models become virtually indistinguishable. By comparing the predictions of the inversion methods with the values extracted from the simulation one can immediately notice that the aforementioned trends of axisymmetric model are too extreme and $g(v_\s{min})$ is better approximated by the Eddington inversion. This is a consequence of the fact that the axisymmetric method tends to under-predict the velocity dispersion along the azimuthal direction at intermediate galactocentric distances, as previously noted in \citesec{sec:results_velocity}. However, this discrepancy is rather small and does not amount to relative errors grater than a few per cent. On the other hand, at $v_\s{min} \gtrsim 300 \, \s{km/s}$, where the inversion methods lead to nearly identical predictions, the relative differences with respect to the simulation can become significantly larger, reaching up to $50 \%$. The later is most likely sourced by non-equilibrium features, such as DM substructures or debris flow, which also manifest themselves as a additional peaks in the velocity distributions at large $v$ (see, e.g., the middle panel of \citefig{fig:mochima_vel}). In the case of $h(v_\s{min})$ the differences between the axisymmetric and Eddington inversion are reversed -- at small $v_\s{min}$ the axisymmetric method leads to larger values of $h(v_\s{min})$, while the opposite is true at intermediate values of $v_\s{min}$. Since this astrophysical factor is more sensitive to the high-velocity tail of the $P_\st{LAB}(\textbf{v})$, the axisymmetric method leads to slightly more accurate prediction at low $v_\s{min}$, however, the relative difference with respect to the Eddington inversion is below the per cent level. At large $v_\s{min}$ the inversion methods again result in very similar values for $h(v_\s{min})$, while the relative difference with respect to the simulation grows up to $\sim 50 \%$ due to the same reason as in the case of $g(v_\s{min})$.
	
	\begin{figure}[H]
		\centering
		\includegraphics[width=\textwidth]{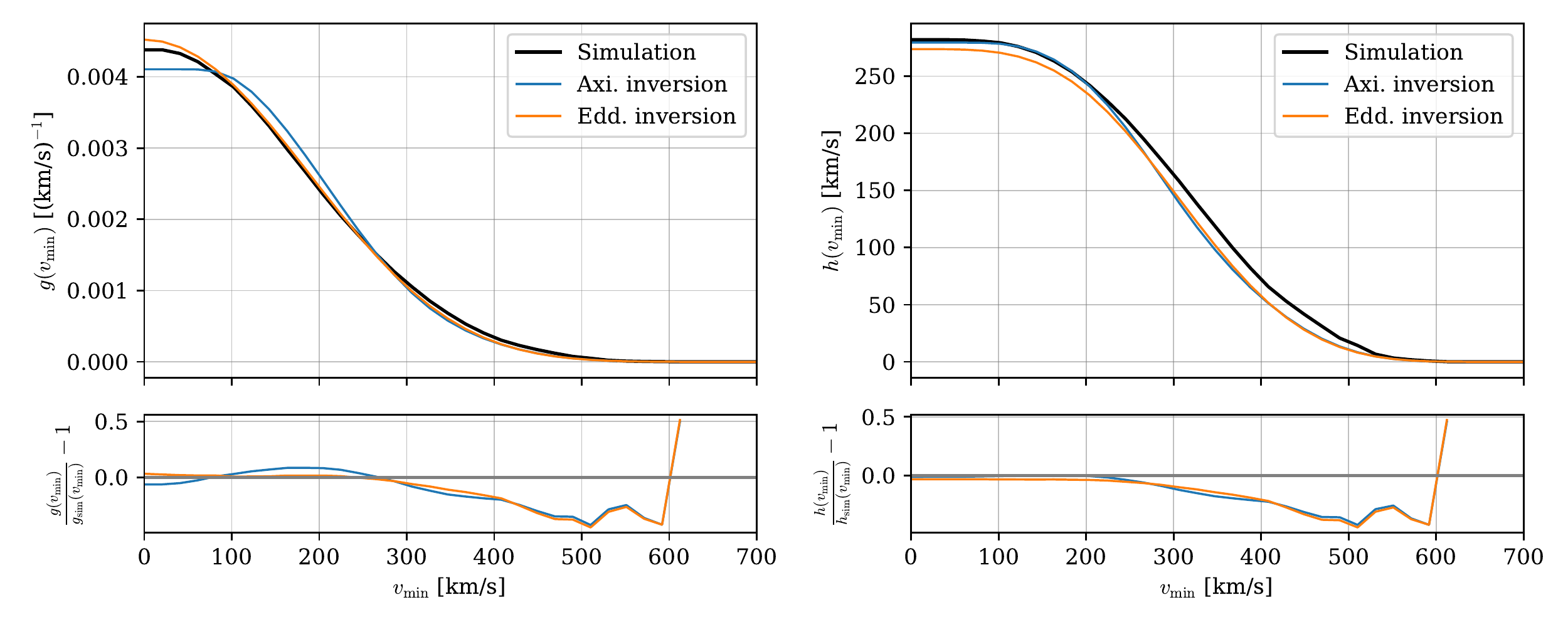}
		\caption{Comparison of direct detection astrophysical factors extracted directly from the Mochima simulation with the corresponding predictions of the inversion methods. The lower pannels show the relative differences between predictions and the true values.}
		\label{fig:mochima_dd}
	\end{figure}

	Similar trends can be observed in the case of Halo B and Halo C simulations, for which the analogous plots of astrophysical factors can be found in \citefig{fig:halob_dd} and \citefig{fig:haloc_dd} of \citeapp{app:halo_bc}. In both objects $g(v_\s{min})$ is more accurately predicted the Eddington inversion, while the axisymmetric method can lead to even larger relative errors than those observed in the case of Mochima. Furthermore, we generally find a better agreement between the predictions of the Eddington inversion and the simulations also for $h(v_\s{min})$. Both of this findings are most likely related to the fact that the Eddington approach tends to predict the individual components of velocity distribution more accurately around $R_\odot$, as can be seen from \citefig{fig:halob_entropy} and \citefig{fig:haloc_entropy}. Similarly as in the case of Mochima, the relative errors in the predictions of the inversion methods are relatively small (below $15\%$) at low $v_\s{min}$, but can exceed $50 \%$ at large $v_\s{min}$.
	
	The above results highlight several issues that hinder our ability to accurately determine the astrophysical factors which enter the interpretation of DD experiments. While the axisymmetric inversion method is generally expected to perform better in describing the DM velocity distribution in galaxies with massive baryonic discs, this is not necessary the case. In particular, the complex dynamics of galaxy formation can lead to significant deviations from the assumed axial symmetry, which in turn causes the axisymmetric method to be less accurate than the simpler Eddington inversion. Further errors in the predictions can also be sourced by the unaccounted rotation of the DM halo.
	On the other hand, our results show that both methods lead to relative errors smaller than $15 \%$ at low $v_\s{min}$, however, they can become significantly larger with increasing $v_\s{min}$. The latter is most likely related to the presence of numerous non-equilibrium structures which can not be accounted for within a framework based on dynamical equilibrium. As a consequence, this can source appreciable uncertainties in the scattering rate of light DM candidates or the high-energy part of the recoil energy spectrum.
	
	\subsection{Indirect detection}
	\label{sec:searches_indirect}
	
	If DM particles can annihilate in SM states, which is generically true for thermal relic candidates, the associated emissions could be detected through various messengers, ranging from $\gamma$-rays, neutrinos to cosmic rays. Currently the dominant limits on most annihilation channels come from $\gamma$-ray observations of the galactic center~\cite{collaboration_search_2018,chang_search_2018,fermi-lat_collaboration_search_2019,abazajian_strong_2020} and dwarf satellite galaxies~\cite{collaboration_dark_2014,ackermann_searching_2015,archambault_dark_2017,boddy_model-independent_2018,hoof_global_2018,the_hawc_collaboration_search_2020,rico_gamma-ray_2020,alvarez_dark_2020}, however, important complementary bounds can be obtained from Cherenkov and neutrino telescopes -- for a review on the topic see, e.g.,~\cite{lavalle_dark_2012,gaskins_review_2016, hinton_multi-messenger_2020}.
	
	The expected flux due to pair annihilation of DM into SM particles, for a given DM halo with a PSDF $f(\textbf{x}, \textbf{v})$  integrated over angular acceptance $\Delta \Omega$, is given by:
	\begin{align}
	\label{eqn:gen_annihilation_flux}
	\frac{\s{d} \Phi}{\s{d}E} = \frac{1}{8 \pi} \frac{\sv_0}{m^2_{\chi}} \frac{\s{d}N}{\s{d}E} \, \int_{\Delta \Omega} \s{d} \Omega \int_{\textrm{l.o.s.}} \s{d} \ell \int \s{d}^3 v_1 f (\textbf{x}, \textbf{v}_1) \int \s{d}^3 v_2 \;  f(\textbf{x}, \textbf{v}_2) \, S( | \textbf{v}_{\textrm{rel}}|) \,,
	\end{align}
	where the DM particle, $\chi$, is assumed to be its own antiparticle (otherwise an extra factor of 1/2 is needed), $m_{\chi}$ is its mass and $dN/dE$ the energy spectrum of the emitted radiation. The above formula is applicable to the general case in which the pair annihilation cross-section $\sv$ has a non-trivial dependence on the modulus of the relative velocity $v_\s{rel} = |\textbf{v}_1 - \textbf{v}_2|$, with $\textbf{v}_1$ and $\textbf{v}_2$ being the velocities of two annihilating particles, namely $\sv$ is factorised into the velocity independent term $\sv_0$ and a dimensionless factor fully comprising its dependence on relative velocity, $\sv = \sv_0 \cdot S(v_\s{rel})$.
	
	By isolating the astrophysical contribution in \citeeq{eqn:gen_annihilation_flux}, one can define:
	\begin{align}
	\label{eqn:j_factor}
	J \equiv & \int_{\Delta \Omega} \s{d} \Omega \int_{\textrm{l.o.s.}} \s{d} \ell \int \s{d}^3 v_1 f_\st{DM} (\textbf{x}, \textbf{v}_1) \int \s{d}^3 v_2 \;  f_\st{DM}(\textbf{x}, \textbf{v}_2) \, S(v_\s{rel}) \nonumber\\
	= & \int_{\Delta\Omega} \s{d} \Omega \int_{\rm l.o.s.} \s{d} \ell \, \rho^2(\textbf{x}) \ \langle S (v_\s{rel}) \rangle (\textbf{x}) \, .
	\end{align}
	This definition is in analogy to what is usually denoted in the literature as ``$J$-factor'', which is typically limited to the standard lore of s-wave annihilations, in which $\sv$ is velocity independent.
	In the latter case, the velocity boost factor can be omitted, i.e. $\langle S (v_\s{rel}) \rangle = 1$, and the $J$-factor simply depends on the DM density distribution along the line of sight, making the predictions of inversion methods irrelevant.
	However, there exist several well-motivated DM models in which s-wave annihilations are forbidden or severely suppressed, and hence p-wave becomes the dominant contribution to the annihilation cross-section~\cite{hagelin_perhaps_1984,kim_minimal_2007,pospelov_secluded_2008,lee_singlet_2008}, leading to $S(v_\s{rel}) \propto v_\s{rel}^2$. Alternatively, non-perturbative effects due to long-range interactions in the non-relativistic limit, commonly known as Sommerfeld enhancement, can introduce an additional velocity dependence, which can be in certain limiting cases well approximated by inverse powers of $v_{\textrm{rel}}$~\cite{iengo_sommerfeld_2009,dent_thermal_2010,slatyer_sommerfeld_2010,tulin_beyond_2013}. More precisely, under the assumption of the Yukawa coupling one finds $S(v_\s{rel}) \propto v_\s{rel}^{-1}$ in the Coulomb regime (i.e. for vanishing mediator mass) and as $S(v_\s{rel}) \propto v_\s{rel}^{-2}$ in the resonant regime, which occurs for particular values of the associated particle physics parameters. This motivates us to consider four different power-law scalings, namely $S(v_\s{rel}) \propto v_\s{rel}^{n}$ for $n \in [-2,-1,1,2]$, corresponding to all possible combinations of the aforementioned velocity dependences.
	
	In \citefig{fig:mochima_rel_moments} we present our results for the four moments of the relative velocity as a function of the radial distance along the galactic plane for the Mochima simulation. In the left-hand side panel we show the values of $\langle v_\s{rel}^{n} \rangle$ extracted from the simulation, as well as the predictions of the axisymmetric and Eddington inversion methods, while in the right-hand side panel we show the corresponding relative differences between the predictions and the simulation. Not surprisingly, the accuracy of the predictions for $\langle v_\s{rel}^{n} \rangle$ is fairly similar to the one obtained for $\langle v^n \rangle$, presented in \citesec{sec:results_moments}. The relative differences between the simulation and inversion methods are also in this case smaller than 7\% (15\%) for first (second) moments over the entire radial range, with the exception of $\langle v_\s{rel}^{-2} \rangle$ at $R \gtrsim 30 \, \s{kpc}$. In the inner $\sim 30 \, \s{kpc}$ the inversion methods again systematically over predict the negative and under predict the positive moments, while at $R \gtrsim 30 \, \s{kpc}$ we observe a smaller, $\mc{O}(10\%)$, offset towards lower values only in the case of $\langle v_\s{rel}^{-2} \rangle$. The axisymmetric method tends to lead to more accurate predictions in the range $2 \, \s{kpc} \lesssim R \lesssim 10 \s{kpc}$. While at larger radii the difference between the two methods gradually diminishes, the innermost part of the galaxy again shows systematic under (over) prediction of negative (positive) velocity moments due to the miss-modelling of the central DM density slope. 
	
	Similar conclusions can be made for the moments of relative velocity distribution in Halo B and Halo C, for which the analogous plots are shown in Figures~\ref{fig:halob_rel_mom} and~\ref{fig:haloc_rel_mom} of \citeapp{app:halo_bc}. The key differences with respect to Mochima are the aforementioned discrepancies in the central parts of the haloes. In the case of Halo B, which has a significant mismatch between the centre of DM halo and barycentre, the axisymmetric method leads to significantly more accurate results than the Eddington inversion, while the opposite is true in the case of Halo C. Apart from the central mismatch of one of the methods, the typical relative errors amount to less than 10\% over most of the radial range. Furthermore, we find that the negative (positive) moments tend to be over (under) predicted by the inversion methods, which agrees with our observations based on the Mochima simulation.
	
	In conclusion, both inversion methods tend to provide fairly accurate estimates for the first two positive and negative moments of the relative velocity distribution. While the axisymmetric method performs notably better in Mochima and Halo B simulations, this is not necessarily true for all the objects, as seen in the case of Halo C. However, since the typical relative errors do not exceed $\sim 20\%$, apart from possible larger deviations in the case of $\langle v_\s{rel}^{-2} \rangle$, we expect that the main uncertainty in the indirect searches will continue to be sourced by the poorly determined DM density profiles of the observed galaxies. In particular, it turns out to be very difficult to properly constrain the central slope of the DM density profile which, however, crucially determines the strength of the expected annihilation signal, but can also noticeably affect the predictions of the inversion methods, as demonstrated on the example of single-particle velocity moments in the \citeapp{app:comparison_zhao}.

	\begin{figure}[H]
		\centering
		\includegraphics[width=\textwidth]{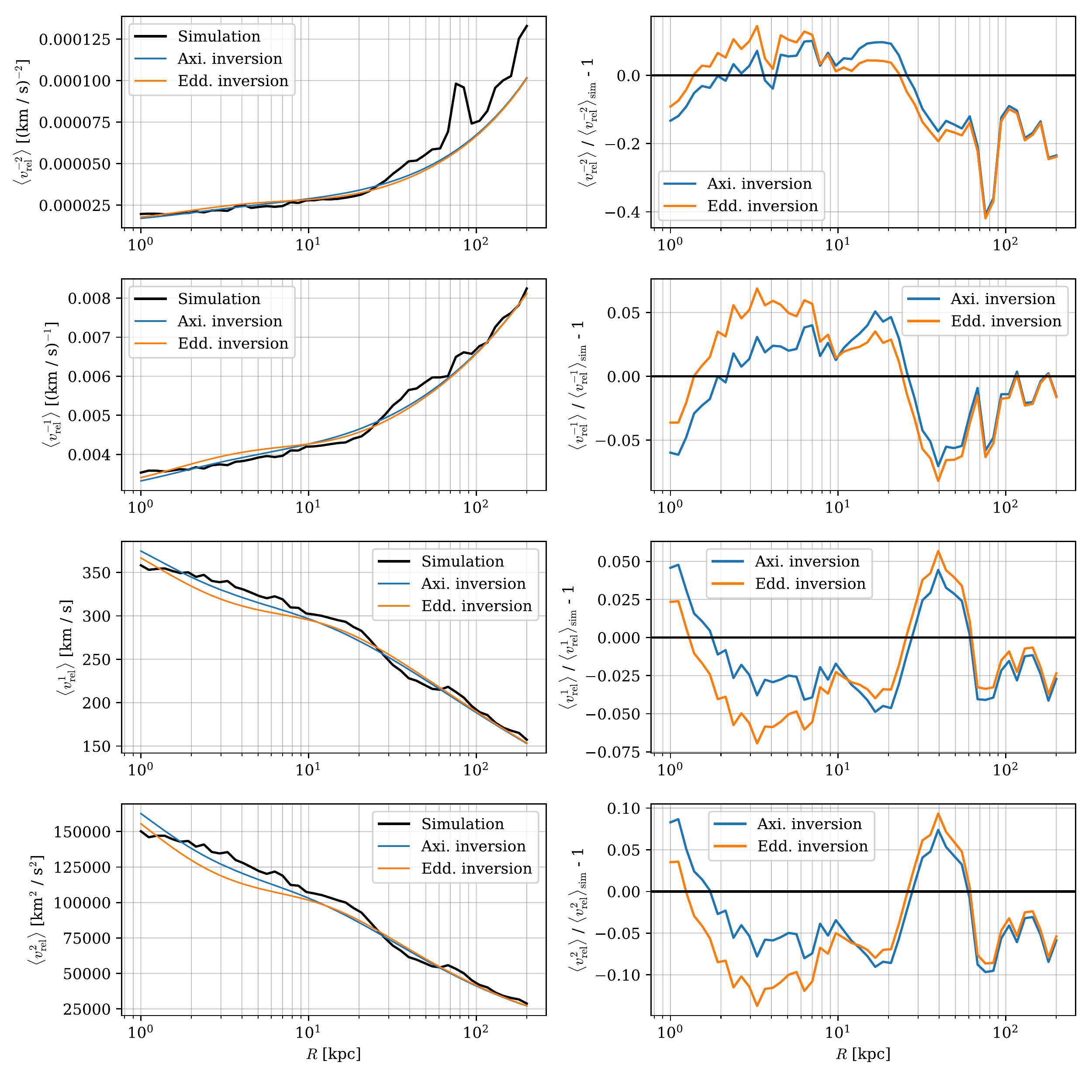}
		\caption{Values of the first two positive and negative moments of the relative speed distribution as a function of the radial distance along the galactic plane for the Mochima simulation. The left-hand side panels shows the values extracted directly from the simulation as well as the corresponding predictions of the inversion methods, while the right-hand side panels show the relative differences between the predictions and the true values.}
		\label{fig:mochima_rel_moments}
	\end{figure}
	
	\section{Summary and conclusions}
	\label{sec:conclusions}
	
	In this work, we have compared the accuracy of Eddington and axisymmetric inversion methods in predicting the DM phase-space distribution within galaxies. The two methods most notably differ in the assumed level of symmetry, with the Eddington formulation being limited to spherically symmetric objects, which represent merely a subclass of systems allowed by the more general axisymmetric approach. The comparison was performed on high-resolution hydrodynamical simulations, which make it possible to adequately sample the DM distribution throughout the DM halo. Since the DM density profile serves as an input quantity for the inversions, we focused our attention on the velocity distributions, which fully characterize the corresponding predictions for the phase-space distribution. In the following, we summarize the main results of this work.
	
	\paragraph{Fits of the baryonic gravitational potential and the DM density distribution}
	We began our analysis by fitting simple parametric functions for the key input quantities of the inversion methods to the simulations. For the baryonic gravitational potential, we adopted a minimalistic model that was composed of a Hernquist bulge and a single Miyamoto-Nagai disc, which nonetheless resulted in good fits with typical relative errors below 10\%. On the other hand, regarding the DM density distribution, several comments are in place. Most importantly, to avoid the need of computing the corresponding gravitational potential numerically and circumvent possible non-physical solutions of the axisymmetric method, we decided to adopt the two-parameter NFW and Burkert density profiles. The latter provided a good fit to the enclosed DM mass in the case of Halo B and Halo C, leading to residuals at the level of 10\%. Conversely, the NFW profile turned out to be a better choice for the Mochima simulation, however, resulted in residuals below 10\% only beyond the inner 5 kpc due to a particularly steep central cusp. This could be improved by allowing for more general DM density profile, e.g. the Zhao parametrization. However, as demonstrated in \citeapp{app:comparison_zhao}, this has only a few per cent impact on the resulting moments of DM velocity distribution and becomes negligible beyond the inner $\sim 2 \, \s{kpc}$. Finally, it has to be noted that the high-resolution simulations contain	significant amount of resolved DM substructures, which can not be captured by the standard parametric functions used for approximating the DM density distribution.
	
	\paragraph{Velocity distribution} After obtaining satisfying fits for the DM density profile and the baryonic potential gravitational, we were able to compute the corresponding PSDFs and compare their predictions with the simulations. Firstly, we provided a qualitative comparison by inspecting the probability density functions for velocity magnitude, meridional velocity and azimuthal velocity at distinctive radial distances along the galactic plane. For two out of three analysed objects, our results show that the axisymmetric method can lead to more accurate predictions for the speed distribution, particularly in the inner part of the galaxy, where the baryonic disc has a strong impact on the dynamics. The improvements manifest themselves as more accurate predictions for the peak as well as the general shape of the speed distribution. On the other hand, for Halo C the Eddington inversion method performed better, especially in the innermost few kpc. As the main reason for the failure of the axisymmetric approach, we identify the misalignment between the angular momentum of DM halo and the principal axis defined by the baryonic disc. Due to the same reason, the axisymmetric method does not always perform better in predicting the individual (meridional and azimuthal) components of the velocity distribution, which can be, however, mitigated by choosing a principal axis parallel to the halo's net angular momentum. Secondly, to quantify the match between the predicted velocity distributions with the ones inferred from the simulations over the entire radial range, we adopted the relative entropy metric. This confirmed our qualitative findings, clearly showing that the axisymmetric method leads to more accurate predictions for the speed distribution in two of the analysed simulations at all radii (with exception of few local fluctuations), while Eddington inversion performed notably better in the central part of Halo C. For the meridional and azimuthal velocity distributions, we observed significant variations in the accuracy within different simulations, which mainly stem from the aforementioned misalignment between the symmetry axes of DM and baryons. If the principal axis of the system was chosen perpendicular to the galactic disc, the predictions of the Eddington inversion often outperformed those of the axisymmetric method. On the other hand, if the principal axis was aligned with the angular momentum vector of the DM halo, the axisymmetric method generally leads to slightly more accurate predictions. As a final note, in the outskirts of all the analysed DM haloes we observed gradual degradation of the performance for both of the inversion methods due to an increasing amount of DM substructures and/or tidal debris.
	
	\paragraph{Moments of velocity distribution}
	Subsequently, we turned our attention to the first two positive and negative moments of the speed distributions, which are (unlike the relative entropy) particularly sensitive to the tails of the distribution. For the velocity moments, we find notably better agreement between the predictions of the axisymmetric method and the Mochima simulation at $2 \, \s{kpc} \lesssim R \lesssim 10 \, \s{kpc}$, where the impact of the baryonic disc is significant. However, it is important to note that in this radial range both inversion methods generically over predict the negative and underpredict the positive velocity moments by $\mc{O}(10\%)$, which was observed in all the studied objects. At larger $R$ the differences between inversion methods diminish, while the velocity moments extracted from the simulations become significantly noisier due to the presence of numerous DM substructures, which drastically worsens the agreement with the predictions. In the innermost part of the DM halo of Mochima simulation, our predictions tend to degrade due to the aforementioned mismatch in the DM density slope. On the other hand, in the case of Halo B (Halo C) we find very good agreement between the predictions of the axisymmetric (Eddington) method with the true velocity moments down to the very centre -- the better performance of Eddington inversion within Halo C is mostly driven by a large, i.e. $53^\circ$, misalignment of the DM halo's angular momentum with respect to the galactic disc, while in Halo B the axisymmetric method performs significantly better due to the presence of a prominent baryonic disc. In conclusion, both inversion methods (with the aforementioned exceptions in central parts of Halo B and Halo C) lead to fairly accurate predictions for the velocity moments over the majority of the explored radial range, with typical relative errors well below 20\%.
	
	\paragraph{Velocity anisotropy}
	Another interesting comparison, which is made possible by the high-resolution simulations, can be performed by examining the orbital anisotropy of DM. While in the case of Eddington inversion the resulting phase-space distribution is isotropic by construction (or the anisotropy profile has to be specified beforehand in its anisotropic generalizations), the axisymmetric method provides us with a prediction for the orbital anisotropy from a given density-potential pair. By comparing the anisotropy along the galactic plane, we found that the axisymmetric method correctly predicts radially biased orbits (i.e. $\beta_{\hat{R}} > 0$), however, quantitative agreement becomes poor at large galactocentric distances. In particular, the axisymmetric method leads to an orbital anisotropy profile which peaks around the disc scale length, while in simulations we find gradually increasing $\beta_{\hat{R}}(R)$ up to $R \sim 100 \, \s{kpc}$. The magnitude of the anisotropy approximately agrees with the prediction for the Mochima simulation at $R \lesssim 10 \, \s{kpc}$, while at larger radii the true values are substantially above the predicted ones. The agreement is even worse for Halo B and Halo C, where the axisymmetric inversion method overshoots the true values in the inner parts of the haloes and under predicts the anisotropy in the outskirts.
	
	\paragraph{Implications for direct and indirect searches}
	As a final comparison, we have contrasted the predictions of inversion methods to the true values of astrophysical factors that are needed for interpreting the results of direct and indirect DM searches. In the context of direct detection, our results show that both inversion methods perform reasonably well, with the typical relative errors in the astrophysical factors below 15\% for $v_\s{min} \lesssim 300 \, \s{km/s}$. The axisymmetric method often performs worse than the Eddington inversion due to the aforementioned difficulties in predicting individual components of velocity distributions. On the other hand, at large $v_\s{min}$ the differences between the predictions of the two inversion methods become smaller than the typical relative errors with respect to the true values, as the latter can surge up to $50 \%$. The reason for this is most likely the presence of non-equilibrium structures within the DM halo, which can not be captured by the models based on dynamical equilibrium.
	In the context of indirect searches, we find that both inversion methods provide relatively robust predictions for the moments of the relative velocity distribution that enter the predictions for velocity-dependent DM annihilations. Similarly to the single-particle velocity moments, the typical relative errors are below 20\% over most of the studied radial range. While the axisymmetric method performs better in the case of Mochima and Halo B simulations, the opposite in true for Halo C. Finally, we note that the differences between inversion methods, as well as residuals with respect to the simulations, are relatively small compared to the typical uncertainties associated with the observational determinations of the DM density profiles.
	
	In conclusion, the inversion methods provide us with an indispensable tool for reconstructing the phase-space distribution of the galactic DM. Both, the Eddington inversion and its axisymmetric generalization, allow us to accurately predict the DM velocity distribution over a broad range of galactocentric distances, as well as the corresponding astrophysical factors that needed for the interpretation of various DM searches. The axisymmetric approach generally leads to more accurate results, especially in the central parts of galaxies with massive baryonic discs. On the other hand, it is subject to additional systematic errors due to the assumptions related to the principal axis of the system as well as rotational properties of the DM halo. While the latter is most cases negligible, substantial misalignment between the principal axis of DM halo and the baryonic disc can spoil the predictions of the axisymmetric method and more accurate results are obtained through the simpler Eddington inversion. A further source of inaccuracy, common to both of the inversion methods, is the assumption that all of the galactic DM is smoothly distributed and has reached dynamical equilibrium. This is in stark contrast with the results of simulations, which predict a large amount of DM substructure that becomes increasingly prominent with the distance from the galactic centre. Nonetheless, the inversion methods manage to predict the moments of DM velocity distribution, which are of prime interest for DM searches, with typical relative errors well below $20 \%$.
	
	An interesting question, which we leave open for future work, is how the inversion methods compare to more general approaches of constructing DM distribution functions, such as action-angle modelling, where the restraining symmetry assumptions can be further relaxed. This could, for example, allow for misaligned principal axes of different galactic components, enable the use of more realistic triaxial description of the central bulge/bar regions of the galaxies or accommodate the large radial anisotropies of the DM particles in the outskirts of the halos, as predicted by the simulations. On the other hand, we expect that more flexible models could give rise to further systematic errors due to the additional degrees of freedom that can not be constrained through the existing observations. Beyond the scope of equilibrium models, there is also a pressing need for improvements in stochastic modelling of possible local and/or global non-equilibrium DM components, which can substantially affect the interpretation of DM searches. The development of such corrections for the equilibrium models most likely requires further theoretical work as well as novel insights provided by the high-resolution simulations of galaxy formation and evolution.
	
	\section*{Acknowledgments}
	
	We thank Thomas Lacroix, Benoit Famaey and other members of the GaDaMa initiative for stimulating discussions during the preparation of this work. Centre de Calcul Intensif d’Aix-Marseille is acknowledged for granting access to its high performance computing resources. The authors acknowledge partial support from the ANR project ANR-18-CE31-0006. This project has been partly supported by the European Union’s Horizon 2020 research and innovation program under the Marie Skłodowska -Curie grant agreement No 860881-HIDDeN. It has also benefited from funding from the CNRS-INSU programs PNHE and PNCG.
	
	\begin{appendices}
		
		\section{Dependence of the DM velocity moments on the assumptions regarding the DM density profile}
		\label{app:comparison_zhao}
		
		In our work we have adopted a simplifying assumption that the DM density profiles of simulated objects can be adequately approximated either by NFW~\cite{navarro_universal_1997} or Burkert~\cite{burkert_structure_1995} density profile. On the other hand, Lacroix et al.~\cite{lacroix_predicting_2020} performed the analysis of the same simulations using a more flexible Zhao parametrization~\cite{zhao_analytical_1996}. Even though the latter provides a somewhat better fit to the actual DM distribution found within the simulations, it does not have a significant impact on the moments of the velocity distribution of DM, which are of the prime interest for DM search. In \citefig{fig:moments_comparison} we show the comparison of DM velocity moments obtained in this work with those of Lacroix et al. for the Mochima simulation. As can be seen from the plots, the more flexible parametrization provides better agreement with the true velocity moments in the very central parts of the halo, since it correctly accounts for the particularly steep central cusp -- Lacroix et al. report that their best fit value for the central DM density slope is $\gamma = 1.718$, while the NFW parametrization implicitly assumes $\gamma = 1$. On the other, it is surprising to see that our results become more accurate at $R \gtrsim 2$ kpc. This is perhaps related to the fact that the two analyses also differ in the parametrization of the baryonic gravitational potential. At even larger galactocentric distances, i.e. $R \gtrsim 10$ kpc, the predictions of the Eddington inversion method in both works become nearly identical.
		
		\begin{figure}[H]
			\includegraphics[width=0.9\textwidth]{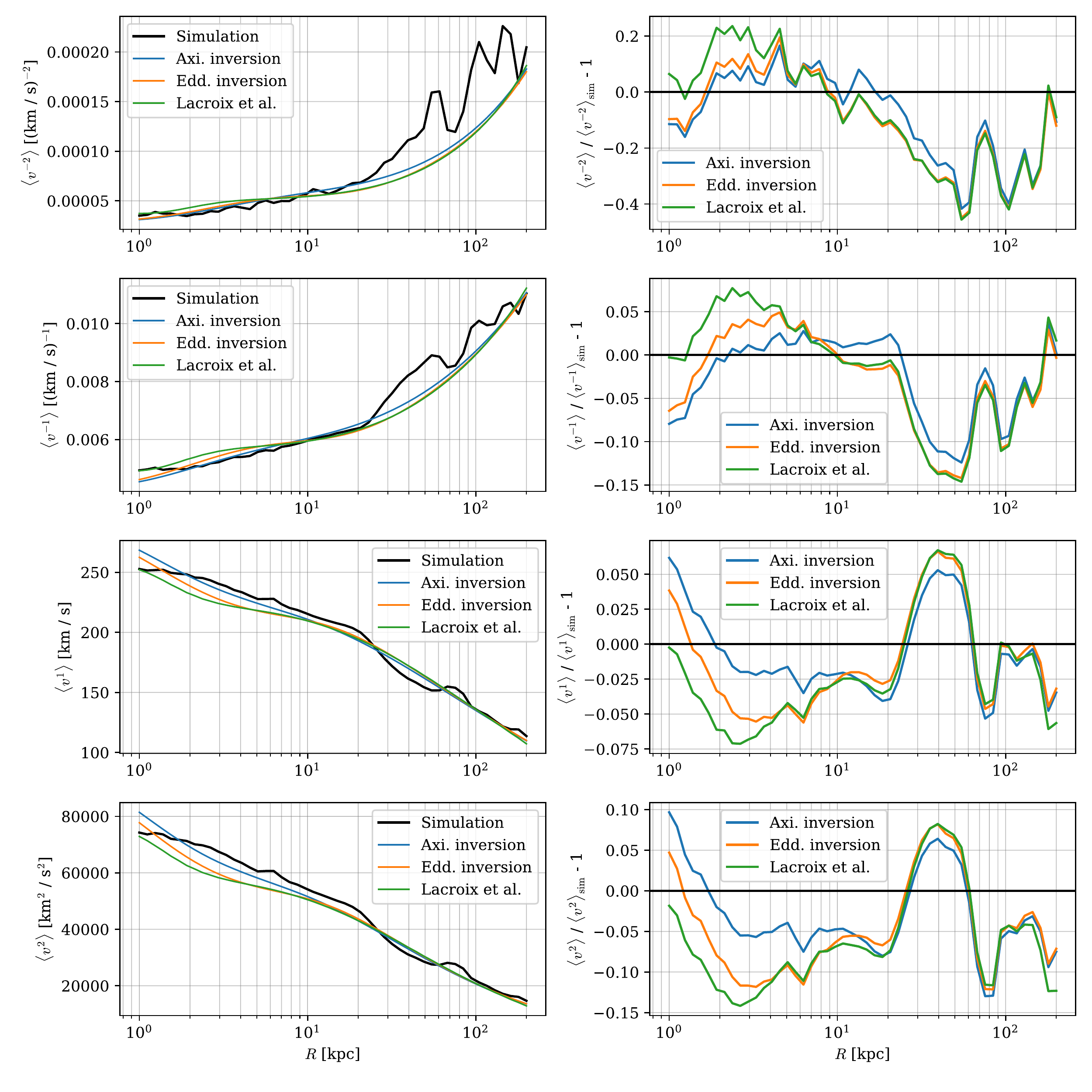}
			\caption{Same as in \citefig{fig:mochima_mom}, but additionally showing the results of Lacroix et al.~\cite{lacroix_predicting_2020}, who performed the Eddington inversion using a more accurate fit of the DM density distribution by adopting the Zhao profile as well as a more flexible approach of approximating the baryonic gravitational potential.}
			\label{fig:moments_comparison}
		\end{figure}
	
		\section{Systematic uncertainties associated with the choice of principal axis}
		\label{app:misalignment}
		
		To address the issue of misalignment between the principal axes of DM and baryons we repeat the analysis of Mochima simulation under the assumption that the $z$-axis coincides with the net angular momentum of the DM halo. The latter differs by $21^\circ$ from the normal vector defined by the galactic plane, which was used as the principal axis in the main text. This naturally leads to slightly different best-fit parameters for the baryonic gravitational potential, while the parameters of the DM density profile remain essentially unchanged since the halo is nearly spherical. On the other hand, the velocity distributions extracted from the simulations are noticeably different, particularly in the case of meridional and azimuthal components as their definitions clearly depend on the choice of the coordinate system. In \citefig{fig:mochima_velocity_tilted} we show the corresponding speed distribution, as well as the meridional and azimuthal velocity distributions, at three distinctive galactocentric radii. The solid lines correspond to the principal axis aligned with the angular momentum vector of the DM halo, while the dashed lines show the results obtained when $z$-axis was chosen perpendicular to the galactic disc. As can be seen from the plots, at $R = 3 \, \s{kpc}$ this leads to a better agreement of the axisymmetric method's predictions with the true values of the meridional and azimuthal velocity distribution, while the speed distribution is less affected by the change of principal axis. As a further comparison, we show in \citefig{fig:mochima_entropy_tilted} the corresponding values of $D_\st{KL}$, as defined by \citeeq{eqn:relative_entropy}. The axisymmetric method can be seen to outperform the Eddington inversion over most of the radial range, with the exception of few local fluctuations. In comparison with the case where the principal axis was chosen perpendicular to the galactic plane, the improvement is particularly significant for the meridional and azimuthal velocity distributions, while some improvement can also be seen in predictions for the speed distribution at small $R$.
		
		\begin{figure}[H]
			\centering
			\includegraphics[width=0.95\linewidth]{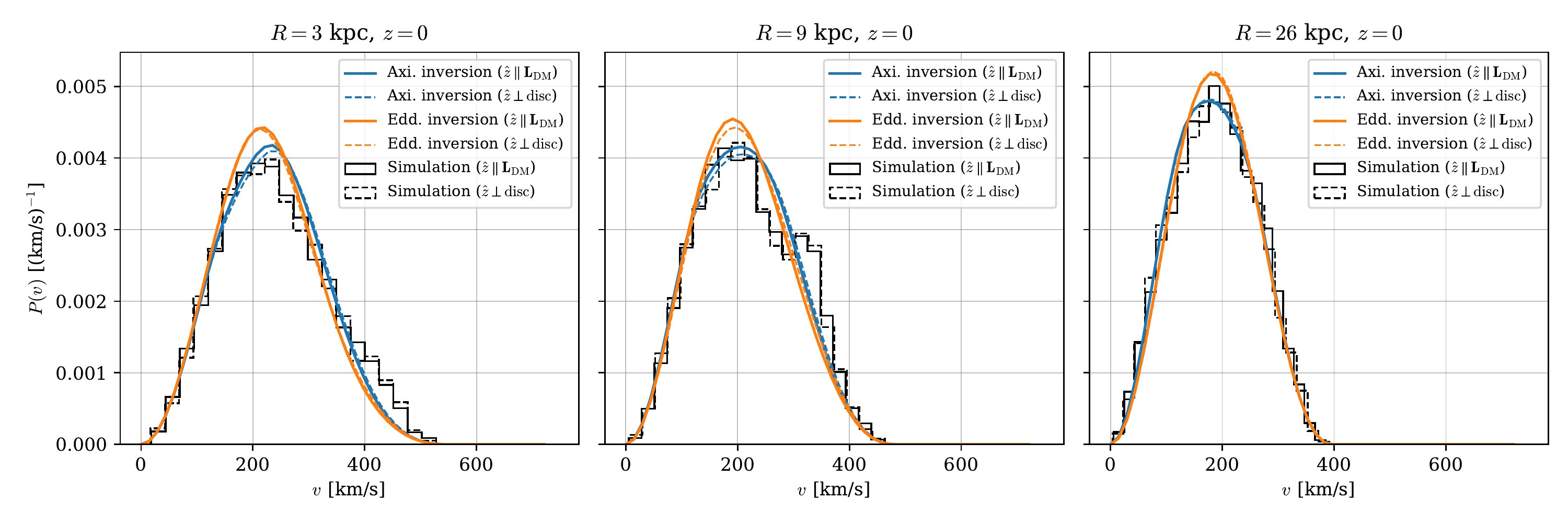}
			\includegraphics[width=0.95\linewidth]{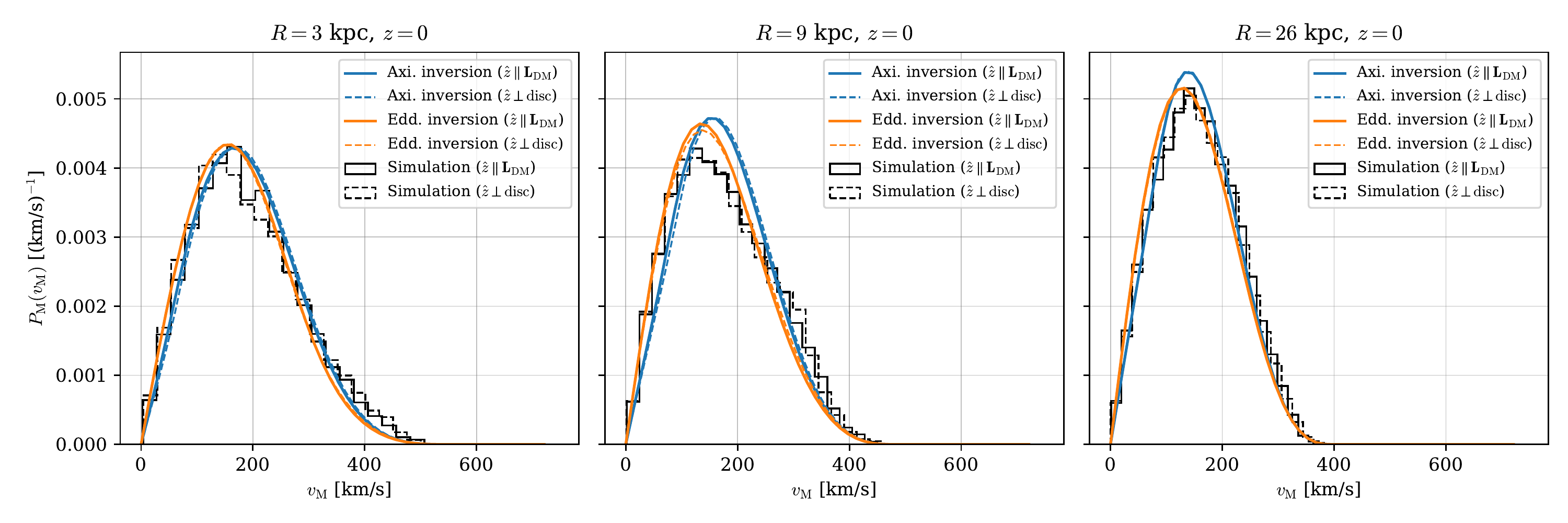}
			\includegraphics[width=0.95\linewidth]{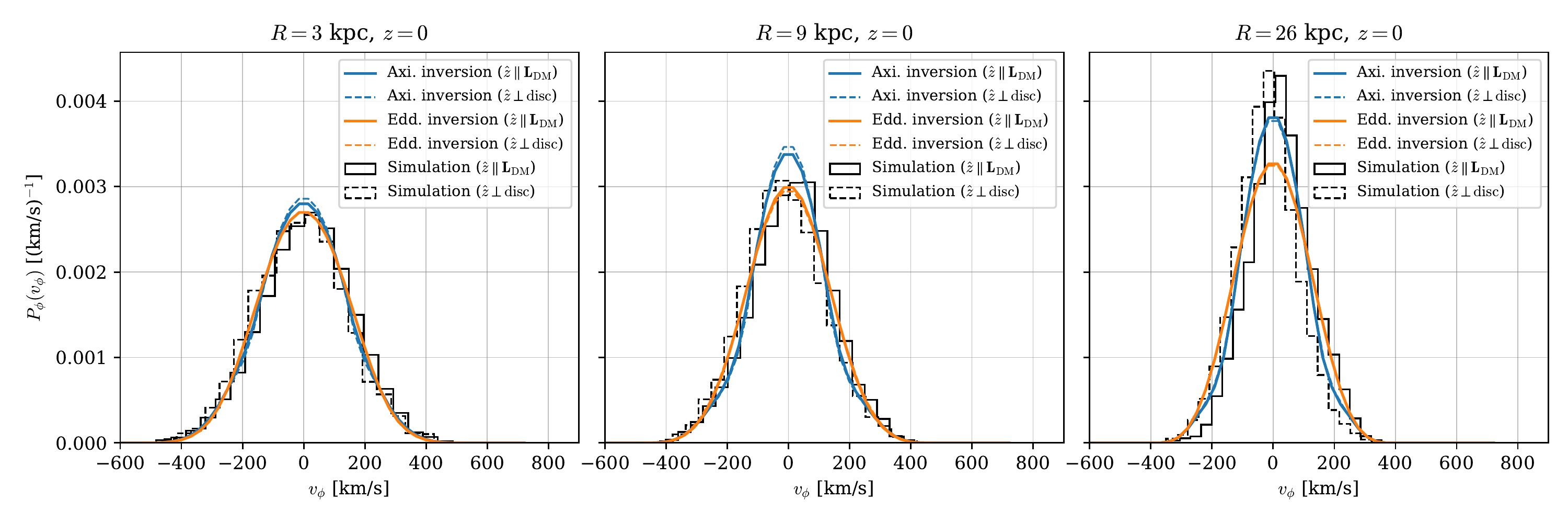}
			\caption{Speed distribution (top row), meridional velocity distribution (middle row) and azimuthal velocity distribution (bottom row) for Mochima simulation. The results are shown for the two different choices of $z$-axis alignment, i.e. parallel to the DM angular momentum ($\textbf{L}_\st{DM}$) and perpendicular to the baryonic disc.}
			\label{fig:mochima_velocity_tilted}
		\end{figure}
	
		\begin{figure}[H]
			\centering
			\includegraphics[width=\linewidth]{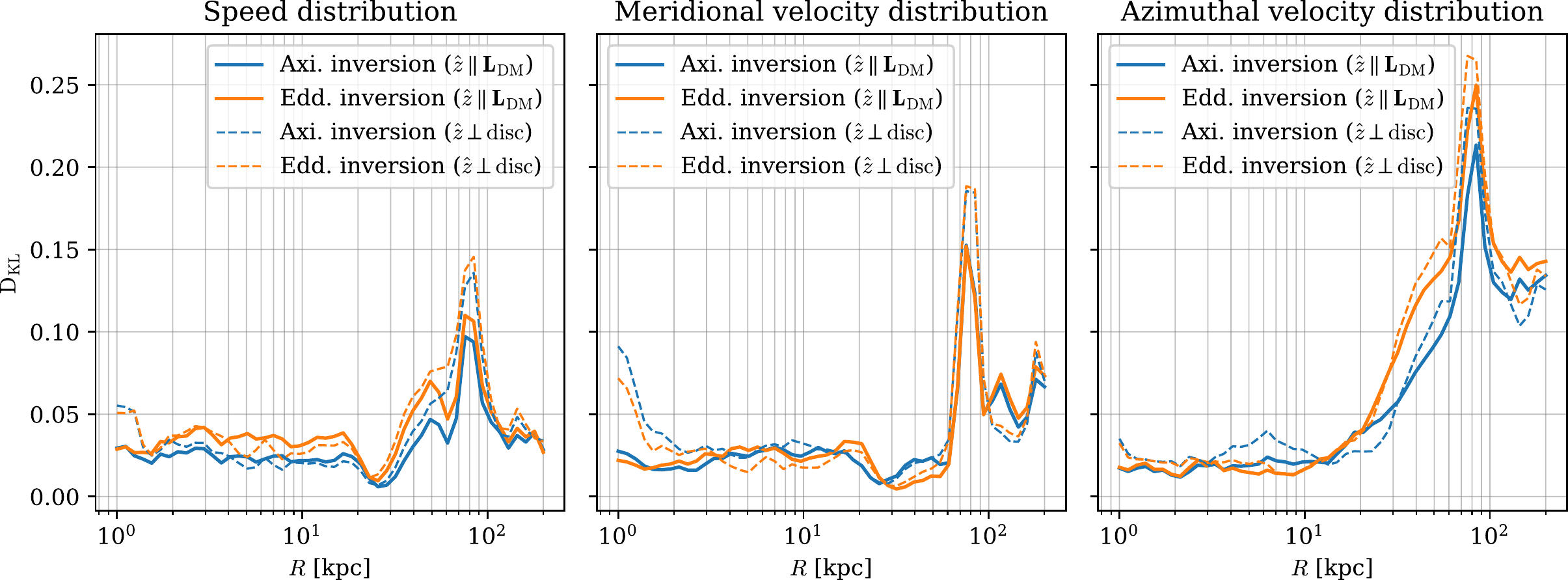}
			\caption{Relative entropy between the true velocity distributions and the predictions of axiymmetric and Eddington inversion methods as a function of the radial distance for Mochima simulation. The results are shown for the two different choices of $z$-axis alignment, i.e. parallel to the DM angular momentum ($\textbf{L}_\st{DM}$) and perpendicular to the baryonic disc.}
			\label{fig:mochima_entropy_tilted}
		\end{figure}
		
		\section{Results for Halo B and Halo C}
		\label{app:halo_bc}
		
		\subsection{Baryonic gravitational potential}
		
		\begin{figure}[H]
			\centering
			\includegraphics[width=0.49\linewidth]{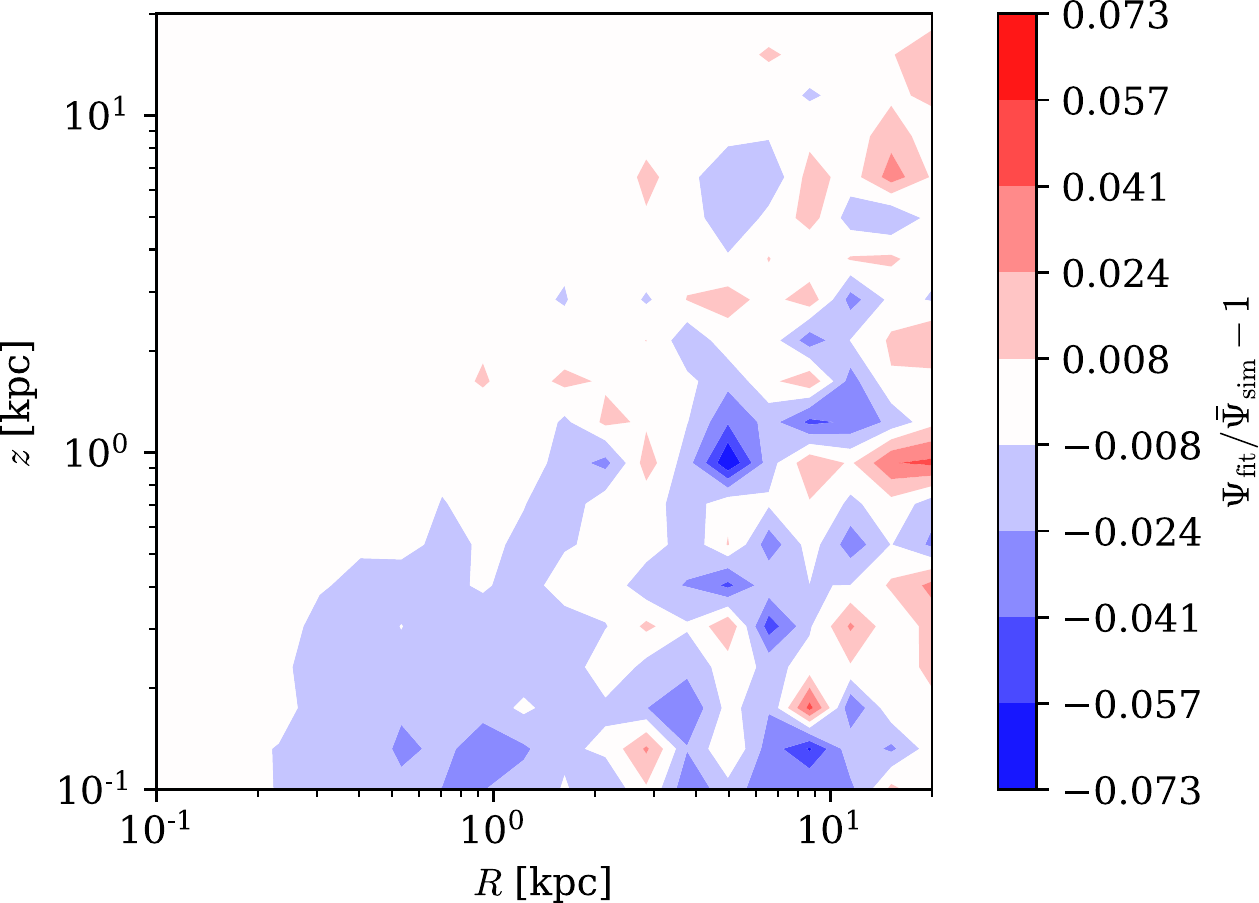}
			\includegraphics[width=0.49\linewidth]{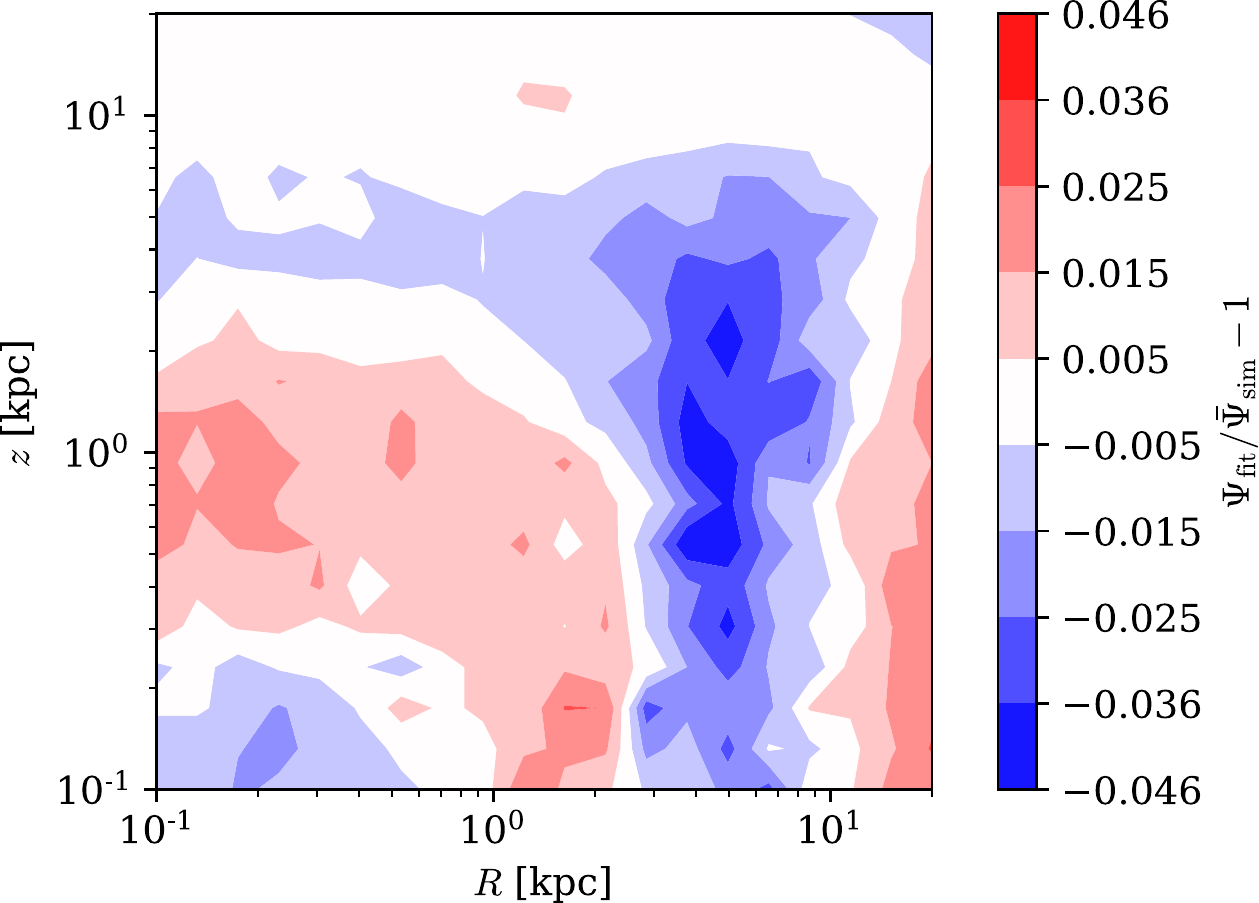}
			\caption{Relative difference between our best-fit and the azimuthally-averaged baryonic gravitational potential for Halo B (left-hand side panel) and Halo C (right-hand side panel) simulations.}
			\label{fig:halobc_potential}
		\end{figure}
		
		\subsection{Mass profiles}
		
		\begin{figure}[H]
			\centering
			\includegraphics[width=0.4\linewidth]{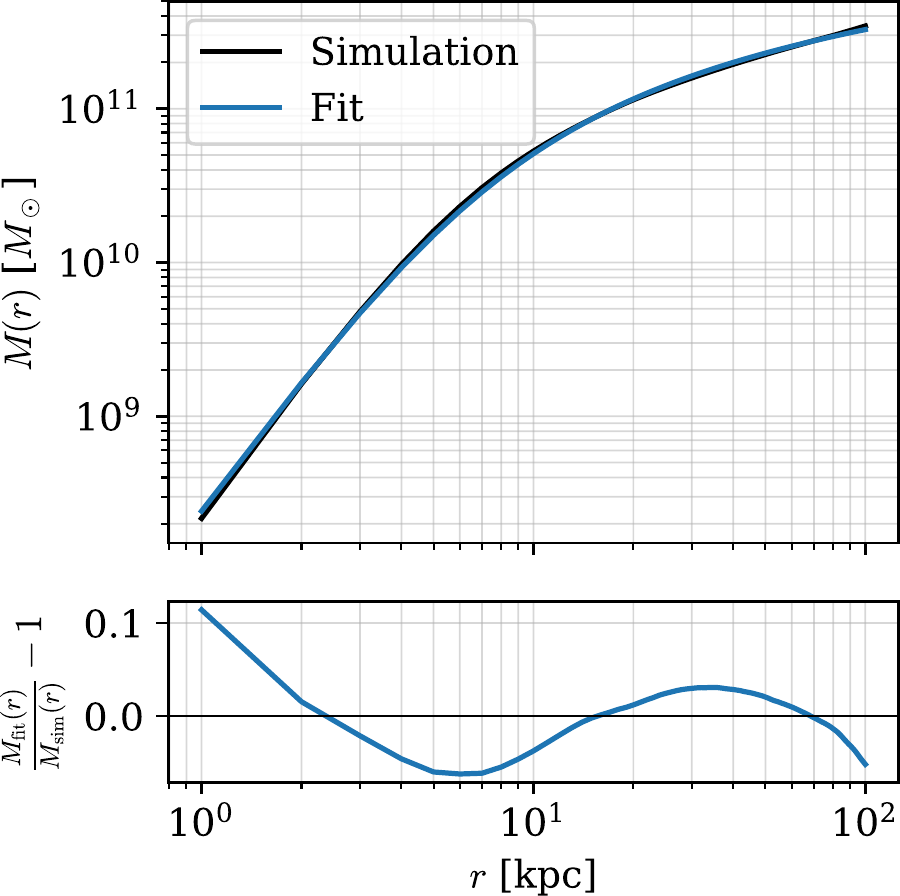}
			\includegraphics[width=0.4\linewidth]{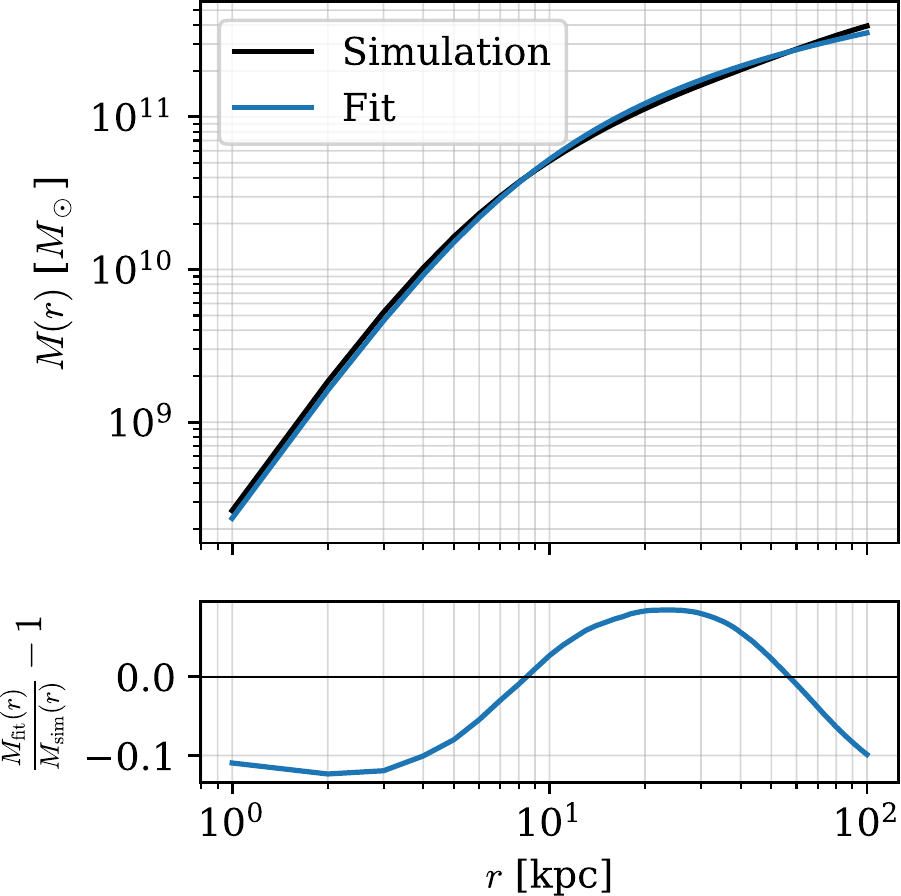}
			\caption{The best-fit DM mass profiles for Halo B (left-hand side panel) and Halo C (right-hand side panel) simulations, obtained under the assumption of Burkert parametrization.}
			\label{fig:halobc_mass}
		\end{figure}
		
		\subsection{Velocity distributions}
		
		\begin{figure}[H]
			\centering
			\includegraphics[width=\textwidth]{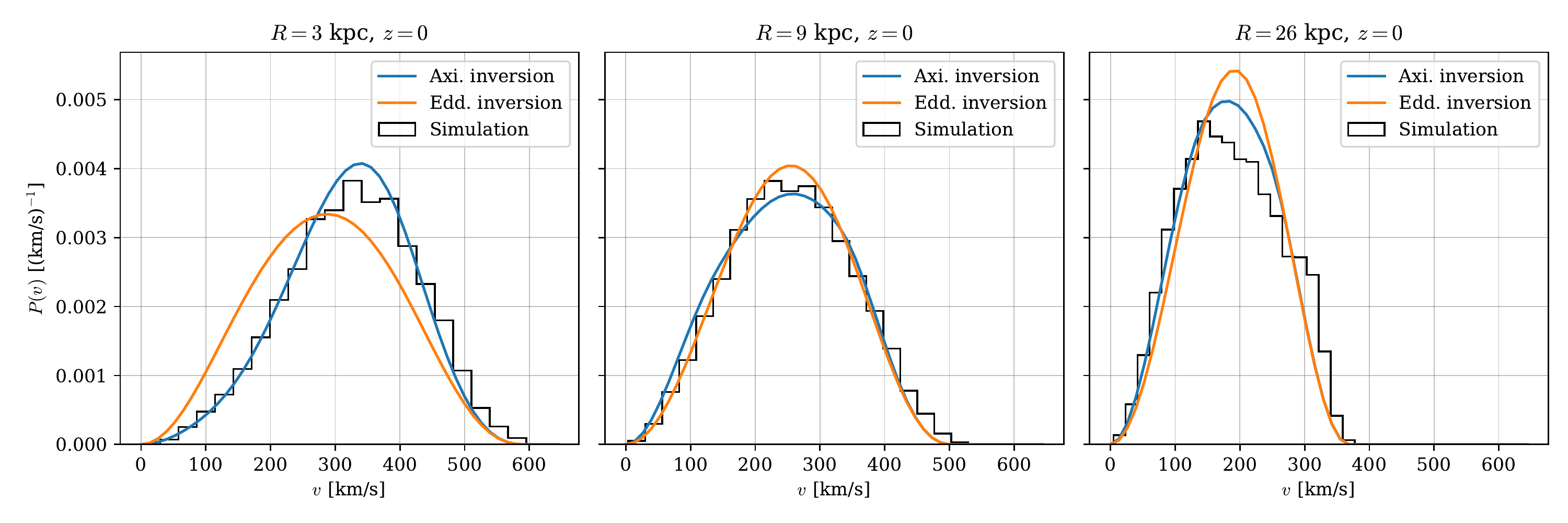}
			\includegraphics[width=\textwidth]{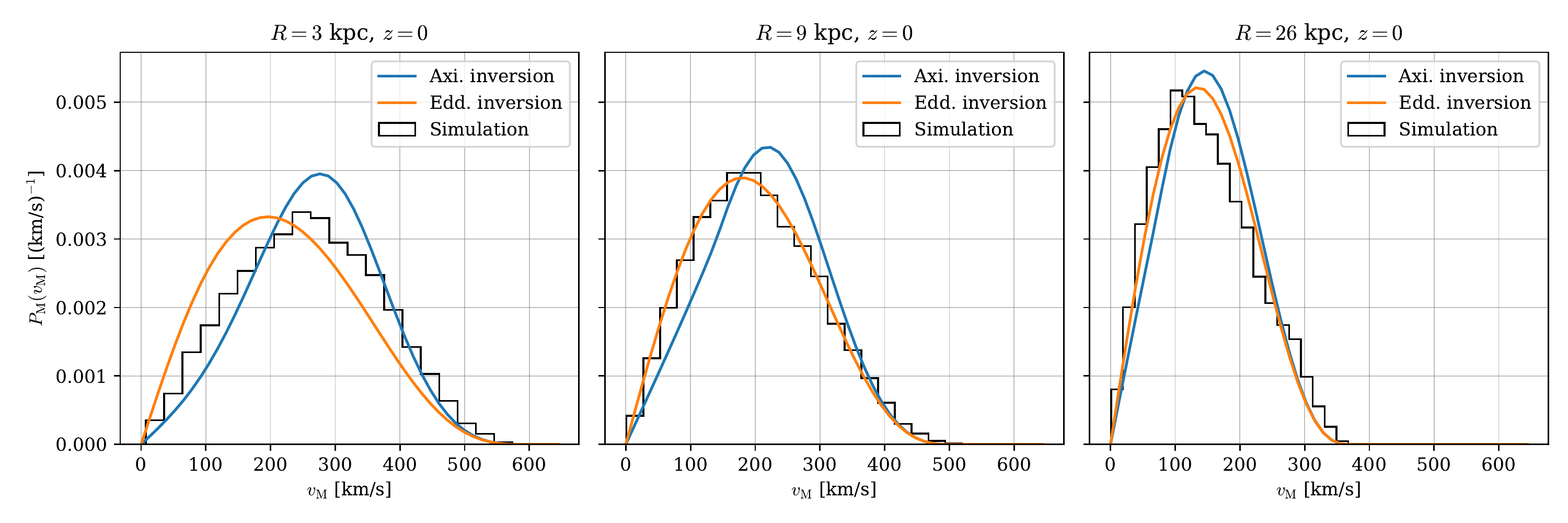}
			\includegraphics[width=\textwidth]{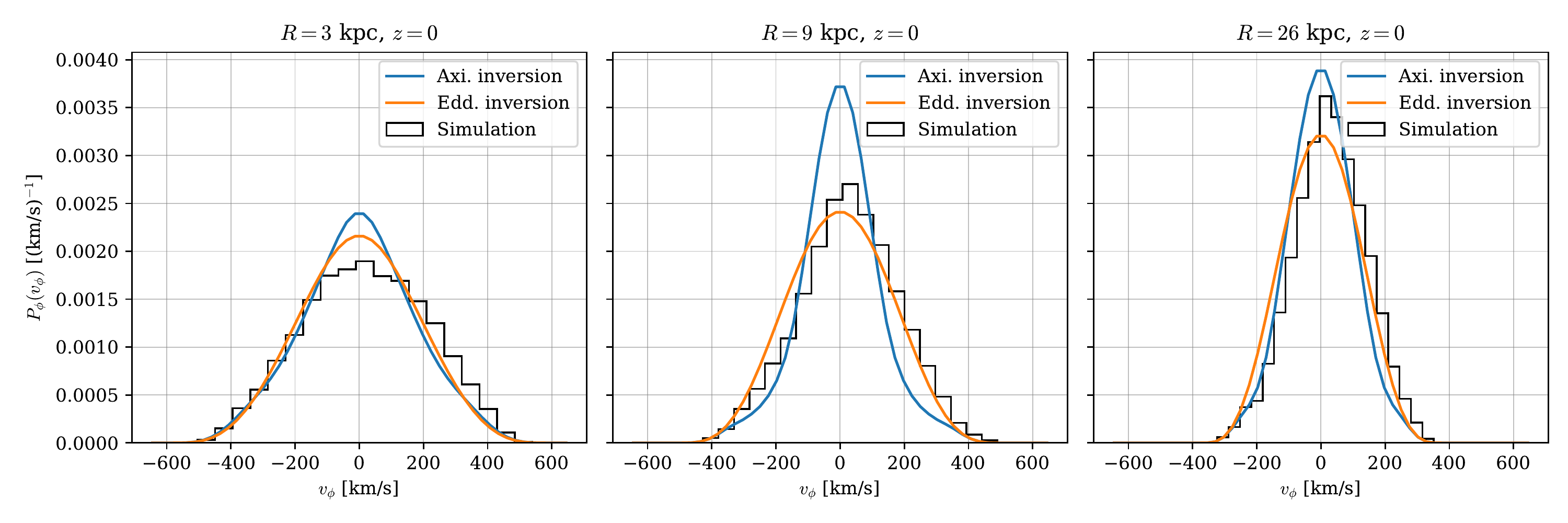}
			\caption{Speed distribution (top row), meridional velocity distribution (middle row) and azimuthal velocity distribution (bottom row) for Halo B simulation at three distinctive galactocentric distances along the disc.}
			\label{fig:halob_vel}
		\end{figure}
		
		\begin{figure}[H]
			\centering
			\includegraphics[width=\textwidth]{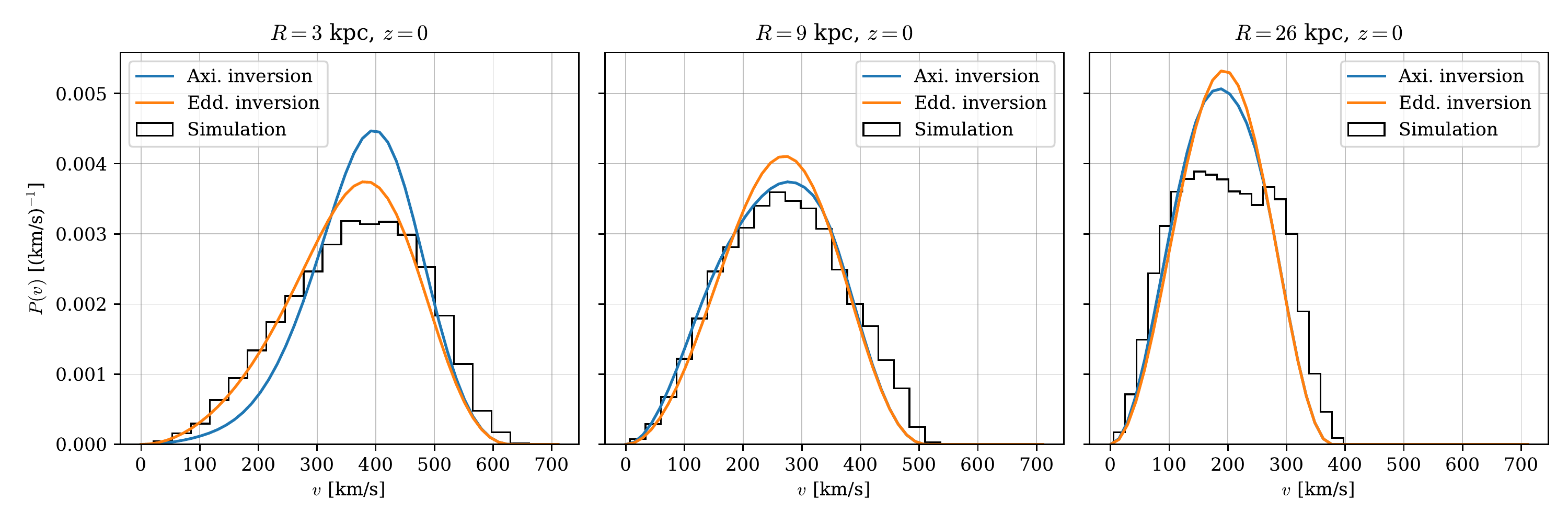}
			\includegraphics[width=\textwidth]{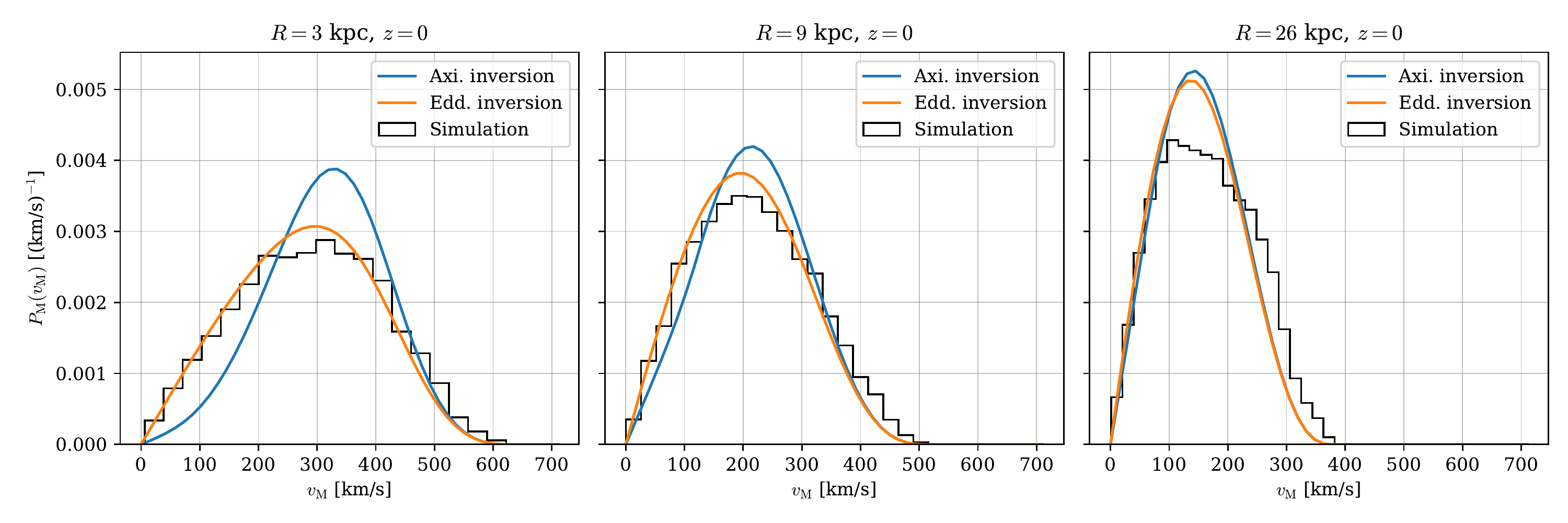}
			\includegraphics[width=\textwidth]{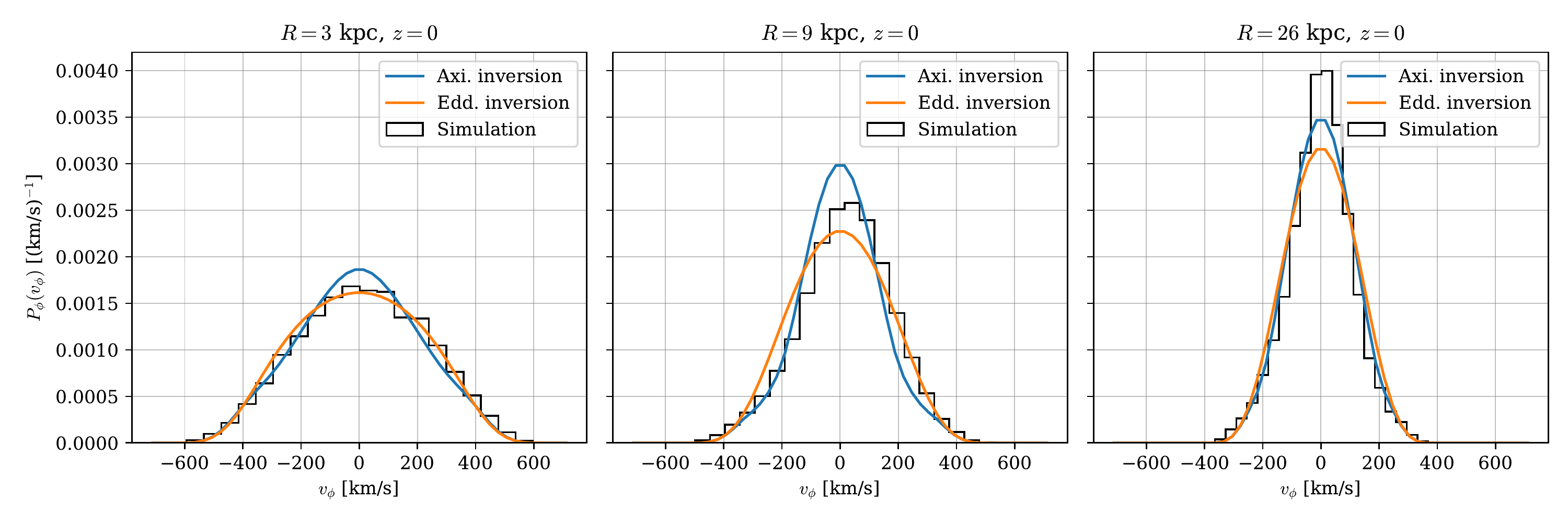}
			\caption{Speed distribution (top row), meridional velocity distribution (middle row) and azimuthal velocity distribution (bottom row) for Halo C simulation at three distinctive galactocentric distances along the disc.}
			\label{fig:haloc_vel}
		\end{figure}
	
		\begin{figure}[H]
			\centering
			\includegraphics[width=\textwidth]{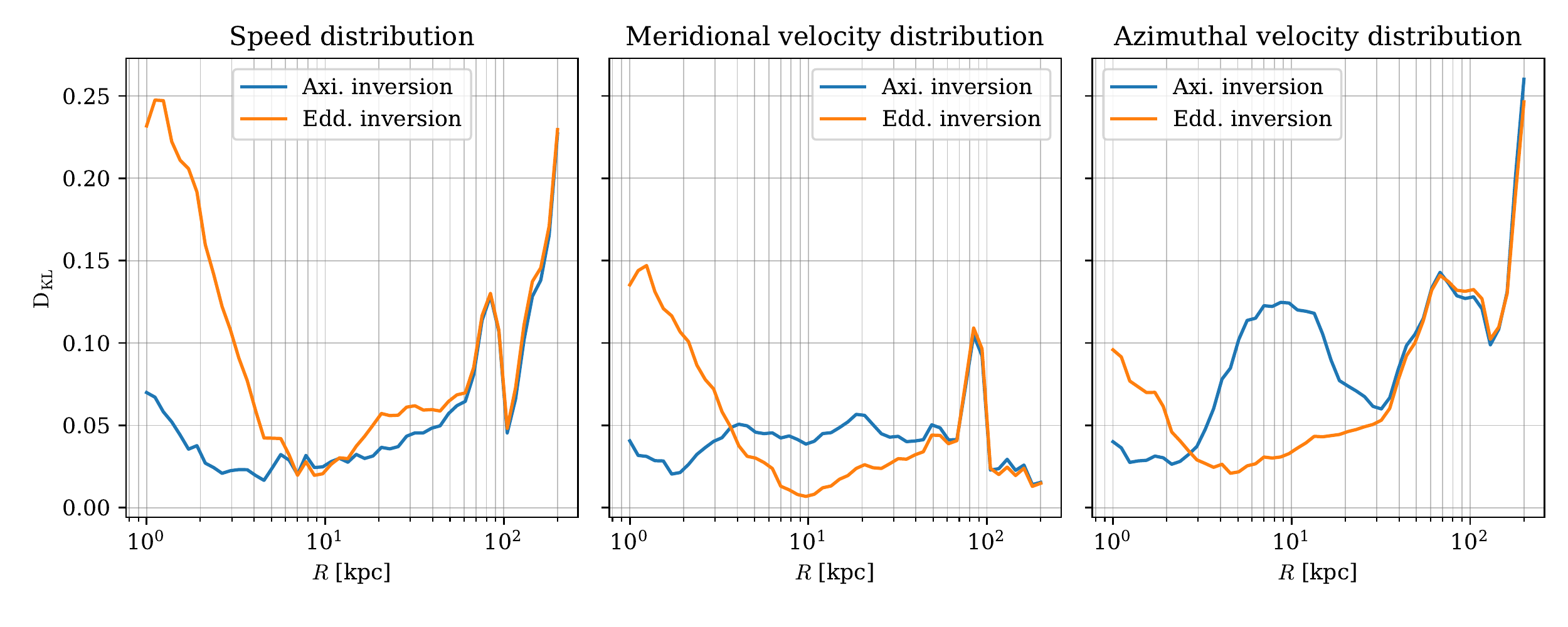}
			\caption{Same as in \citefig{fig:mochima_entropy}, but for Halo B simulation.}
			\label{fig:halob_entropy}
		\end{figure}
		
		\begin{figure}[H]
			\centering
			\includegraphics[width=\textwidth]{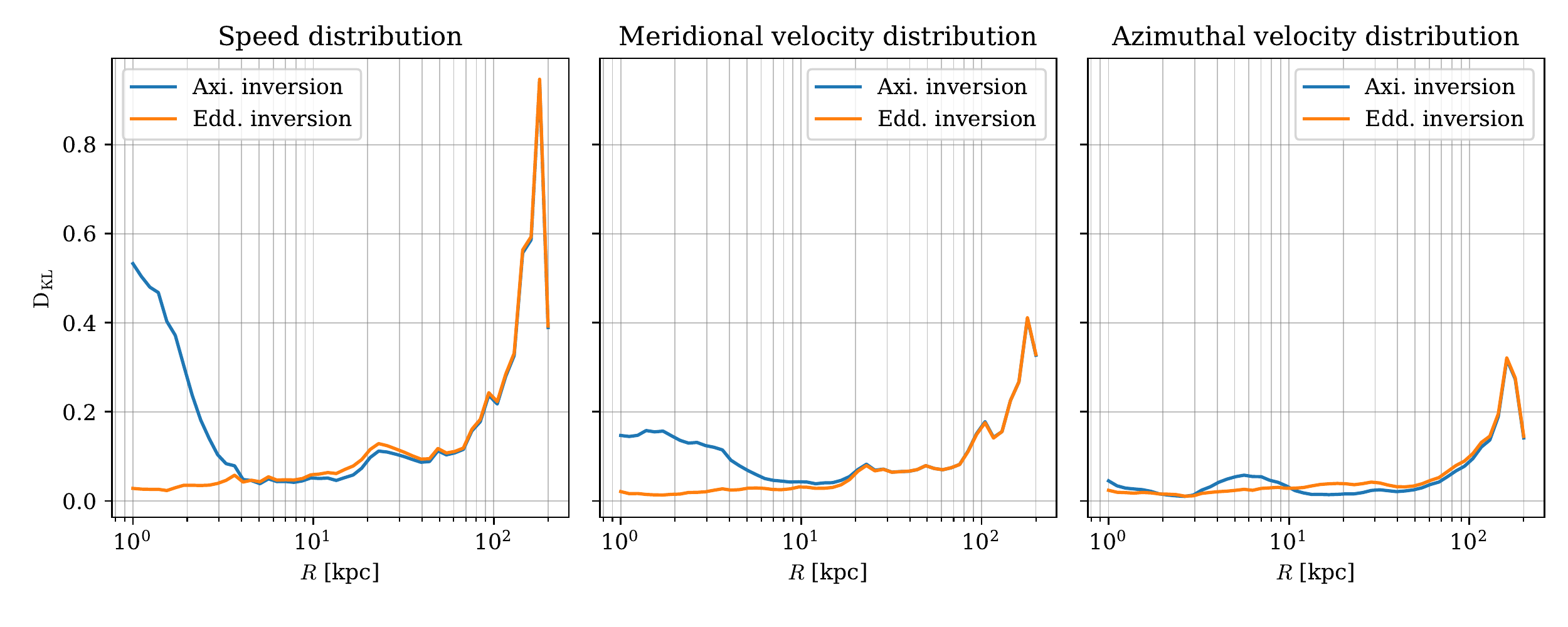}
			\caption{Same as in \citefig{fig:mochima_entropy}, but for Halo C simulation.}
			\label{fig:haloc_entropy}
		\end{figure}
	
		\subsection{Velocity moments}
		
		\begin{figure}[H]
			\centering
			\includegraphics[width=\textwidth]{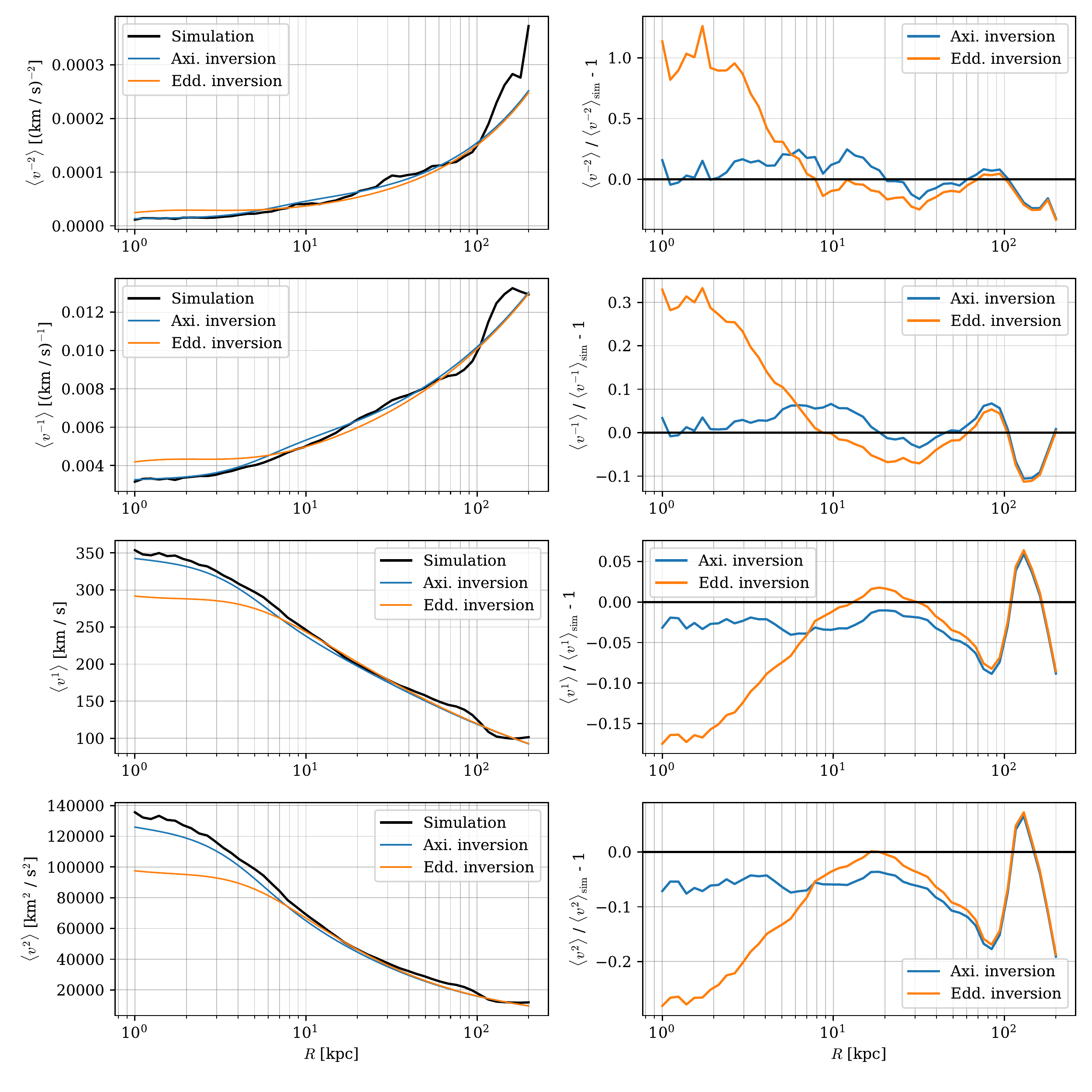}
			\caption{Same as in \citefig{fig:mochima_mom}, but for Halo B simulation.}
			\label{fig:halob_mom}
		\end{figure}
		
		\begin{figure}[H]
			\centering
			\includegraphics[width=\textwidth]{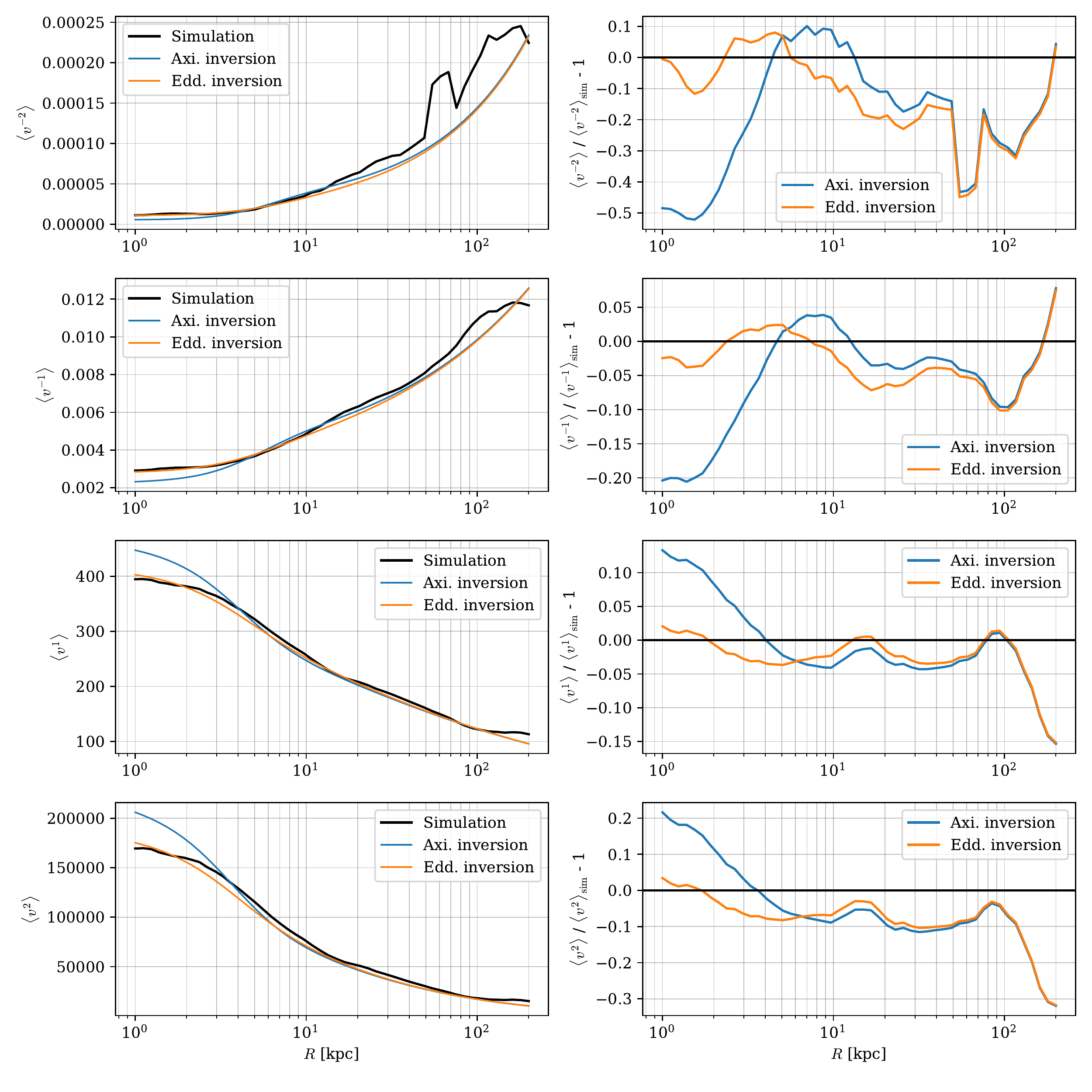}
			\caption{Same as in \citefig{fig:mochima_mom}, but for Halo C simulation.}
			\label{fig:haloc_mom}
		\end{figure}
	
		\subsection{Velocity anisotropy}
		
		\begin{figure}[H]
			\centering
			\includegraphics[width=0.49\textwidth]{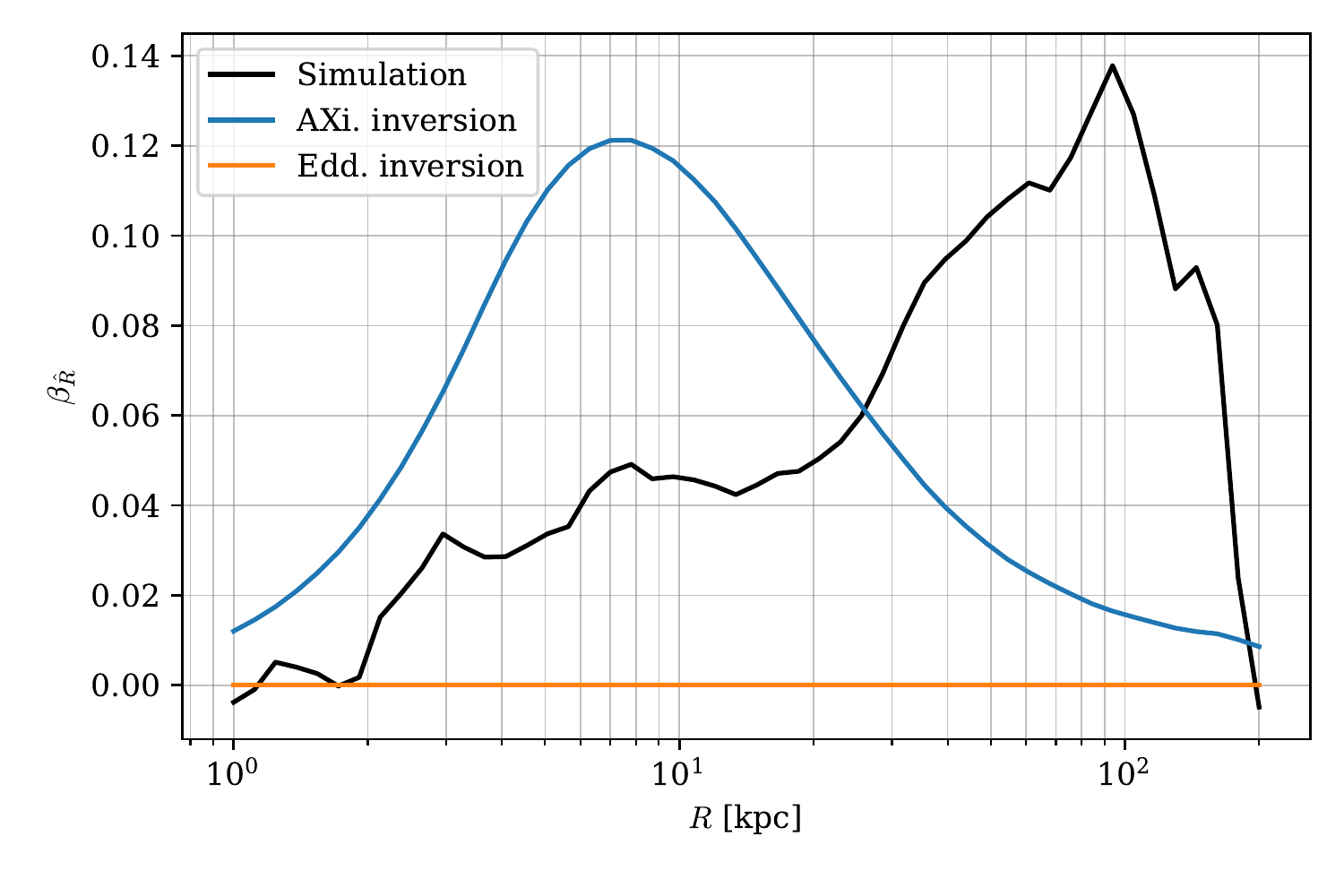}
			\includegraphics[width=0.49\textwidth]{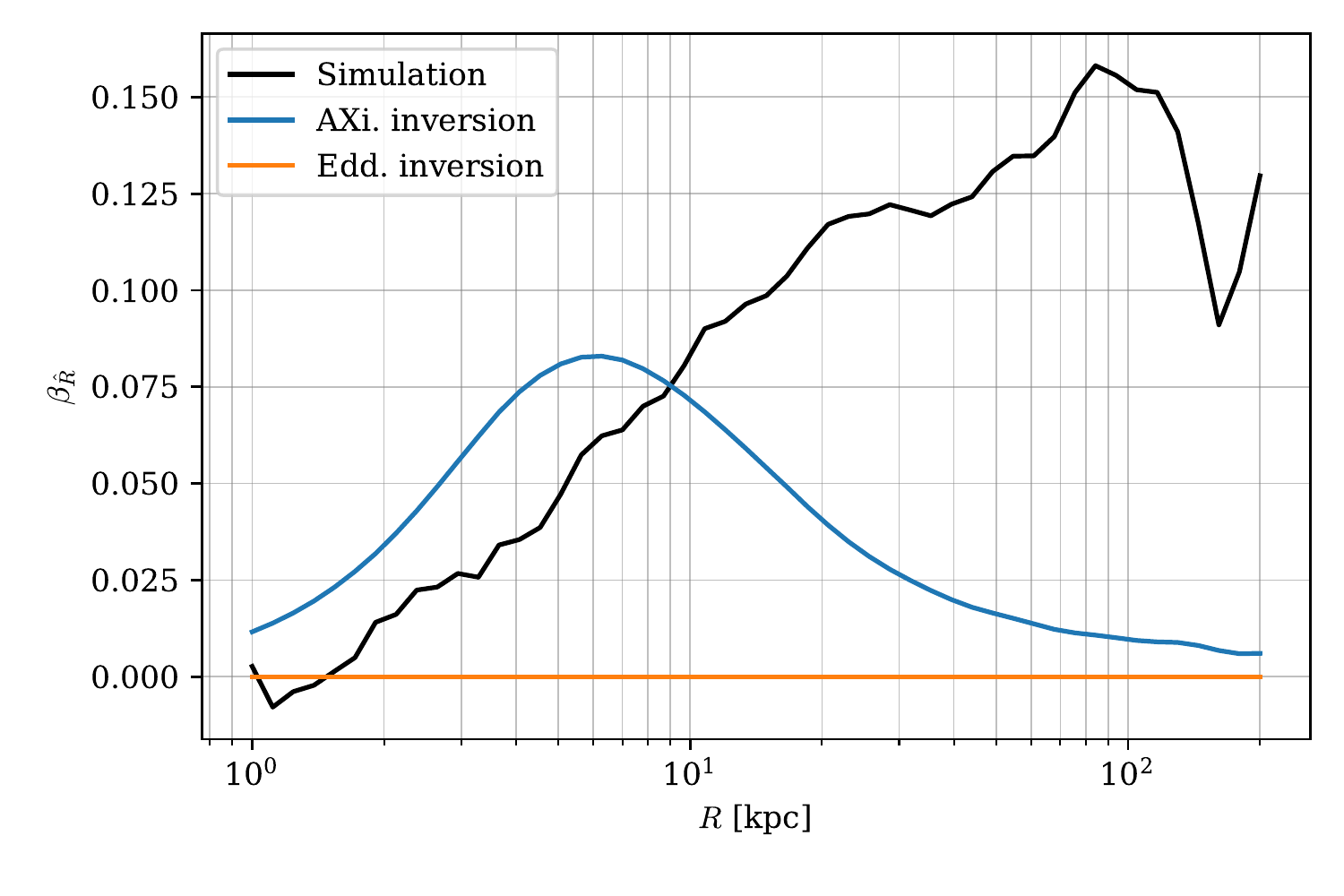}
			\caption{Same as in \citefig{fig:mochima_aniso}, but for Halo B simulation in the left-hand side panel and Halo C simulation in the right-hand side panel.}
			\label{fig:halobc_aniso}
		\end{figure}
	
		\subsection{DM searches}
		
		\begin{figure}[H]
			\centering
			\includegraphics[width=0.95\textwidth]{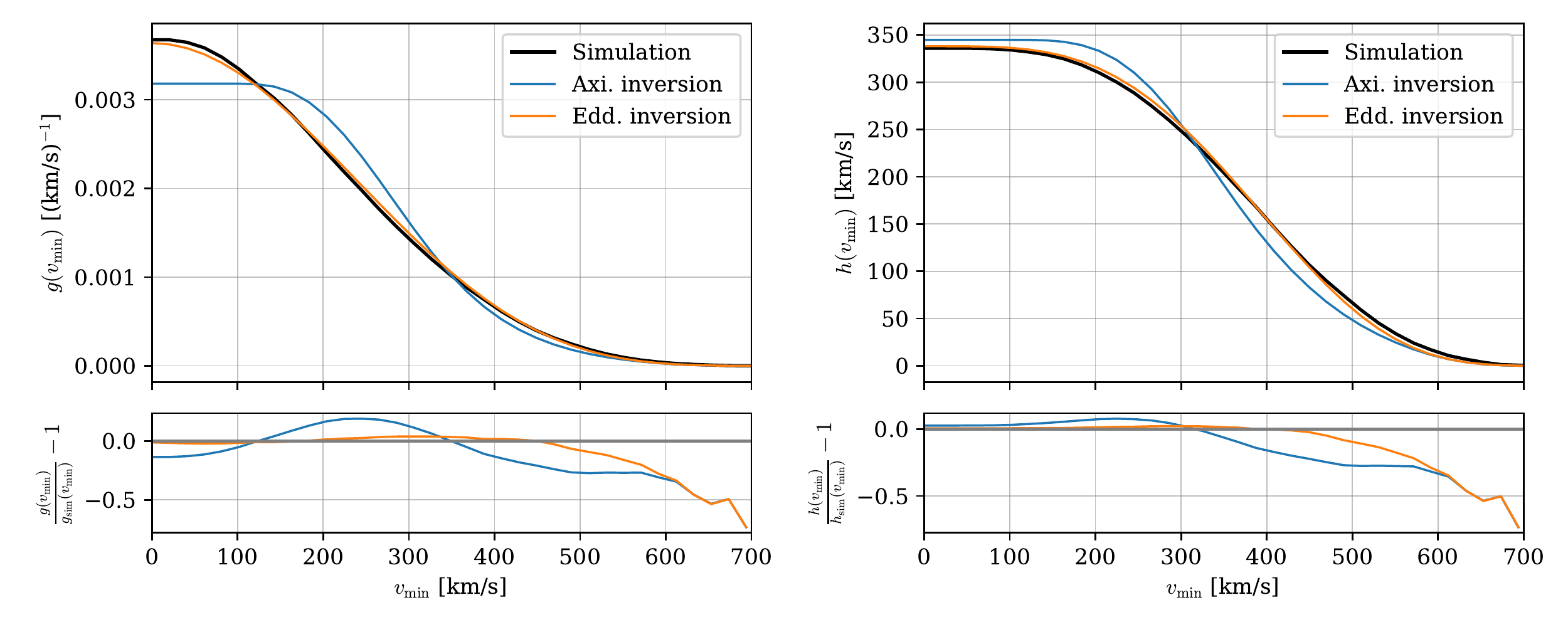}
			\caption{Same as in \citefig{fig:mochima_dd}, but for Halo B simulation.}
			\label{fig:halob_dd}
		\end{figure}
	
		\begin{figure}[H]
			\centering
			\includegraphics[width=0.95\textwidth]{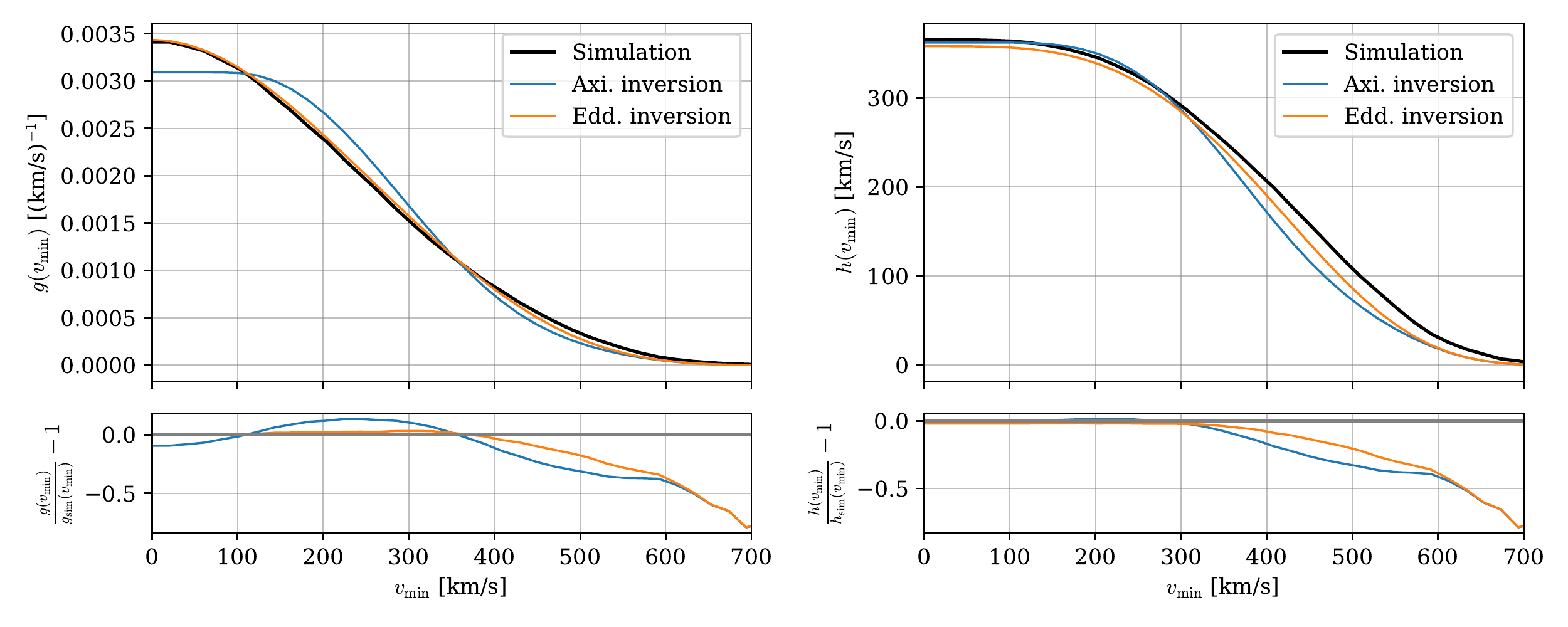}
			\caption{Same as in \citefig{fig:mochima_dd}, but for Halo C simulation.}
			\label{fig:haloc_dd}
		\end{figure}
		
		\begin{figure}[H]
			\centering
			\includegraphics[width=\textwidth]{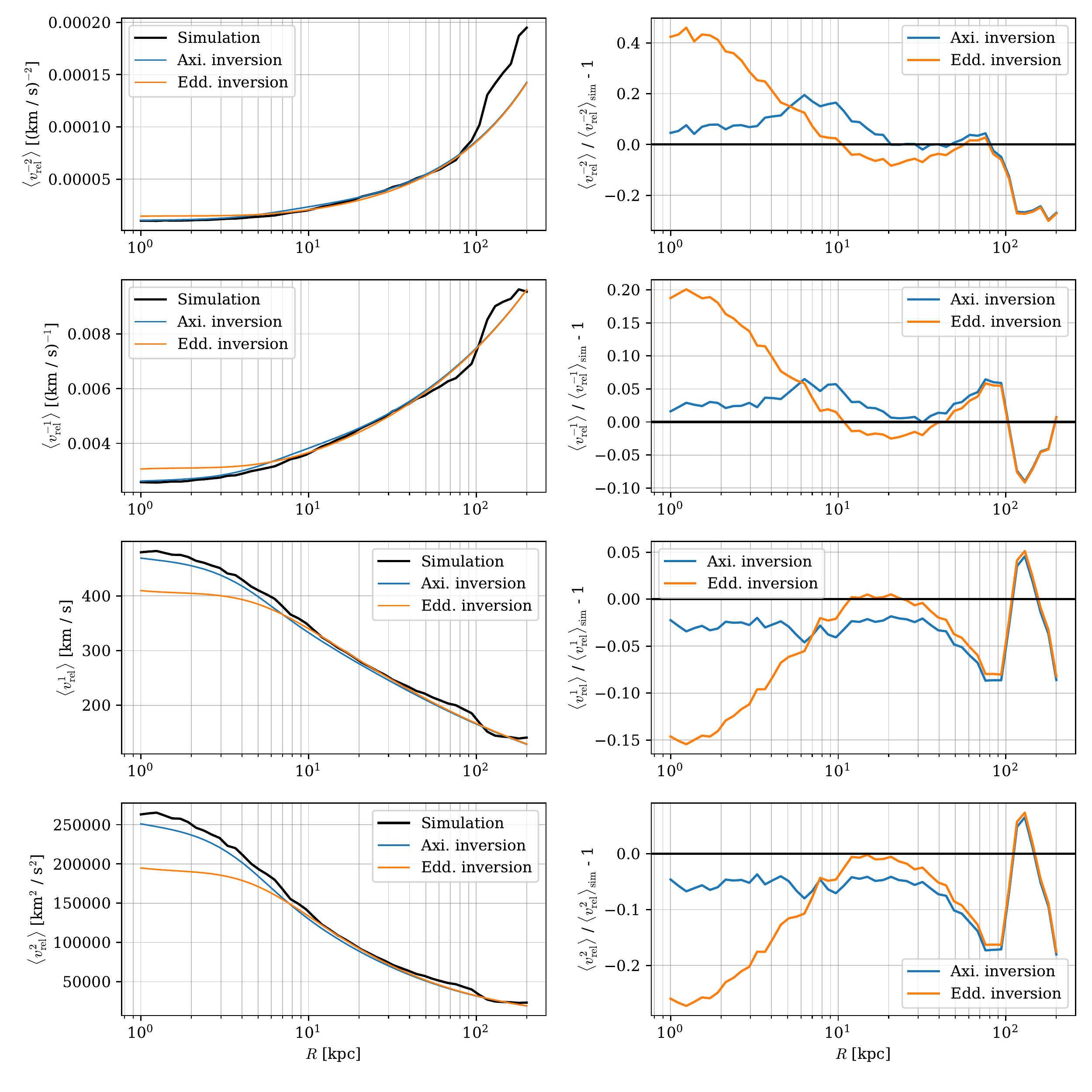}
			\caption{Same as in \citefig{fig:mochima_rel_moments}, but for Halo B simulation.}
			\label{fig:halob_rel_mom}
		\end{figure}
		
		\begin{figure}[H]
			\centering
			\includegraphics[width=\textwidth]{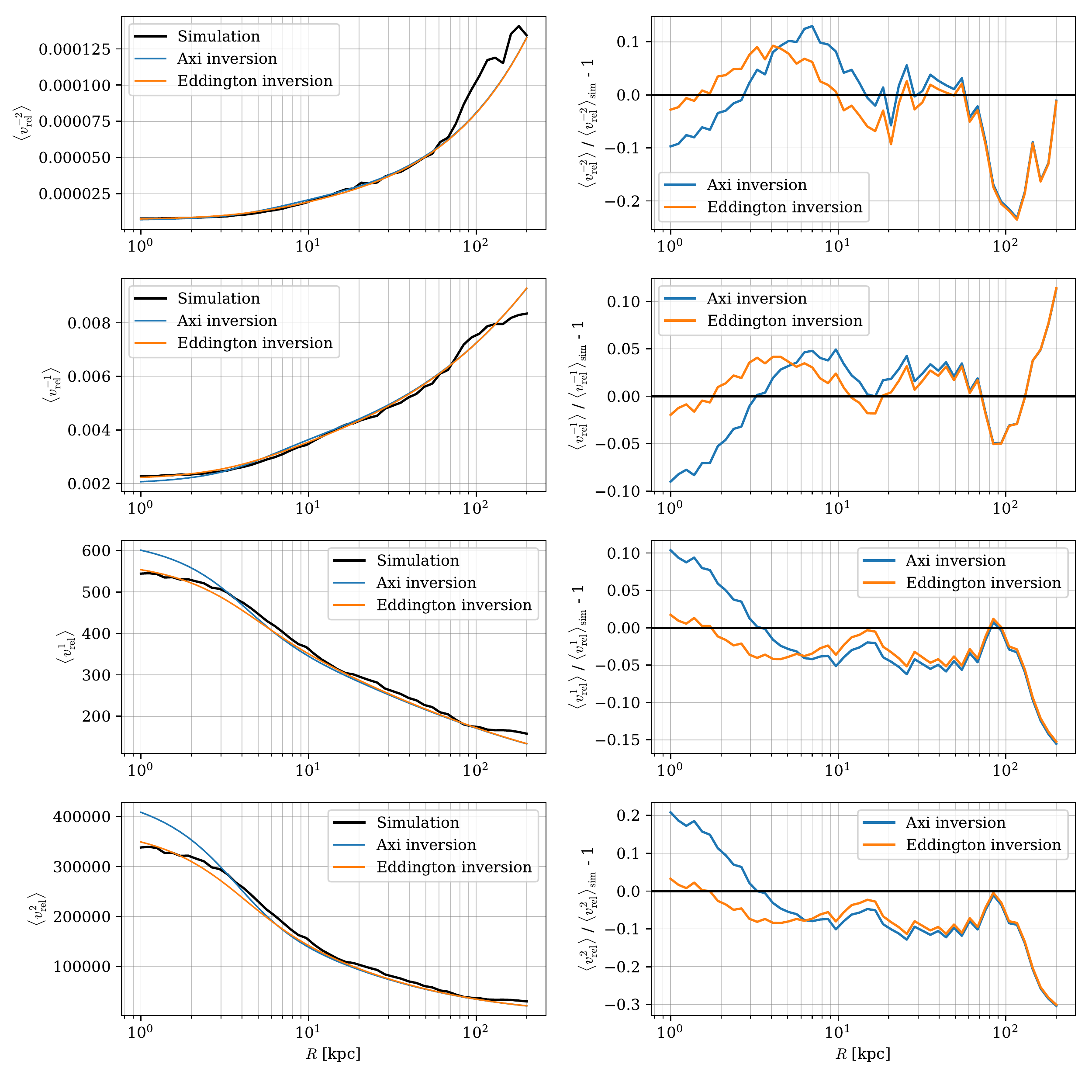}
			\caption{Same as in \citefig{fig:mochima_rel_moments}, but for Halo C simulation.}
			\label{fig:haloc_rel_mom}
		\end{figure}
		
	\end{appendices}
	
\bibliographystyle{JHEP}
\bibliography{references}	

\providecommand{\href}[2]{#2}\begingroup\raggedright\begin{thebibliography}{100}

\bibitem{ZeldovichEtAl1975}
I.B.~Zeldovich and I.D.~Novikov, \emph{Structure and evolution of the
  universe.}, Moscow, Izdatel'stvo Nauka (1975).

\bibitem{LeeEtAl1977a}
B.W.~{Lee} and S.~{Weinberg}, \emph{{Cosmological lower bound on heavy-neutrino
  masses}}, \href{https://doi.org/10.1103/PhysRevLett.39.165}{\emph{\prl}
  {\bfseries 39} (1977) 165}.

\bibitem{GunnEtAl1978}
J.E.~{Gunn}, B.W.~{Lee}, I.~{Lerche}, D.N.~{Schramm} and G.~{Steigman},
  \emph{{Some astrophysical consequences of the existence of a heavy stable
  neutral lepton.}}, \href{https://doi.org/10.1086/156335}{\emph{\apj}
  {\bfseries 223} (1978) 1015}.

\bibitem{DolgovEtAl1981}
A.D.~{Dolgov} and Y.B.~{Zeldovich}, \emph{{Cosmology and elementary
  particles}}, \href{https://doi.org/10.1103/RevModPhys.53.1}{\emph{Reviews of
  Modern Physics} {\bfseries 53} (1981) 1}.

\bibitem{BinetruyEtAl1984a}
P.~{Bin{\'e}truy}, G.~{Girardi} and P.~{Salati}, \emph{{Cosmological
  constraints on unstable heavy neutrinos}},
  \href{https://doi.org/10.1016/0370-2693(84)90665-8}{\emph{Physics Letters B}
  {\bfseries 134} (1984) 174}.

\bibitem{SrednickiEtAl1988}
M.~{Srednicki}, R.~{Watkins} and K.A.~{Olive}, \emph{{Calculations of relic
  densities in the early universe}},
  \href{https://doi.org/10.1016/0550-3213(88)90099-5}{\emph{Nuclear Physics B}
  {\bfseries 310} (1988) 693}.

\bibitem{SteigmanEtAl2012}
G.~{Steigman}, B.~{Dasgupta} and J.F.~{Beacom}, \emph{{Precise relic WIMP
  abundance and its impact on searches for dark matter annihilation}},
  \href{https://doi.org/10.1103/PhysRevD.86.023506}{\emph{\prd} {\bfseries 86}
  (2012) 023506} [\href{https://arxiv.org/abs/1204.3622}{{\ttfamily
  1204.3622}}].

\bibitem{PrimackEtAl1988}
J.R.~{Primack}, D.~{Seckel} and B.~{Sadoulet}, \emph{{Detection of cosmic dark
  matter}},
  \href{https://doi.org/10.1146/annurev.ns.38.120188.003535}{\emph{Annual
  Review of Nuclear and Particle Science} {\bfseries 38} (1988) 751}.

\bibitem{jungman_supersymmetric_1996}
G.~Jungman, M.~Kamionkowski and K.~Griest, \emph{Supersymmetric dark matter},
  \href{https://doi.org/10.1016/0370-1573(95)00058-5}{\emph{Physics Reports}
  {\bfseries 267} (1996) 195}.

\bibitem{ArcadiEtAl2017}
G.~Arcadi, M.~Dutra, P.~Ghosh, M.~Lindner, Y.~Mambrini, M.~Pierre et~al.,
  \emph{{The waning of the WIMP? A review of models, searches, and
  constraints}},
  \href{https://doi.org/10.1140/epjc/s10052-018-5662-y}{\emph{Eur. Phys. J. C}
  {\bfseries 78} (2018) 203}
  [\href{https://arxiv.org/abs/1703.07364}{{\ttfamily 1703.07364}}].

\bibitem{LeaneEtAl2018}
R.K.~Leane, T.R.~Slatyer, J.F.~Beacom and K.C.Y.~Ng, \emph{{GeV-scale thermal
  WIMPs: Not even slightly ruled out}},
  \href{https://doi.org/10.1103/PhysRevD.98.023016}{\emph{Phys. Rev. D}
  {\bfseries 98} (2018) 023016}
  [\href{https://arxiv.org/abs/1805.10305}{{\ttfamily 1805.10305}}].

\bibitem{goodman_detectability_1985}
M.W.~Goodman and E.~Witten, \emph{Detectability of certain dark-matter
  candidates}, \href{https://doi.org/10.1103/PhysRevD.31.3059}{\emph{Physical
  Review D} {\bfseries 31} (1985) 3059}.

\bibitem{DrukierEtAl1986}
A.K.~{Drukier}, K.~{Freese} and D.N.~{Spergel}, \emph{{Detecting cold
  dark-matter candidates}},
  \href{https://doi.org/10.1103/PhysRevD.33.3495}{\emph{\prd} {\bfseries 33}
  (1986) 3495}.

\bibitem{FreeseEtAl2013}
K.~{Freese}, M.~{Lisanti} and C.~{Savage}, \emph{{Colloquium: Annual modulation
  of dark matter}},
  \href{https://doi.org/10.1103/RevModPhys.85.1561}{\emph{Reviews of Modern
  Physics} {\bfseries 85} (2013) 1561}
  [\href{https://arxiv.org/abs/1209.3339}{{\ttfamily 1209.3339}}].

\bibitem{PressEtAl1985}
W.H.~Press and D.N.~Spergel, \emph{{Capture by the sun of a galactic population
  of weakly interacting massive particles}},
  \href{https://doi.org/10.1086/163485}{\emph{\apj} {\bfseries 296} (1985)
  679}.

\bibitem{KraussEtAl1985}
L.M.~{Krauss}, K.~{Freese}, D.N.~{Spergel} and W.H.~{Press}, \emph{{Cold dark
  matter candidates and the solar neutrino problem}},
  \href{https://doi.org/10.1086/163767}{\emph{\apj} {\bfseries 299} (1985)
  1001}.

\bibitem{SalatiEtAl1989}
P.~Salati and J.~Silk, \emph{A stellar probe of dark matter annihilation in
  galactic nuclei}, \href{https://doi.org/10.1086/167177}{\emph{\apj}
  {\bfseries 338} (1989) 24}.

\bibitem{LopesEtAl2021}
J.~Lopes, T.~Lacroix and I.~Lopes, \emph{Towards a more rigorous treatment of
  uncertainties on the velocity distribution of dark matter particles for
  capture in stars},
  \href{https://doi.org/10.1088/1475-7516/2021/01/073}{\emph{Journal of
  Cosmology and Astroparticle Physics} {\bfseries 2021} (2021) 073}
  [\href{https://arxiv.org/abs/2007.15927}{{\ttfamily 2007.15927}}].

\bibitem{Bergstroem2000}
L.~{Bergstr{\"o}m}, \emph{{Non-baryonic dark matter: observational evidence and
  detection methods}},
  \href{https://doi.org/10.1088/0034-4885/63/5/2r3}{\emph{Reports on Progress
  in Physics} {\bfseries 63} (2000) 793}
  [\href{https://arxiv.org/abs/hep-ph/0002126}{{\ttfamily hep-ph/0002126}}].

\bibitem{BringmannEtAl2012c}
T.~{Bringmann} and C.~{Weniger}, \emph{{Gamma ray signals from dark matter:
  Concepts, status and prospects}},
  \href{https://doi.org/10.1016/j.dark.2012.10.005}{\emph{Physics of the Dark
  Universe} {\bfseries 1} (2012) 194}
  [\href{https://arxiv.org/abs/1208.5481}{{\ttfamily 1208.5481}}].

\bibitem{LavalleEtAl2012}
J.~{Lavalle} and P.~{Salati}, \emph{{Dark matter indirect signatures}},
  \href{https://doi.org/10.1016/j.crhy.2012.05.001}{\emph{Comptes Rendus
  Physique} {\bfseries 13} (2012) 740}
  [\href{https://arxiv.org/abs/1205.1004}{{\ttfamily 1205.1004}}].

\bibitem{McDonaldEtAl2001}
P.~McDonald, R.J.~Scherrer and T.P.~Walker, \emph{Cosmic microwave background
  constraint on residual annihilations of relic particles},
  \href{https://doi.org/10.1103/PhysRevD.63.023001}{\emph{\prd} {\bfseries 63}
  (2001) 023001} [\href{https://arxiv.org/abs/astro-ph/0008134}{{\ttfamily
  astro-ph/0008134}}].

\bibitem{EssigEtAl2013a}
R.~{Essig}, E.~{Kuflik}, S.D.~{McDermott}, T.~{Volansky} and K.M.~{Zurek},
  \emph{{Constraining light dark matter with diffuse X-ray and gamma-ray
  observations}}, \href{https://doi.org/10.1007/JHEP11(2013)193}{\emph{Journal
  of High Energy Physics} {\bfseries 11} (2013) 193}
  [\href{https://arxiv.org/abs/1309.4091}{{\ttfamily 1309.4091}}].

\bibitem{BoddyEtAl2018}
K.K.~Boddy, J.~Kumar and L.E.~Strigari, \emph{The effective j-factor of the
  galactic center for velocity-dependent dark matter annihilation},
  \href{https://doi.org/10.1103/PhysRevD.98.063012}{\emph{\prd} {\bfseries 98}
  (2018) 063012} [\href{https://arxiv.org/abs/1805.08379}{{\ttfamily
  1805.08379}}].

\bibitem{BoudaudEtAl2019a}
M.~Boudaud, T.~Lacroix, M.~Stref and J.~Lavalle, \emph{Robust cosmic-ray
  constraints on $p$-wave annihilating mev dark matter},
  \href{https://doi.org/10.1103/PhysRevD.99.061302}{\emph{\prd} {\bfseries 99}
  (2019) 061302} [\href{https://arxiv.org/abs/1810.01680}{{\ttfamily
  1810.01680}}].

\bibitem{LiuEtAl2020}
H.~Liu, W.~Qin, G.W.~Ridgway and T.R.~Slatyer, \emph{Lyman-$\alpha$ constraints
  on cosmic heating from dark matter annihilation and decay}, {\emph{arXiv
  e-prints} (2020) arXiv:2008.01084}
  [\href{https://arxiv.org/abs/2008.01084}{{\ttfamily 2008.01084}}].

\bibitem{HisanoEtAl2004}
J.~{Hisano}, S.~{Matsumoto} and M.M.~{Nojiri}, \emph{{Explosive Dark Matter
  Annihilation}},
  \href{https://doi.org/10.1103/PhysRevLett.92.031303}{\emph{\prl} {\bfseries
  92} (2004) 031303} [\href{https://arxiv.org/abs/hep-ph/0307216}{{\ttfamily
  hep-ph/0307216}}].

\bibitem{HisanoEtAl2005}
J.~{Hisano}, S.~{Matsumoto}, M.M.~{Nojiri} and O.~{Saito},
  \emph{{Nonperturbative effect on dark matter annihilation and gamma ray
  signature from the galactic center}},
  \href{https://doi.org/10.1103/PhysRevD.71.063528}{\emph{\prd} {\bfseries 71}
  (2005) 063528} [\href{https://arxiv.org/abs/hep-ph/0412403}{{\ttfamily
  hep-ph/0412403}}].

\bibitem{Paczynski1986}
B.~Paczynski, \emph{Gravitational microlensing by the galactic halo},
  \href{https://doi.org/10.1086/164140}{\emph{\apj} {\bfseries 304} (1986) 1}.

\bibitem{Griest1991}
K.~Griest, \emph{Galactic microlensing as a method of detecting massive compact
  halo objects}, \href{https://doi.org/10.1086/169575}{\emph{\apj} {\bfseries
  366} (1991) 412}.

\bibitem{green_astrophysical_2017}
A.M.~Green, \emph{Astrophysical uncertainties on the local dark matter
  distribution and direct detection experiments},
  \href{https://doi.org/10.1088/1361-6471/aa7819}{\emph{Journal of Physics G:
  Nuclear and Particle Physics} {\bfseries 44} (2017) 084001}.

\bibitem{LisantiEtAl2012}
M.~{Lisanti} and D.N.~{Spergel}, \emph{{Dark matter debris flows in the Milky
  Way}}, \href{https://doi.org/10.1016/j.dark.2012.10.007}{\emph{Physics of the
  Dark Universe} {\bfseries 1} (2012) 155}
  [\href{https://arxiv.org/abs/1105.4166}{{\ttfamily 1105.4166}}].

\bibitem{Herzog-ArbeitmanEtAl2018}
J.~Herzog-Arbeitman, M.~Lisanti, P.~Madau and L.~Necib, \emph{Empirical
  determination of dark matter velocities using metal-poor stars},
  \href{https://doi.org/10.1103/PhysRevLett.120.041102}{\emph{\prl} {\bfseries
  120} (2018) 041102} [\href{https://arxiv.org/abs/1704.04499}{{\ttfamily
  1704.04499}}].

\bibitem{binney_galactic_2008}
J.~Binney and S.~Tremaine, \emph{Galactic {Dynamics}: {Second} {Edition}},
  Princeton University Press (2008).

\bibitem{Green2003}
A.M.~{Green}, \emph{{Effect of realistic astrophysical inputs on the phase and
  shape of the weakly interacting massive particles annual modulation signal}},
  \href{https://doi.org/10.1103/PhysRevD.68.023004}{\emph{\prd} {\bfseries 68}
  (2003) 023004} [\href{https://arxiv.org/abs/astro-ph/0304446}{{\ttfamily
  astro-ph/0304446}}].

\bibitem{VogelsbergerEtAl2008}
M.~{Vogelsberger}, S.D.M.~{White}, A.~{Helmi} and V.~{Springel}, \emph{{The
  fine-grained phase-space structure of cold dark matter haloes}},
  \href{https://doi.org/10.1111/j.1365-2966.2007.12746.x}{\emph{Monthly Notices
  of the Royal Astronomical Society} {\bfseries 385} (2008) 236}
  [\href{https://arxiv.org/abs/0711.1105}{{\ttfamily 0711.1105}}].

\bibitem{lacroix_predicting_2020}
T.~Lacroix, A.~Núñez-Castiñeyra, M.~Stref, J.~Lavalle and E.~Nezri,
  \emph{Predicting the dark matter velocity distribution in galactic
  structures: tests against hydrodynamic cosmological simulations},
  {\emph{arXiv:2005.03955 [astro-ph, physics:hep-ph]} (2020) }.

\bibitem{LingEtAl2010}
F.-S.~{Ling}, E.~{Nezri}, E.~{Athanassoula} and R.~{Teyssier}, \emph{{Dark
  matter direct detection signals inferred from a cosmological N-body
  simulation with baryons}},
  \href{https://doi.org/10.1088/1475-7516/2010/02/012}{\emph{Journal of
  Cosmology and Astroparticle Physics} {\bfseries 2} (2010) 12}
  [\href{https://arxiv.org/abs/0909.2028}{{\ttfamily 0909.2028}}].

\bibitem{BozorgniaEtAl2016}
N.~{Bozorgnia}, F.~{Calore}, M.~{Schaller}, M.~{Lovell}, G.~{Bertone},
  C.S.~{Frenk} et~al., \emph{{Simulated Milky Way analogues: implications for
  dark matter direct searches}},
  \href{https://doi.org/10.1088/1475-7516/2016/05/024}{\emph{Journal of
  Cosmology and Astroparticle Physics} {\bfseries 5} (2016) 024}
  [\href{https://arxiv.org/abs/1601.04707}{{\ttfamily 1601.04707}}].

\bibitem{Strigari2013}
L.E.~{Strigari}, \emph{{Galactic searches for dark matter}},
  \href{https://doi.org/10.1016/j.physrep.2013.05.004}{\emph{\physrep}
  {\bfseries 531} (2013) 1} [\href{https://arxiv.org/abs/1211.7090}{{\ttfamily
  1211.7090}}].

\bibitem{GaiaCollab2016}
{The Gaia Collaboration}, T.~{Prusti}, J.H.J.~{de Bruijne}, A.G.A.~{Brown},
  A.~{Vallenari}, C.~{Babusiaux} et~al., \emph{{The Gaia mission}},
  \href{https://doi.org/10.1051/0004-6361/201629272}{\emph{\aap} {\bfseries
  595} (2016) A1} [\href{https://arxiv.org/abs/1609.04153}{{\ttfamily
  1609.04153}}].

\bibitem{GaiaCollab2016a}
{Gaia Collaboration}, A.G.A.~Brown, A.~Vallenari, T.~Prusti, J.H.J.~de~Bruijne,
  F.~Mignard et~al., \emph{Gaia data release 1. summary of the astrometric,
  photometric, and survey properties},
  \href{https://doi.org/10.1051/0004-6361/201629512}{\emph{\aap} {\bfseries
  595} (2016) A2} [\href{https://arxiv.org/abs/1609.04172}{{\ttfamily
  1609.04172}}].

\bibitem{GaiaCollab2018}
{\scshape Gaia Collaboration} collaboration, \emph{Gaia data release 2. summary
  of the contents and survey properties},
  \href{https://doi.org/10.1051/0004-6361/201833051}{\emph{\aap} {\bfseries
  616} (2018) A1} [\href{https://arxiv.org/abs/1804.09365}{{\ttfamily
  1804.09365}}].

\bibitem{GaiaCollab2020}
{Gaia Collaboration}, A.G.A.~Brown, A.~Vallenari, T.~Prusti, J.H.J.~de~Bruijne,
  C.~Babusiaux et~al., \emph{Gaia early data release 3: Summary of the contents
  and survey properties}, {\emph{arXiv e-prints} (2020) arXiv:2012.01533}
  [\href{https://arxiv.org/abs/2012.01533}{{\ttfamily 2012.01533}}].

\bibitem{CatenaEtAl2010a}
R.~{Catena} and P.~{Ullio}, \emph{{A novel determination of the local dark
  matter density}},
  \href{https://doi.org/10.1088/1475-7516/2010/08/004}{\emph{Journal of
  Cosmology and Astroparticle Physics} {\bfseries 8} (2010) 4}
  [\href{https://arxiv.org/abs/0907.0018}{{\ttfamily 0907.0018}}].

\bibitem{McMillan2011}
P.J.~{McMillan}, \emph{{Mass models of the Milky Way}},
  \href{https://doi.org/10.1111/j.1365-2966.2011.18564.x}{\emph{Monthly Notices
  of the Royal Astronomical Society} {\bfseries 414} (2011) 2446}
  [\href{https://arxiv.org/abs/1102.4340}{{\ttfamily 1102.4340}}].

\bibitem{CatenaEtAl2012}
R.~{Catena} and P.~{Ullio}, \emph{{The local dark matter phase-space density
  and impact on WIMP direct detection}},
  \href{https://doi.org/10.1088/1475-7516/2012/05/005}{\emph{Journal of
  Cosmology and Astroparticle Physics} {\bfseries 5} (2012) 5}
  [\href{https://arxiv.org/abs/1111.3556}{{\ttfamily 1111.3556}}].

\bibitem{Read2014}
J.I.~{Read}, \emph{{The local dark matter density}},
  \href{https://doi.org/10.1088/0954-3899/41/6/063101}{\emph{Journal of Physics
  G Nuclear Physics} {\bfseries 41} (2014) 063101}
  [\href{https://arxiv.org/abs/1404.1938}{{\ttfamily 1404.1938}}].

\bibitem{PifflEtAl2014}
T.~{Piffl}, J.~{Binney}, P.J.~{McMillan}, M.~{Steinmetz}, A.~{Helmi},
  R.F.G.~{Wyse} et~al., \emph{{Constraining the Galaxy's dark halo with RAVE
  stars}}, \href{https://doi.org/10.1093/mnras/stu1948}{\emph{Monthly Notices
  of the Royal Astronomical Society} {\bfseries 445} (2014) 3133}
  [\href{https://arxiv.org/abs/1406.4130}{{\ttfamily 1406.4130}}].

\bibitem{FornasaEtAl2014}
M.~{Fornasa} and A.M.~{Green}, \emph{{Self-consistent phase-space distribution
  function for the anisotropic dark matter halo of the Milky{\^A} Way}},
  \href{https://doi.org/10.1103/PhysRevD.89.063531}{\emph{\prd} {\bfseries 89}
  (2014) 063531} [\href{https://arxiv.org/abs/1311.5477}{{\ttfamily
  1311.5477}}].

\bibitem{McMillan2017}
P.J.~{McMillan}, \emph{{The mass distribution and gravitational potential of
  the Milky Way}}, \href{https://doi.org/10.1093/mnras/stw2759}{\emph{Monthly
  Notices of the Royal Astronomical Society} {\bfseries 465} (2017) 76}
  [\href{https://arxiv.org/abs/1608.00971}{{\ttfamily 1608.00971}}].

\bibitem{CautunEtAl2020}
M.~Cautun, A.~Ben{\'\i}tez-Llambay, A.J.~Deason, C.S.~Frenk, A.~Fattahi,
  F.A.~G{\'o}mez et~al., \emph{The milky way total mass profile as inferred
  from gaia dr2}, \href{https://doi.org/10.1093/mnras/staa1017}{\emph{Monthly
  Notices of the Royal Astronomical Society} {\bfseries 494} (2020) 4291}
  [\href{https://arxiv.org/abs/1911.04557}{{\ttfamily 1911.04557}}].

\bibitem{petac_equilibrium_2020}
M.~{Peta{\v{c}}}, \emph{{Equilibrium axisymmetric halo model for the Milky Way
  and its implications for direct and indirect dark matter searches}},
  \href{https://doi.org/10.1103/PhysRevD.102.123028}{\emph{\prd} {\bfseries
  102} (2020) 123028} [\href{https://arxiv.org/abs/2008.11172}{{\ttfamily
  2008.11172}}].

\bibitem{eddington_distribution_1916}
A.S.~Eddington, \emph{The {Distribution} of {Stars} in {Globular} {Clusters}},
  \href{https://doi.org/10.1093/mnras/76.7.572}{\emph{Monthly Notices of the
  Royal Astronomical Society} {\bfseries 76} (1916) 572}.

\bibitem{Widrow2000}
L.M.~{Widrow}, \emph{{Distribution Functions for Cuspy Dark Matter Density
  Profiles}}, \href{https://doi.org/10.1086/317367}{\emph{\apjs} {\bfseries
  131} (2000) 39} [\href{https://arxiv.org/abs/astro-ph/0003302}{{\ttfamily
  astro-ph/0003302}}].

\bibitem{UllioEtAl2001a}
P.~{Ullio} and M.~{Kamionkowski}, \emph{{Velocity distributions and
  annual-modulation signatures of weakly-interacting massive particles}},
  \href{https://doi.org/10.1088/1126-6708/2001/03/049}{\emph{Journal of High
  Energy Physics} {\bfseries 3} (2001) 49}
  [\href{https://arxiv.org/abs/hep-ph/0006183}{{\ttfamily hep-ph/0006183}}].

\bibitem{VergadosEtAl2003}
J.D.~{Vergados} and D.~{Owen}, \emph{{New Velocity Distribution for Cold Dark
  Matter in the Context of the Eddington Theory}},
  \href{https://doi.org/10.1086/368350}{\emph{\apj} {\bfseries 589} (2003) 17}
  [\href{https://arxiv.org/abs/astro-ph/0203293}{{\ttfamily
  astro-ph/0203293}}].

\bibitem{FerrerEtAl2013}
F.~{Ferrer} and D.R.~{Hunter}, \emph{{The impact of the phase-space density on
  the indirect detection of dark matter}},
  \href{https://doi.org/10.1088/1475-7516/2013/09/005}{\emph{Journal of
  Cosmology and Astroparticle Physics} {\bfseries 9} (2013) 5}
  [\href{https://arxiv.org/abs/1306.6586}{{\ttfamily 1306.6586}}].

\bibitem{lacroix_anatomy_2018}
T.~Lacroix, M.~Stref and J.~Lavalle, \emph{Anatomy of {Eddington}-like
  inversion methods in the context of dark matter searches},
  \href{https://doi.org/10.1088/1475-7516/2018/09/040}{\emph{Journal of
  Cosmology and Astroparticle Physics} {\bfseries 2018} (2018) 040}.

\bibitem{osipkov_spherical_1979}
L.P.~Osipkov, \emph{Spherical systems of gravitating bodies with an ellipsoidal
  velocity distribution}, {\emph{Pisma v Astronomicheskii Zhurnal} {\bfseries
  5} (1979) 77}.

\bibitem{merritt_spherical_1985}
D.~Merritt, \emph{Spherical stellar systems with spheroidal velocity
  distributions}, \href{https://doi.org/10.1086/113810}{\emph{The Astronomical
  Journal} {\bfseries 90} (1985) 1027}.

\bibitem{Cuddeford1991}
P.~{Cuddeford}, \emph{{An analytic inversion for anisotropic spherical
  galaxies}}, \href{https://doi.org/10.1093/mnras/253.3.414}{\emph{Monthly
  Notices of the Royal Astronomical Society} {\bfseries 253} (1991) 414}.

\bibitem{BozorgniaEtAl2013}
N.~{Bozorgnia}, R.~{Catena} and T.~{Schwetz}, \emph{{Anisotropic dark matter
  distribution functions and impact on WIMP direct detection}},
  \href{https://doi.org/10.1088/1475-7516/2013/12/050}{\emph{Journal of
  Cosmology and Astroparticle Physics} {\bfseries 12} (2013) 50}
  [\href{https://arxiv.org/abs/1310.0468}{{\ttfamily 1310.0468}}].

\bibitem{BinneyEtAl2015}
J.~{Binney} and T.~{Piffl}, \emph{{The distribution function of the Galaxy's
  dark halo}}, \href{https://doi.org/10.1093/mnras/stv2225}{\emph{Monthly
  Notices of the Royal Astronomical Society} {\bfseries 454} (2015) 3653}
  [\href{https://arxiv.org/abs/1509.06877}{{\ttfamily 1509.06877}}].

\bibitem{SandersEtAl2016}
J.L.~{Sanders} and J.~{Binney}, \emph{{A review of action estimation methods
  for galactic dynamics}},
  \href{https://doi.org/10.1093/mnras/stw106}{\emph{Monthly Notices of the
  Royal Astronomical Society} {\bfseries 457} (2016) 2107}
  [\href{https://arxiv.org/abs/1511.08213}{{\ttfamily 1511.08213}}].

\bibitem{PifflEtAl2015}
T.~{Piffl}, Z.~{Penoyre} and J.~{Binney}, \emph{{Bringing the Galaxy's dark
  halo to life}}, \href{https://doi.org/10.1093/mnras/stv938}{\emph{Monthly
  Notices of the Royal Astronomical Society} {\bfseries 451} (2015) 639}
  [\href{https://arxiv.org/abs/1502.02916}{{\ttfamily 1502.02916}}].

\bibitem{CallinghamEtAl2020}
T.M.~Callingham, M.~Cautun, A.J.~Deason, C.S.~Frenk, R.J.J.~Grand, F.~Marinacci
  et~al., \emph{The orbital phase space of contracted dark matter halos},
  {\emph{arXiv e-prints} (2020) arXiv:2001.07742}
  [\href{https://arxiv.org/abs/2001.07742}{{\ttfamily 2001.07742}}].

\bibitem{HattoriEtAl2020}
K.~Hattori, M.~Valluri and E.~Vasiliev, \emph{Action-based distribution
  function modelling for constraining the shape of the galactic dark matter
  halo}, {\emph{arXiv e-prints} (2020) arXiv:2012.03908}
  [\href{https://arxiv.org/abs/2012.03908}{{\ttfamily 2012.03908}}].

\bibitem{hunter_two-integral_1993}
C.~Hunter and E.~Qian, \emph{Two-integral distribution functions for
  axisymmetric galaxies},
  \href{https://doi.org/10.1093/mnras/262.2.401}{\emph{Monthly Notices of the
  Royal Astronomical Society} {\bfseries 262} (1993) 401}.

\bibitem{petac_two-integral_2019}
M.~Peta{\v c} and P.~Ullio, \emph{Two-integral distribution functions in
  axisymmetric galaxies: {Implications} for dark matter searches},
  \href{https://doi.org/10.1103/PhysRevD.99.043003}{\emph{Physical Review D}
  {\bfseries 99} (2019) 043003}.

\bibitem{mollitor_baryonic_2015}
P.~Mollitor, E.~Nezri and R.~Teyssier, \emph{Baryonic and dark matter
  distribution in cosmological simulations of spiral galaxies},
  \href{https://doi.org/10.1093/mnras/stu2466}{\emph{Monthly Notices of the
  Royal Astronomical Society} {\bfseries 447} (2015) 1353}.

\bibitem{nunez-castineyra_cosmological_2021}
A.~Nuñez-Castiñeyra, E.~Nezri, J.~Devriendt and R.~Teyssier,
  \emph{Cosmological simulations of the same spiral galaxy: the impact of
  baryonic physics},
  \href{https://doi.org/10.1093/mnras/staa3233}{\emph{Monthly Notices of the
  Royal Astronomical Society} {\bfseries 501} (2021) 62}.

\bibitem{petac_velocity-dependent_2018}
M.~Petač, P.~Ullio and M.~Valli, \emph{On velocity-dependent dark matter
  annihilations in dwarf satellites},
  \href{https://doi.org/10.1088/1475-7516/2018/12/039}{\emph{Journal of
  Cosmology and Astroparticle Physics} {\bfseries 2018} (2018) 039}.

\bibitem{wojtak_distribution_2008}
R.~{Wojtak}, E.L.~{{\L}okas}, G.A.~{Mamon}, S.~{Gottl{\"o}ber}, A.~{Klypin} and
  Y.~{Hoffman}, \emph{{The distribution function of dark matter in massive
  haloes}},
  \href{https://doi.org/10.1111/j.1365-2966.2008.13441.x}{\emph{\mnras}
  {\bfseries 388} (2008) 815}
  [\href{https://arxiv.org/abs/0802.0429}{{\ttfamily 0802.0429}}].

\bibitem{qian_axisymmetric_1995}
E.E.~Qian, P.T.~de~Zeeuw, R.P.~van~der Marel and C.~Hunter, \emph{Axisymmetric
  galaxy models with central black holes, with an application to {M}32},
  \href{https://doi.org/10.1093/mnras/274.2.602}{\emph{Mon.Not.Roy.Astron.Soc.}
  {\bfseries 274} (1995) 602}.

\bibitem{lynden-bell_stellar_1962}
D.~Lynden-Bell, \emph{Stellar dynamics: {Exact} solution of the
  self-gravitation equation},
  \href{https://doi.org/10.1093/mnras/123.5.447}{\emph{Monthly Notices of the
  Royal Astronomical Society} {\bfseries 123} (1962) 447}.

\bibitem{bullock_universal_2001}
J.S.~Bullock, A.~Dekel, T.S.~Kolatt, A.V.~Kravtsov, A.A.~Klypin, C.~Porciani
  et~al., \emph{A {Universal} {Angular} {Momentum} {Profile} for {Galactic}
  {Halos}}, \href{https://doi.org/10.1086/321477}{\emph{The Astrophysical
  Journal} {\bfseries 555} (2001) 240}.

\bibitem{sharma_angular_2005}
S.~Sharma and M.~Steinmetz, \emph{The {Angular} {Momentum} {Distribution} of
  {Gas} and {Dark} {Matter} in {Galactic} {Halos}},
  \href{https://doi.org/10.1086/430660}{\emph{The Astrophysical Journal}
  {\bfseries 628} (2005) 21}.

\bibitem{teklu_connecting_2015}
A.F.~Teklu, R.-S.~Remus, K.~Dolag, A.M.~Beck, A.~Burkert, A.S.~Schmidt et~al.,
  \emph{Connecting {Angular} {Momentum} and {Galactic} {Dynamics}: {The}
  {Complex} {Interplay} between {Spin}, {Mass}, and {Morphology}},
  \href{https://doi.org/10.1088/0004-637X/812/1/29}{\emph{The Astrophysical
  Journal} {\bfseries 812} (2015) 29}.

\bibitem{zavala_link_2016}
J.~Zavala, C.S.~Frenk, R.~Bower, J.~Schaye, T.~Theuns, R.A.~Crain et~al.,
  \emph{The link between the assembly of the inner dark matter halo and the
  angular momentum evolution of galaxies in the {EAGLE} simulation},
  \href{https://doi.org/10.1093/mnras/stw1286}{\emph{Monthly Notices of the
  Royal Astronomical Society} {\bfseries 460} (2016) 4466}.

\bibitem{zjupa_angular_2017}
J.~Zjupa and V.~Springel, \emph{Angular momentum properties of haloes and their
  baryon content in the {Illustris} simulation},
  \href{https://doi.org/10.1093/mnras/stw2945}{\emph{Monthly Notices of the
  Royal Astronomical Society} {\bfseries 466} (2017) 1625}.

\bibitem{teyssier_cosmological_2002}
R.~Teyssier, \emph{Cosmological hydrodynamics with adaptive mesh refinement.
  {A} new high resolution code called {RAMSES}},
  \href{https://doi.org/10.1051/0004-6361:20011817}{\emph{Astronomy and
  Astrophysics} {\bfseries 385} (2002) 337}.

\bibitem{hahn_multi-scale_2011}
O.~Hahn and T.~Abel, \emph{Multi-scale initial conditions for cosmological
  simulations},
  \href{https://doi.org/10.1111/j.1365-2966.2011.18820.x}{\emph{Monthly Notices
  of the Royal Astronomical Society} {\bfseries 415} (2011) 2101}.

\bibitem{onorbe_how_2014}
J.~Oñorbe, S.~Garrison-Kimmel, A.H.~Maller, J.S.~Bullock, M.~Rocha and
  O.~Hahn, \emph{How to zoom: bias, contamination and {Lagrange} volumes in
  multimass cosmological simulations},
  \href{https://doi.org/10.1093/mnras/stt2020}{\emph{Monthly Notices of the
  Royal Astronomical Society} {\bfseries 437} (2014) 1894}.

\bibitem{hernquist_analytical_1990}
L.~Hernquist, \emph{An analytical model for spherical galaxies and bulges},
  \href{https://doi.org/10.1086/168845}{\emph{The Astrophysical Journal}
  {\bfseries 356} (1990) 359}.

\bibitem{miyamoto_three-dimensional_1975}
M.~Miyamoto and R.~Nagai, \emph{Three-dimensional models for the distribution
  of mass in galaxies.}, {\emph{Publications of the Astronomical Society of
  Japan} {\bfseries 27} (1975) 533}.

\bibitem{navarro_universal_1997}
J.F.~Navarro, C.S.~Frenk and S.D.M.~White, \emph{A {Universal} {Density}
  {Profile} from {Hierarchical} {Clustering}},
  \href{https://doi.org/10.1086/304888}{\emph{The Astrophysical Journal}
  {\bfseries 490} (1997) 493}.

\bibitem{burkert_structure_1995}
A.~Burkert, \emph{The {Structure} of {Dark} {Matter} {Halos} in {Dwarf}
  {Galaxies}}, \href{https://doi.org/10.1086/309560}{\emph{The Astrophysical
  Journal} {\bfseries 447} (1995) }.

\bibitem{zhao_analytical_1996}
H.~Zhao, \emph{Analytical models for galactic nuclei},
  \href{https://doi.org/10.1093/mnras/278.2.488}{\emph{Monthly Notices of the
  Royal Astronomical Society} {\bfseries 278} (1996) 488}.

\bibitem{kullback_information_1951}
S.~Kullback and R.A.~Leibler, \emph{On {Information} and {Sufficiency}},
  \href{https://doi.org/10.1214/aoms/1177729694}{\emph{The Annals of
  Mathematical Statistics} {\bfseries 22} (1951) 79}.

\bibitem{feng_dark_2010}
J.L.~Feng, \emph{Dark {Matter} {Candidates} from {Particle} {Physics} and
  {Methods} of {Detection}},
  \href{https://doi.org/10.1146/annurev-astro-082708-101659}{\emph{Annual
  Review of Astronomy and Astrophysics} {\bfseries 48} (2010) 495}.

\bibitem{lin_tasi_2019}
T.~Lin, \emph{{TASI} lectures on dark matter models and direct detection},
  {\emph{arXiv e-prints} {\bfseries 1904} (2019) arXiv:1904.07915}.

\bibitem{fitzpatrick_effective_2013}
A.L.~Fitzpatrick, W.~Haxton, E.~Katz, N.~Lubbers and Y.~Xu, \emph{The effective
  field theory of dark matter direct detection},
  \href{https://doi.org/10.1088/1475-7516/2013/02/004}{\emph{Journal of
  Cosmology and Astroparticle Physics} {\bfseries 2013} (2013) 004}.

\bibitem{anand_model-independent_2013}
N.~Anand, A.L.~Fitzpatrick and W.C.~Haxton, \emph{Model-independent {WIMP}
  {Scattering} {Responses} and {Event} {Rates}: {A} {Mathematica} {Package} for
  {Experimental} {Analysis}}, {\emph{arXiv e-prints} (2013) arXiv:1308.6288}.

\bibitem{dent_general_2015}
J.B.~Dent, L.M.~Krauss, J.L.~Newstead and S.~Sabharwal, \emph{General analysis
  of direct dark matter detection: {From} microphysics to observational
  signatures}, \href{https://doi.org/10.1103/PhysRevD.92.063515}{\emph{Physical
  Review D} {\bfseries 92} (2015) 063515}.

\bibitem{bishara_chiral_2017}
F.~Bishara, J.~Brod, B.~Grinstein and J.~Zupan, \emph{Chiral effective theory
  of dark matter direct detection},
  \href{https://doi.org/10.1088/1475-7516/2017/02/009}{\emph{Journal of
  Cosmology and Astroparticle Physics} {\bfseries 2017} (2017) 009}.

\bibitem{collaboration_search_2018}
H.E.S.S.~Collaboration, H.~Abdallah, A.~Abramowski, F.~Aharonian,
  F.A.~Benkhali, E.O.~Angüner et~al., \emph{Search for
  \${\textbackslash}gamma\$-ray line signals from dark matter annihilations in
  the inner {Galactic} halo from ten years of observations with
  {H}.{E}.{S}.{S}}, {\emph{arXiv:1805.05741 [astro-ph]} (2018) }.

\bibitem{chang_search_2018}
L.J.~Chang, M.~Lisanti and S.~Mishra-Sharma, \emph{A {Search} for {Dark}
  {Matter} {Annihilation} in the {Milky} {Way} {Halo}},
  \href{https://doi.org/10.1103/PhysRevD.98.123004}{\emph{Physical Review D}
  {\bfseries 98} (2018) 123004}.

\bibitem{fermi-lat_collaboration_search_2019}
{Fermi-LAT Collaboration}, C.~Johnson, R.~Caputo, C.~Karwin, S.~Murgia, S.~Ritz
  et~al., \emph{Search for gamma-ray emission from \$p\$-wave dark matter
  annihilation in the {Galactic} {Center}},
  \href{https://doi.org/10.1103/PhysRevD.99.103007}{\emph{Physical Review D}
  {\bfseries 99} (2019) 103007}.

\bibitem{abazajian_strong_2020}
K.N.~{Abazajian}, S.~{Horiuchi}, M.~{Kaplinghat}, R.E.~{Keeley} and
  O.~{Macias}, \emph{{Strong constraints on thermal relic dark matter from
  Fermi-LAT observations of the Galactic Center}}, {\emph{arXiv e-prints}
  (2020) arXiv:2003.10416} [\href{https://arxiv.org/abs/2003.10416}{{\ttfamily
  2003.10416}}].

\bibitem{collaboration_dark_2014}
T.F.-L.~Collaboration, M.~Ackermann, A.~Albert, B.~Anderson, L.~Baldini,
  J.~Ballet et~al., \emph{Dark {Matter} {Constraints} from {Observations} of 25
  {Milky} {Way} {Satellite} {Galaxies} with the {Fermi} {Large} {Area}
  {Telescope}},
  \href{https://doi.org/10.1103/PhysRevD.89.042001}{\emph{Physical Review D}
  {\bfseries 89} (2014) }.

\bibitem{ackermann_searching_2015}
M.~Ackermann and {others}, \emph{Searching for dark matter annihilation from
  {Milky} {Way} dwarf spheroidal galaxies with six years of {Fermi} {Large}
  {Area} {Telescope} data}, {\emph{Physical review letters} {\bfseries 115}
  (2015) 231301}.

\bibitem{archambault_dark_2017}
S.~Archambault, A.~Archer, W.~Benbow, R.~Bird, E.~Bourbeau, T.~Brantseg et~al.,
  \emph{Dark matter constraints from a joint analysis of dwarf {Spheroidal}
  galaxy observations with {VERITAS}},
  \href{https://doi.org/10.1103/PhysRevD.95.082001}{\emph{Physical Review D}
  {\bfseries 95} (2017) 082001}.

\bibitem{boddy_model-independent_2018}
K.K.~Boddy, J.~Kumar, D.~Marfatia and P.~Sandick, \emph{Model-independent
  constraints on dark matter annihilation in dwarf spheroidal galaxies},
  \href{https://doi.org/10.1103/PhysRevD.97.095031}{\emph{Physical Review D}
  {\bfseries 97} (2018) 095031}.

\bibitem{hoof_global_2018}
S.~Hoof, A.~Geringer-Sameth and R.~Trotta, \emph{A {Global} {Analysis} of
  {Dark} {Matter} {Signals} from 27 {Dwarf} {Spheroidal} {Galaxies} using {Ten}
  {Years} of {Fermi}-{LAT} {Observations}}, {\emph{arXiv e-prints} (2018)
  arXiv:1812.06986}.

\bibitem{the_hawc_collaboration_search_2020}
T.H.~Collaboration, \emph{Search for {Gamma}-ray {Spectral} {Lines} from {Dark}
  {Matter} {Annihilation} in {Dwarf} {Galaxies} with the {High}-{Altitude}
  {Water} {Cherenkov} {Observatory}},
  \href{https://doi.org/10.1103/PhysRevD.101.103001}{\emph{Physical Review D}
  {\bfseries 101} (2020) 103001}.

\bibitem{rico_gamma-ray_2020}
J.~Rico, \emph{Gamma-{Ray} {Dark} {Matter} {Searches} in {Milky} {Way}
  {Satellites} -- {A} {Comparative} {Review} of {Data} {Analysis} {Methods} and
  {Current} {Results}},
  \href{https://doi.org/10.3390/galaxies8010025}{\emph{Galaxies} {\bfseries 8}
  (2020) 25}.

\bibitem{alvarez_dark_2020}
A.~Alvarez, F.~Calore, A.~Genina, J.I.~Read, P.D.~Serpico and B.~Zaldivar,
  \emph{Dark matter constraints from dwarf galaxies with data-driven
  {J}-factors}, {\emph{arXiv:2002.01229 [astro-ph]} (2020) }.

\bibitem{lavalle_dark_2012}
J.~Lavalle and P.~Salati, \emph{Dark {Matter} {Indirect} {Signatures}},
  \href{https://doi.org/10.1016/j.crhy.2012.05.001}{\emph{Comptes Rendus
  Physique} {\bfseries 13} (2012) 740}.

\bibitem{gaskins_review_2016}
J.M.~Gaskins, \emph{A review of indirect searches for particle dark matter},
  \href{https://doi.org/10.1080/00107514.2016.1175160}{\emph{Contemporary
  Physics} {\bfseries 57} (2016) 496}.

\bibitem{hinton_multi-messenger_2020}
J.~Hinton and E.~Ruiz-Velasco, \emph{Multi-messenger astronomy with
  very-high-energy gamma-ray observations}, .

\bibitem{hagelin_perhaps_1984}
J.S.~Hagelin, G.L.~Kane and S.~Raby, \emph{Perhaps scalar neutrinos are the
  lightest supersymmetric partners},
  \href{https://doi.org/10.1016/0550-3213(84)90064-6}{\emph{Nuclear Physics B}
  {\bfseries 241} (1984) 638}.

\bibitem{kim_minimal_2007}
Y.G.~Kim and K.Y.~Lee, \emph{Minimal model of fermionic dark matter},
  \href{https://doi.org/10.1103/PhysRevD.75.115012}{\emph{Physical Review D}
  {\bfseries 75} (2007) 115012}.

\bibitem{pospelov_secluded_2008}
M.~Pospelov, A.~Ritz and M.~Voloshin, \emph{Secluded {WIMP} dark matter},
  \href{https://doi.org/10.1016/j.physletb.2008.02.052}{\emph{Physics Letters
  B} {\bfseries 662} (2008) 53}.

\bibitem{lee_singlet_2008}
K.Y.~Lee, Y.G.~Kim and S.~Shin, \emph{Singlet fermionic dark matter},
  \href{https://doi.org/10.1088/1126-6708/2008/05/100}{\emph{Journal of High
  Energy Physics} {\bfseries 2008} (2008) 100}.

\bibitem{iengo_sommerfeld_2009}
R.~{Iengo}, \emph{{Sommerfeld enhancement: general results from field theory
  diagrams}},
  \href{https://doi.org/10.1088/1126-6708/2009/05/024}{\emph{Journal of High
  Energy Physics} {\bfseries 2009} (2009) 024}
  [\href{https://arxiv.org/abs/0902.0688}{{\ttfamily 0902.0688}}].

\bibitem{dent_thermal_2010}
J.B.~{Dent}, S.~{Dutta} and R.J.~{Scherrer}, \emph{{Thermal relic abundances of
  particles with velocity-dependent interactions}},
  \href{https://doi.org/10.1016/j.physletb.2010.03.018}{\emph{Physics Letters
  B} {\bfseries 687} (2010) 275}
  [\href{https://arxiv.org/abs/0909.4128}{{\ttfamily 0909.4128}}].

\bibitem{slatyer_sommerfeld_2010}
T.R.~{Slatyer}, \emph{{The Sommerfeld enhancement for dark matter with an
  excited state}},
  \href{https://doi.org/10.1088/1475-7516/2010/02/028}{\emph{Journal of
  Cosmology and Astroparticle Physics} {\bfseries 2010} (2010) 028}
  [\href{https://arxiv.org/abs/0910.5713}{{\ttfamily 0910.5713}}].

\bibitem{tulin_beyond_2013}
S.~Tulin, H.-B.~Yu and K.M.~Zurek, \emph{Beyond collisionless dark matter:
  {Particle} physics dynamics for dark matter halo structure},
  \href{https://doi.org/10.1103/PhysRevD.87.115007}{\emph{Physical Review D}
  {\bfseries 87} (2013) 115007}.

\end{thebibliography}\endgroup

\end{document}